\documentclass[10pt,preprintnumbers,pra,aps,
nofootinbib,tightenlines,floatfix, twocolumn,superscriptaddress,longbibliography]{revtex4-2}

\usepackage[colorlinks=true]{hyperref}
\hypersetup{
  linkcolor  = TealBlue!90!Black,
  citecolor  = YellowGreen!90!Black,
  urlcolor   = Peach!90!Black,
  colorlinks = true,
}
\usepackage[all]{hypcap}
\usepackage{amsmath,amssymb}
\usepackage{graphicx}
\usepackage{balance}
\usepackage[utf8]{inputenc}
\usepackage{bm}
\usepackage{qcircuit}
\usepackage{csquotes}
\usepackage{empheq}
\usepackage{comment}
\usepackage[dvipsnames]{xcolor}
\usepackage[customcolors]{hf-tikz}
\usepackage{multirow}
\usepackage{enumitem}
\usepackage[top=1.8cm,bottom=1.8cm,left=1.45cm,right=1.45cm]{geometry}

\allowdisplaybreaks
\makeatletter
\renewcommand\onecolumngrid{
\do@columngrid{one}{\@ne}
\def\set@footnotewidth{\onecolumngrid}
\def\footnoterule{\kern-6pt\hrule width 1.5in\kern6pt}%
}
\makeatother

\newcommand\scalemath[2]{\scalebox{#1}{\mbox{\ensuremath{\displaystyle #2}}}}
\newcommand{\Tr}{\ensuremath{\text{Tr}}}

\newcount\hour \newcount\hourminute \newcount\minute
\hour=\time \divide \hour by 60
\hourminute=\hour \multiply \hourminute by 60
\minute=\time \advance \minute by -\hourminute
\newcommand{\mydate}{\ \today \ - \number\hour :\number\minute}

\begin{document}

\title{\texorpdfstring{Entanglement Structures in Quantum Field Theories: \\ Negativity Cores and Bound Entanglement in the Vacuum}{Entanglement Structures in Quantum Field Theories: Negativity Cores and Bound Entanglement in the Vacuum}}
\author{Natalie Klco}
\email{natklco@caltech.edu}
\affiliation{Institute for Quantum Information and Matter and Walter Burke Institute for Theoretical Physics,
California Institute of Technology, Pasadena CA 91125, USA}
\author{D.~H.~Beck}
\email{dhbeck@illinois.edu}
\affiliation{Department of Physics and Illinois Quantum Information Science and Technology (IQUIST) Center, University of Illinois at Urbana-Champaign,  Urbana, IL 61801, USA}
\author{Martin J.~Savage}
\email{mjs5@uw.edu}
\thanks{On leave from the Institute for Nuclear Theory.}
\affiliation{InQubator for Quantum Simulation (IQuS), Department of Physics, University of Washington, Seattle, WA 98195, USA}

\preprint{IQuS@UW-21-012}

\begin{abstract}
The many-body entanglement between two finite (size-$d$) disjoint vacuum regions of non-interacting lattice scalar field theory in one spatial dimension---a $(d_A \times d_B)_{\rm mixed}$ Gaussian continuous variable system---is locally transformed into a tensor-product \emph{core} of $(1_A \times 1_B)_{\rm mixed}$ entangled pairs.
Accessible entanglement within these core pairs exhibits an exponential hierarchy, and as such 
identifies the structure of dominant region modes 
from which vacuum entanglement could be extracted into a spatially separated pair of quantum detectors. 
Beyond the core, remaining modes of the \emph{halo} are determined to be AB-separable in isolation, as well as separable from the core.
However, state preparation protocols that distribute entanglement in the form of $(1_A \times 1_B)_{\rm mixed}$ core pairs are found to require additional entanglement in the halo that is obscured by classical correlations.
This inaccessible (bound) halo entanglement is found to mirror the accessible entanglement, but with a step behavior as the continuum is approached.  
It remains possible that alternate initialization protocols that do not utilize the exponential hierarchy of core-pair entanglement may require less inaccessible entanglement.
Entanglement consolidation is expected to persist in higher dimensions and may aid classical and quantum simulations of asymptotically free gauge field theories, 
such as quantum chromodynamics.
\end{abstract}
\date{\mydate}
\maketitle

{
\tableofcontents
}

\section{Introduction}

Quantum correlations are a signature characteristic and valuable resource for quantum computation, communication, sensing, information, and simulation~\cite{Feynman1982}.
Studies of  quantum correlations are 
providing unique insights~\cite{Ho:2015rga,Kharzeev:2017qzs,Baker:2017wtt,Cervera-Lierta:2017tdt,Berges:2018cny,Beane:2018oxh,GortonJohnson2019a,Beane:2019loz,Tu:2019ouv,Beane:2020wjl,Beane:2021zvo,Kharzeev:2021yyf,Robin:2020aeh,low2021symmetry,Gong:2021bcp,Roggero:2021asb,Mueller:2021gxd} and capabilities to robustly explore physical systems from the Standard Model and its quantum field theories (QFTs)~\cite{Banuls:2019bmf,Alexeev:2020xrq,Klco:2021lap,Zohar:2021nyc} to gravitation.
Conversely, because quantum fields naturally distribute entanglement, both dynamically and at spacelike separations~\cite{Halvorson:1999pz,Reeh1961,summers1985vacuum,summers1987bell1,summers1987bell2,VALENTINI1991321,Reznik:2002fz,Reznik:2003mnx}, their quantum correlations have long been recognized as a  resource in quantum information protocols.
For these reasons, studies of correlation structures in such systems---including entanglement entropies and distillable and bound entanglement---are at the forefront of research.

That the quantum field vacuum is entangled at spacelike separations is a simple consequence of the gradient coupling.
For example, in the idealized scenario of a constant field (i.e., vanishing gradient), local measurement establishes the global value of the field, and therefore collapses the wavefunction throughout the spatial volume.
Such a non-local response of measurement statistics is a sufficient condition for the identification of entanglement.
In general, the gradient operator in the Hamiltonian results in low-energy wavefunctions whose smoothness properties cause local measurements to modify measurement statistics throughout the volume.   This collapse, however, is partial with diminishing effect at increasing distance from the measurement location. 
Although diminishing, the effect on the statistics, and thus the entanglement, is non-zero throughout the volume.

Following decades of remarkable progress in unraveling the entanglement structures of QFTs~\cite{Srednicki:1993im,Holzhey:1994we,Callan:1994py,Audenaert:2002xfl,2004PhRvA..70e2329B,Calabrese:2004eu,Ryu:2006bv,kofler2006entanglement,Marcovitch:2008sxc,Casini:2009sr,Nishioka:2009un,Calabrese:2009ez,Calabrese:2009qy,Casini:2011kv,Calabrese:2012ew,Calabrese:2012nk,MohammadiMozaffar:2017nri,Coser_2017,Ruggiero:2018hyl,DiGiulio:2019cxv,Klco:2020rga,Roy:2020frd,Klco:2021biu}, leveraging both analytic techniques including the AdS/CFT correspondence and numerical techniques of lattice field theory and harmonic chains,
substantial experimental efforts have begun to quantify entanglement structures in low-dimensional QFTs, e.g., Refs.~\cite{Wang:2018kla,Roy:2020ppa,hung2021quantum}.
However, many quantum correlation properties that may guide simulation protocols, e.g., for scalar fields~\cite{kitaev2009wavefunction,Jordan:2011ci,Jordan:2011ne,Marshall:2015mna,Bermudez:2017yrq,Yeter-Aydeniz:2017ubh,Klco:2018zqz,Macridin:2018oli,Yeter-Aydeniz:2019scalar,Klco:2019yrb,Klco:2020aud,Barata:2020jtq,Macridin:2021uwn,Kurkcuoglu:2021dnw,Deliyannis:2021che} towards gauge
field theories~\cite{PhysRevA.73.022328,Zohar:2012xf,Banerjee:2012xg,Tagliacozzo:2012df,Stannigel:2013zka,Zohar:2013zla,Zohar:2014qma,Kasper:2015cca,Mezzacapo:2015bra,Bazavov:2015kka,Martinez:2016yna,Gonzalez-Cuadra:2017lvz,Zohar:2018cwb,Raychowdhury:2018osk,Stryker:2018efp,Zohar:2019ygc,Zache:2019xkx,Alexandru:2019nsa,Singh:2019uwd,Luo:2019vmi,Surace:2019dtp,Klco:2019evd,Mil:2019pbt,Ott:2020ycj,Haase:2020kaj,Dasgupta:2020itb,Shaw2020quantumalgorithms,Buser:2020cvn,Shen:2020coq,Surace_2020,VanDamme:2020rur,Yang_2020,Kreshchuk:2020dla,Kasper:2020akk,Kreshchuk:2020aiq,Kasper:2020owz,Uwe:CERN_2020,Davoudi:2020yln,Ciavarella:2021nmj,Bauer:2021gup,Klco:2021jxl,Stryker:2021asy,Davoudi:2021ney,Atas:2021ext,ARahman:2021ktn,aidelsburger2021cold}, require further study.

In this work, we describe correlation structures within the simplest quantum field---the time-independent 1D non-interacting lattice scalar field vacuum.
The vacuum state is infinite and pure.
In order to connect to properties and protocols that are experimentally viable in the laboratory with local detectors, we consider the entanglement between a pair of spatially disjoint, finite regions of the field. 
In a discretized representation appropriate for a simulation platform, this simple vacuum state is a set of harmonic oscillators, one at each lattice point, with gradient coupling that may be approximated through nearest-neighbor interactions.   With $d$ sites per region, the pair forms a $(d \times d)$-mode mixed continuous variable (CV) Gaussian quantum state. As such, well-developed tools of Gaussian quantum information are employed to perform the analysis.
With two operations in mind---initial state preparation for simulation, and extraction of entanglement from the vacuum as a resource---we study (numerically and for systems of modest size) the quantum and classical correlations in this vacuum state.  

We now outline the structure of the paper and the key ideas.
Section~\ref{sec:gaussianframework} provides a brief introduction to the scalar field and the spatially discretized representation of its vacuum as an infinite-mode Gaussian quantum system. 
Following this review, Section~\ref{sec:consolidation} develops and  introduces a constructive methodology for creating local unitary operations that concentrate the entirety of the two-region extractable~\footnote{Throughout, the adjectives \emph{extractable} or \emph{accessible} will be used to describe the entanglement that can be localized into $(1_A \times 1_B)_{\rm mixed}$ mode pairs through local operations, and thus viably coupled to low dimensional quantum detectors.} entanglement into a subset of \emph{core} pairs: entangled two-mode Gaussian states spanning the two regions, $(1_{A}\times 1_{B})_{\rm mixed}$.
These modes are composed of linear combinations of those representing lattice sites in each region.

Design of the consolidating local unitary entails transferring structural entanglement information from the full-system partially transposed (PT) basis back to the experimentally relevant physical basis.
The logarithmic negativity of the $n_c$ core pairs is subsequently organized in one-to-one correspondence with the $n_-$ PT symplectic eigenvalues associated with the negative  eigenvalues of the PT density matrix.
As such, the entanglement within the core pairs 
exhibits the same exponential hierarchy as the PT symplectic eigenvalues.

The separability structure of the post-consolidation quantum system is analyzed in Section~\ref{sec:separabilitystructure}.
Through sequential applications of techniques developed in Ref.~\cite{2001PhRvL..87p7904G}, the core is identified to be compatible with a tensor-product state of core pairs, and thus the transformed covariance matrix (CM, denoted by $\sigma$), is determined to be compatible with the form $\sigma_{(1_A \times 1_B)}^{\oplus n_c}\oplus \sigma_h$, up to additional Gaussian classical correlations.
This structure, combined with the localization of extractable entanglement to the core modes, allows  $(1 \times 1)_{\rm mixed}$ interpretations of entanglement measures to be applicable (e.g.,~\cite{Horodecki:1996nc,Vidal:2002zz,Simon:2000zz,Plenio:2005cwa}).
Note that this separability structure provides a valuable rearrangement of quantum correlations for any entanglement measure, i.e., it is not limited to the logarithmic negativity and PT eigensystem that is utilized to guide the local operator design.

The remaining modes after core-identification, $\sigma_h$, form a \emph{halo} space that carries no operationally accessible entanglement.  
In fact, upon tracing of the core modes, the halo system is entirely separable along the region-region bipartition.  
However, by imposing the core-halo separability structure identified above, the halo is constrained to be
a mixed state with distributed entanglement.
The excess entanglement is calculated in Section~\ref{sec:HaloEnt} as that of $\sigma_h$ after core-halo separability identification through the constructive methods of Ref.~\cite{2001PhRvL..87p7904G}. 
This entanglement is found to be commensurate, in both magnitude and decay as a function of disjoint region separation, with the accessible core entanglement and to exhibit approximate plateaus as the continuum is approached.  
While not available as an experimental quantum resource, the presence of this entanglement structure is expected to impact experimental state preparation and quantum simulation design, as discussed in the perspectives of Section~\ref{sec:perspectives}.

In Section~\ref{sec:inaccessibleHaloEnt}, the inaccessibility of the halo entanglement is discussed in the context of entanglement obscured by classical correlations.
Tracing of the volume outside the detection regions modifies the observable and underlying correlations in important ways.
These modifications arise due to the natural linkage between quantum and classical correlations.  
The lattice scalar field vacuum is shown to be a concrete and physically relevant example of uncoupling the relationship between separability and vanishing entanglement.
This type of inaccessible entanglement is present when disjoint regions of a free scalar field are prepared through distribution of $(1_A \times 1_B)_{\rm mixed}$ entangled pairs organized by the PT symplectic eigenvalues. 
Section~\ref{sec:boundEntConnection} discusses the implications of this observation in the context of bound entanglement.

In Section~\ref{sec:extractingSpacelike}, insights are provided for optimizing the distributed interaction profiles that allow extraction of entanglement from a field vacuum via local detectors composed of a few quantum degrees of freedom.
Such explorations also begin to quantify the landscape of many-body entangling operations that arise when arrays of quantum degrees of freedom are exposed to a common field. 
In the context of quantum sensing and error correction, this may yield constraints or guidance for coupling quantum or classical fields to a quantum sensor or simulation platform. 
Bringing these potential applications to fruition
necessitates detailed and operational understanding of the quantum correlation structures within the field and its representation(s) on quantum devices.

The presence of a sign problem in the analysis of the disjoint-region, free-field entanglement structure leads the numerical exploration of this system---a physically ubiquitous system without fine tunings---to be non-trivial.   
As such, the detailed description of this simplest quantum field, and the consequences of its local measurement, are viewed as an important baseline for understanding more complicated (e.g., interacting) systems where analysis tools are not so readily available.
Though the free scalar field is a greatly simplified version of those comprising the Standard Model, it has broad leading-order relevance, from quarks and gluons at high energies to pions at low energies.
This work is expected to contribute to the foundation for future explorations of the entanglement visible by local detectors in more complex fields.

\section{Gaussian Formulation of Lattice Scalar Field Vacuum}
\label{sec:gaussianframework}

The Hamiltonian density of the 1D non-interacting scalar field, $\phi$, of mass $m$ may be written as
\begin{equation}
 \mathcal{H} = \frac{1}{2} \pi^2 + \frac{1}{2} \left( \nabla \phi\right)^2 + \frac{1}{2} m^2 \phi^2 \ \ \ ,
\end{equation}
where $\pi$ is the momentum conjugate to the field operator satisfying the canonical commutation relation,
\begin{equation}
\left[ \phi(x), \pi(y)\right] = i \delta(x-y) \ \ \ .
\end{equation}
Upon latticization of the field and rescaling of the field operators by the (unit) lattice spacing, the dimensionless Hamiltonian density becomes
\begin{equation}
\hat{\mathcal{H}} = \frac{1}{2} \hat{\pi}^2 - \frac{1}{2} \hat{\phi} \hat{\nabla}_a^2 \hat{\phi} + \frac{1}{2} \hat{m}^2 \hat{\phi}^2 \ \ \ ,
\end{equation}
with a discretized Laplacian operator,
\begin{equation}
\hat{\nabla}^2_a \hat{\phi} = \hat{\phi}(x+1)+\hat{\phi}(x-1) - 2 \hat{\phi}(x) \ \ \ ,
\end{equation}
and commutation relation, 
\begin{equation}
\left[ \hat{\phi}(x), \hat{\pi}(y) \right] = i \delta_{xy} \hat{\mathbb{I}} \ \ \ .
\end{equation}
This Hamiltonian describes a chain of harmonic oscillators (one at each lattice point) with gradient couplings approximated by nearest-neighbor interactions, and approaches the free scalar field in the continuum limit.

The ground (vacuum) state of the latticized theory in the field-space basis is 
\begin{eqnarray}
\langle \vec\phi | \psi_0 \rangle = \frac{\det {\bf K}^{1/4}}{\pi^{N/4}} e^{-\frac{1}{2} \vec\phi^{\,T} {\bf K} \vec\phi } \ \ \ ,
\label{eq:groundState}
\end{eqnarray}
where $\vec{\phi} = \{\phi_0, \ldots \phi_{N-1}\}^T$ is the vector of lattice field variables spanning the $N$-site spatial volume.
For a one-dimensional field with periodic boundary conditions, the correlation matrix elements may be calculated as 
\begin{equation}
K_{ij} = \frac{1}{N} \sum_{k = 0}^{N-1} \cos\left(\frac{2 \pi k}{N}(i-j)\right) \sqrt{\hat{m}^2 + \hat{k}^2} \ \ \ ,
\end{equation} 
with lattice momentum $\hat{k} = 2 \sin\left(\frac{\pi k}{N}\right)$ in a finite difference representation of the gradient operator, where $i, j,$ and $k$ are integers.
The dense matrices of field and conjugate-momentum two-point correlators directly follow as
\begin{eqnarray}
G_{ij} &=& \langle \psi_0 | (\hat{\phi}_i - \bar\phi) (\hat{\phi}_j - \bar\phi) | \psi_0 \rangle = \frac{1}{2}\left(K^{-1}\right)_{ij} \ \ \ , 
\label{eq:GG}\\
H_{ij} &=& \langle \psi_0 | (\hat{\pi}_i - \bar\pi) (\hat{\pi}_j - \bar\pi) | \psi_0 \rangle = \frac{1}{2}K_{ij} \ \ \ ,
\label{eq:GH}
\end{eqnarray}
where the vacuum contains vanishing first-moments as well as vanishing $\langle \hat{\phi}_i \hat{\pi}_j\rangle$ expectation values.
Analytic expressions for the two point correlators in finite volume and the $N \rightarrow \infty$ thermodynamic limit, as utilized in the present calculations, have been detailed in, e.g., Refs.~\cite{2004PhRvA..70e2329B,Coser_2017,Klco:2021biu}.

As the free lattice scalar field vacuum is a multi-dimensional Gaussian wavefunction, it can be naturally described by a parameterized family of Gaussian quantum states.
In this formalism, the entanglement properties of CV Gaussian density matrices are entirely characterized by $2n$-dimensional CMs for a lattice of $n$ sites.
Matrix elements of the ground state CM are defined by the symmetric two-point vacuum expectation values,
\begin{equation}
  \sigma_{i,j} = \big\langle \left\{\hat{r}_i-\bar{r}_i, \hat{r}_j - \bar{r}_j \right\}_+ \big\rangle
  \ \ \ ,
  \label{eq:cm2exp}
\end{equation}
where $\hat{\mathbf{r}} = \{ \hat{\phi}_1, \hat{\pi}_1, \hat{\phi}_2, \hat{\pi}_2, \cdots, \hat{\phi}_{n}, \hat{\pi}_{n}\}$ is a vector of field and conjugate momentum operators, with $\bar{\mathbf{r}}$ the vacuum expectation values.
The positive definiteness of density matrices and adherence to the canonical commutation relations is expressed in phase space with the criterion,
\begin{equation}
   {\sigma} + i  {\Omega} \geq 0
  \ \ \ ,
  \label{eq:bonafide}
\end{equation}
a necessary and sufficient condition for a CM to be physical.
The scalar vacuum CM is an interleaved matrix of the form 
\begin{eqnarray}
\frac{1}{2}\sigma =  
\begin{pmatrix}
G_{11} & 0 &  G_{12} & 0 \ldots \\
0  & H_{11} & 0 & H_{12} \ldots \\
G_{21} & 0 & G_{22} & 0 \ldots \\
 & & \vdots & \\
 \end{pmatrix} \ \ \ .
 \label{eq:sigmaGH}
 \end{eqnarray}
Because CMs associated with reduced density matrices are simply sub-blocks of the full CM, 
these indices could represent the entirety of the lattice volume or a chosen spatial subset. 
In the present work, the indices of the vector $\mathbf{r}$ will be those within two disjoint regions of the field.

Transformations of the system corresponding to unitary operations in Hilbert space are effected by symplectic transformations of the CM,
\begin{equation}\sigma \rightarrow S \sigma S^T \quad , \quad {S} \in Sp(2n, \mathbb{R}) \ \ \ ,
\label{eq:symplectic}
\end{equation}
where $S$ is a symplectic operator.
A symplectic matrix, $\Omega$, encodes the commutation relations,
\begin{equation}
  \left[ \mathbf{r}, \mathbf{r}^T \right] = i {\Omega}_{n}
  \quad , \quad
   {\Omega}_{n} = \bigoplus_{j = 1}^{n} i  {\tau_y}
  \ \ \ ,
\end{equation}
with $\tau_y$ the second Pauli matrix.
By construction, these commutators are invariant under symplectic transformations,
\begin{equation}
   {S}  {\Omega}  {S}^T =  {\Omega}  
  \ \ \ .
\end{equation}
As such, all entanglement rearrangements developed in Section~\ref{sec:consolidation} will preserve the physical uncertainty relations.

Through global symplectic operations spanning the space, every CM can be diagonalized to the Williamson normal form~\cite{Williamson1936},
\begin{equation}
   {\sigma} =  {S} \left( \bigoplus_{j = 1}^{n} \nu_j {\mathbb{I}}_2 \right)  {S}^T
  \ \ \ ,
  \label{eq:symplecticnormal}
\end{equation}
where the diagonal elements $\nu_j \geq 1$ are the symplectic eigenvalues: invariants under symplectic transformations characterizing the state.
If $\nu_j = 1 \ \forall j$, the state is pure.
While a constructive determination of the symplectic transformation to the Williamson normal form is possible, the spectrum of symplectic eigenvalues may be calculated as,
\begin{equation}
  \boldsymbol{\nu} = \text{spec}|i {\Omega} {\sigma}| = 2\ \text{spec}\sqrt{ {G} {H}}
  \ \ \ ,
\end{equation}
where the first spectrum contains a doubling of eigenvalues symmetrically distributed about zero prior to the absolute value.

In the following, the logarithmic negativity is used as quantitative guidance in the design of entanglement reorganizations. 
The logarithmic negativity quantifies the deviation of a density matrix from physicality upon local application of the transpose operation~~\cite{Horodecki:1996nc,Vidal:2002zz,Simon:2000zz,Plenio:2005cwa}.
The CV PT operation executes a local reversal of motion~\cite{Simon:2000zz}, negating the conjugate momentum operators on one side of the chosen mode bipartition, $\hat{\pi}_{j} \rightarrow -\hat{\pi}_{j} \ \forall \ j \in \left\{B\right\}$.
For $n_{a,b}$ modes on either side of the bipartition, this may be implemented through the transformation $\sigma^\Gamma = {\Gamma} {\sigma} {\Gamma}$ with
\begin{equation}
   {\Gamma} = \left(\bigoplus_{j = 1}^{n_a} {\mathbb{I}}_{2}\right) \oplus \left( \bigoplus_{j = n_a+1}^{n_a+n_b} \tau_z \right)
  \ \ \ ,
  \label{eq:gammaPTop}
\end{equation}
where $\tau_z = \text{diag}(1,-1)$ is the third Pauli matrix.
In the present application of symmetric disjoint vacuum regions, $n_a = n_b = d$.
Note that ${\Gamma}$ is not symplectic, and thus is not a unitary operation in the Hilbert space.
As such, the PT symplectic spectrum,
\begin{equation}
  \boldsymbol{\nu}^\Gamma = \text{spec}|i {\Omega} {\sigma}^\Gamma| = 2\text{spec}\sqrt{ {G} {H}^\Gamma}
  \ \ \ ,
\end{equation}
is not invariant under this map, i.e., $\boldsymbol{\nu} \neq \boldsymbol{\nu}^{\Gamma}$.

All physical states have non-negative eigenvalues in their density matrix representations.
When the transpose operator is applied to the entirety of a quantum state or to a portion delineated by a separable bipartition, it acts as a positive map, i.e., the density matrix remains physical with positive or vanishing eigenvalues. 
In the CM formalism, the corresponding criteria is that the PT symplectic eigenvalues remain lower bounded by~1.
However, when the transpose operator is applied to a portion delineated by an inseparable bipartition, entanglement can generate negative eigenvalues in the post-PT density matrix.
The corresponding effect on the CM is the presence of PT symplectic eigenvalues less than one, $\nu_j^\Gamma <1$.
The logarithmic negativity is additive in the basis of PT normal modes, and completely characterized by the PT symplectic eigenvalues,
\begin{align}
  \mathcal{N} &= -\sum_{j = 1}^{n} \log_2 \min(\nu_j^\Gamma, 1) \equiv \sum_{j = 1}^{n} \mathcal{N}_j \ \ \ , \\
  &= -\sum_{j = 1}^{n_-} \log_2 \nu_j^\Gamma
  \ \ \ ,
  \label{eq:neg}
\end{align}
where $n_-$ is the number of PT symplectic eigenvalues less than 1.

\section{Local Transformation for Entanglement Simplification}
\label{sec:consolidation}

\begin{figure}
  \includegraphics[width=0.9\columnwidth]{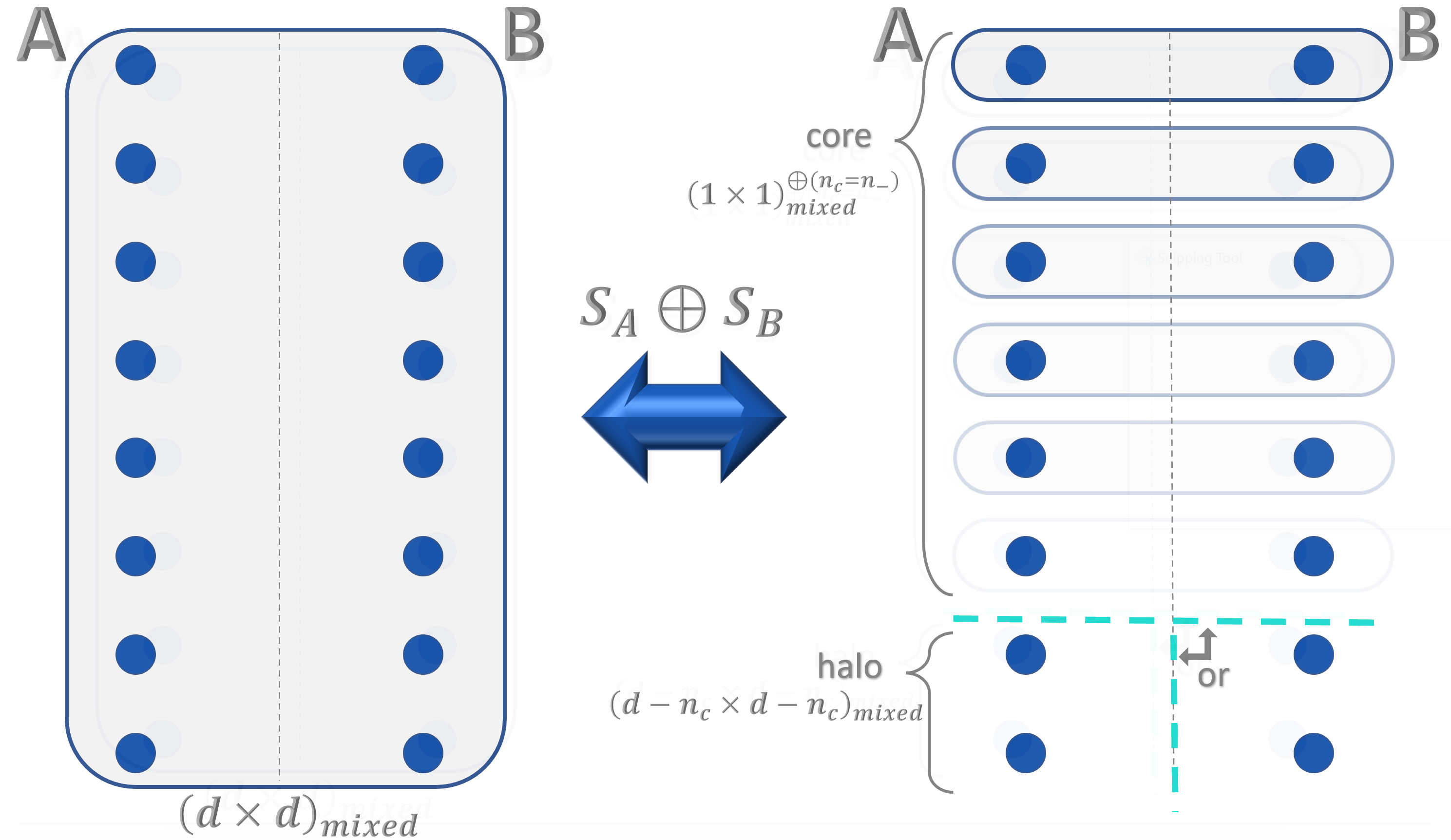}
  \caption{
  Diagrammatic representation of consolidating two regions of the vacuum into a core of accessible entanglement and halo with inaccessible entanglement.
  Prior to consolidation (left), the two regions are entangled with multi-mode $(d\times d)_{\rm mixed}$ entanglement. After consolidation through local transformation, the negativity is captured in a core of $(1\times 1)_{\rm mixed}$ pairs that manifest the exponential hierarchy in entanglement contributions~(\ref{app:structureNegCore}).
  The dashed (teal) lines indicate two partitions that are individually, but not simultaneously, separable.
  }
  \label{fig:corediagram}
\end{figure}

Extracting a pair of regions from the vacuum of a lattice scalar field theory generates a $(d_A \times d_B)_{\rm mixed}$ continuous variable (CV) state parameterized by region size $d$, their separation $\tilde{r}$, and the mass $m$ of the field~\footnote{Numerical values for these quantities are presented in lattice units.}.  
Although a number of entanglement structure reorganizations have been devised for CV systems~\cite{Duan_2000,2000quant.ph..7061G,giedke2001distillability,2001PhRvL..86.3658W,2003PhRvA..67e2311B,2004PhRvA..70e2329B,Serafini_2005,2008PhRvL.100g0505W,2016arXiv161205215L,Serafini2017}, the transformation developed here achieves clear separation and pairwise organization of accessible and inaccessible entanglement.  
This is depicted in Fig.~\ref{fig:corediagram}, where the $A$ and $B$ spaces label the two regions of the vacuum.
The present restructuring of the CM is achieved through {\it local} symplectic transformations that are calculated from a subset of the symplectic eigenvectors of the full PT CM (or product $GH$).
The subset corresponds to the $d$ lowest PT symplectic eigenvalues, where  $n_-$ of the eigenvalues contribute to the negativity with $n_- \leq d$.
Within this subset, components corresponding to a single region are extracted, and the unitarity is repaired using a Gram-Schmidt orthogonalization process modified to incorporate symplectic inner products.
The resulting symplectic operations are applied locally to each region,
\begin{equation}
  \sigma' = (S_A \oplus S_B)\sigma (S_A \oplus S_B)^T \ \ \ ,
\end{equation}
transitioning from the original fully entangled modes of the vacuum regions to an organized \emph{core} of $n_c = n_-$ entangled $(1 \times 1)_{\rm mixed}$ states containing the accessible entanglement and a \emph{halo} space containing additional inaccessible entanglement.
Prior to describing the consolidation transformation in further detail, Sections~\ref{sec:catalog} and~\ref{app:structureNegCore} discuss previous works on entanglement reorganizations and the structure of PT symplectic eigenvalues that will be leveraged in the present consolidation.

\subsection{Known CV Entanglement Reorganizations}
\label{sec:catalog}

In the landscape of Gaussian continuous variable (CV) quantum systems, a number of entanglement structure transformations are known.
For example, any $(\ell \times n)_{\rm mixed}$ state with a permutation symmetry among the modes $\ell$ (monosymmetric) can be transformed to a $(1 \times n)_{\rm mixed}$ state and a set of $(\ell-1)$ unentangled single modes~\cite{2016arXiv161205215L,Serafini2017}.
This reorganization associates $(\ell \times n)_{\rm mixed}$ monosymmetric Gaussian states with the convenient features of $(1 \times n)_{\rm mixed}$ Gaussian states, such as the PPT criterion being both necessary and sufficient for determining separability~\cite{2001PhRvL..86.3658W,Serafini2017}.
In an extension, bi-symmetric $(\ell \times n)_{\rm mixed}$ states, in which the $\ell$ and $n$ modes exhibit independent permutation symmetries, can be locally transformed into a $(1 \times 1)_{\rm mixed}$ state and a set of $(\ell+n-2)$ unentangled single modes~\cite{Serafini_2005}.
Bi-symmetric states thus have the entanglement properties of two-mode Gaussian states~\cite{Simon:2000zz,Duan_2000,2000quant.ph..7061G}.
Without local symmetries, $(\ell \times n)_{\rm pure}$ Gaussian states can be locally transformed to a set of $(1 \times 1)_{\rm pure}$ states and unentangled single modes through local symplectic operations~\cite{2003PhRvA..67e2311B,2004PhRvA..70e2329B}.
This is also possible for $(\ell \times n)_{\rm mixed}$ isotropic states, 
where the matrix of symplectic eigenvalues is proportional to that of pure states, i.e., to the identity~\cite{2004PhRvA..70e2329B}.
For $(\ell \times n)_{\rm mixed}$ Gaussian states without local or global symmetries, a local unitary has been identified that organizes the inter-region two-body entanglement into single-mode A-B pairs~\cite{2008PhRvL.100g0505W}, though higher-body entanglement through intra-region correlations may remain.

While the vacuum state of the free lattice scalar field is not amenable to, or is insufficiently restructured by, the above reorganizations,
in the following we demonstrate the existence of a transformation that analogously simplifies its entanglement structure.
The consolidating transform will allow the total logarithmic negativity of the $(d \times d)_{\rm mixed}$ state to be organized into a set of $(1\times 1)_{\rm mixed}$ mode pair contributions.

\subsection{Partially Transposed Symplectic Structure}
\label{app:structureNegCore}

\begin{figure}
  \includegraphics[width = 0.95\columnwidth]{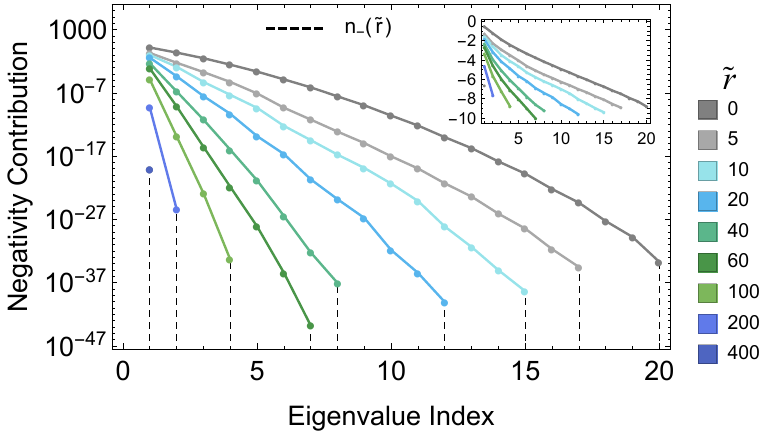}
  \caption{Logarithmic negativity contributions for entanglement between vacuum regions of the one-dimensional massive scalar field with $d = 30$ and $m = 0.003$ as a function of partially transposed symplectic eigenvalue index for a series of separations, $\tilde{r}$.
  The inset shows the same values with scaled y-components as $y' = -\sqrt{-\log{y}}$ such that linear behavior indicates Gaussian decay structure.}
  \label{fig:corestructure}
\end{figure}
As described with constructive techniques in the following, the local operators designed to consolidate the entanglement of the $(d \times d)_{\rm mixed}$ field regions are informed by the PT symplectic eigensystem.
It has been observed~\cite{Klco:2020rga,Klco:2021biu} that the number of PT symplectic eigenvalues contributing to the logarithmic negativity, $n_-$, decreases rapidly with the spatial separation between vacuum regions, $\tilde{r}$.
At large separation between the $\tilde{r}$ boundaries where the $(d \times d)$ and $(d-1 \times d-1)$ vacuum regions become separable~\cite{Audenaert:2002xfl,2004PhRvA..70e2329B,kofler2006entanglement,Marcovitch:2008sxc,Calabrese:2009ez,Calabrese:2012ew,Calabrese:2012nk,MohammadiMozaffar:2017nri,Coser_2017,Klco:2019yrb,DiGiulio:2019cxv,Klco:2021biu}, the logarithmic negativity is characterized by a single PT symplectic eigenvector (e.g., single point for $\tilde{r} = 400$ with eigenvalue index of $1$ in Fig.~\ref{fig:corestructure}) that is entirely delocalized across the modes within each region, i.e., all smaller subsets of these modes are separable~\cite{Klco:2021biu}.
At reduced field separations, the negativity in the PT basis has multiple contributions with an exponential hierarchy, subsequently associated with $(k \times k)$-body entanglement between the field regions for $\ell \leq k \leq d$, with $\ell$ an $\tilde{r}$-dependent constant that becomes equal to one at $\tilde{r}=0$.
These features of the PT basis are shown in Fig.~\ref{fig:corestructure} for a representative system of pixelated vacuum regions.

In light of the exponential hierarchy, it is advantageous to consider the construction of a local symplectic operator that localizes the PT symplectic eigenvector associated with largest negativity contribution into a pair of physical modes, one in each region of the bipartite CV system.
By further building the local operator leveraging the PT basis and a symplectic Gram-Schmidt (sGS) procedure (see Section~\ref{sec:symplecticGS}) to enforce unitarity in the Hilbert space, 
mode pairs of the transformed system can be designed to carry the same entanglement hierarchy as the PT logarithmic negativity contributions.

\subsection{Constructive Consolidation Protocol}
\label{app:explicitconstruction}
Utilizing the CM framework of Section~\ref{sec:gaussianframework}
with 4d-dimensional
CMs for a lattice of d sites in each of two regions, a technique
of entanglement consolidation is now described for the vacuum of lattice scalar field theory.
For the matrices of two-point correlation functions in Eqs.~\eqref{eq:GG} and~\eqref{eq:GH} calculated for a lattice in the limit of infinite spatial volume,
consider the left and right eigenvectors of $GH^\Gamma$ calculated as,
\begin{equation}
  H^\Gamma G \bar{\mathbf{v}}_{\phi,j} = \lambda_j^2 \bar{\mathbf{v}}_{\phi,j} \qquad , \qquad
  GH^\Gamma \bar{\mathbf{v}}_{
  \pi,j} = \lambda_j^2 \bar{\mathbf{v}}_{\pi,j} \ \ \ ,
\end{equation}
where the $\lambda_j$ are the PT symplectic eigenvalues calculated as the eigenvalues of $|i\Omega \sigma^\Gamma|$.
While use of this basis organizes the distribution of negativity, one may choose a standard normalization for concreteness to set the first elements positive $\left(\bar{\mathbf{v}}_{\phi,1}\right)_{1}, \left(\bar{\mathbf{v}}_{\pi,1}\right)_{1} \geq 0$ and,
\begin{equation}
  \bar{\mathbf{v}}_{\phi,j}^T G \bar{\mathbf{v}}_{\phi,j} = \lambda_j \ \  , \ \ 
  \bar{\mathbf{v}}_{\pi,j}^T H^\Gamma \bar{\mathbf{v}}_{\pi,j} = \lambda_j \ \ , \ \ \bar{\mathbf{v}}_{\phi}.\bar{\mathbf{v}}^T_{\pi} = \mathbb{I}
  \ \ \ ,
\end{equation}
where $\bar{\mathbf{v}}_{\phi}, \bar{\mathbf{v}}_{\pi}$ without further indices are stacked matrices with rows corresponding to the PT symplectic eigenvectors.
Subsequently, a local transformation basis is extracted from this eigenvector set as,
\begin{equation}
   {\bar{S}}_A = \begin{pmatrix}
  \begin{pmatrix}
    & \mathbf{\bar{v}}^{\vec{0}}_{\phi,1} &
  \end{pmatrix}_A \\
  \begin{pmatrix}
    & \mathbf{\bar{v}}^{\vec{0}}_{\pi,1} &
  \end{pmatrix}_A \\ \vdots \\
  \begin{pmatrix}
    & \mathbf{\bar{v}}^{\vec{0}}_{\phi,d} &
  \end{pmatrix}_A \\
  \begin{pmatrix}
    & \mathbf{\bar{v}}^{\vec{0}}_{\pi,d} &
  \end{pmatrix}_A
  \end{pmatrix} \ \ \ ,
  \label{eq:sAbarGH}
\end{equation}
where $d$ is the number of modes per region of the field and the subscript indicates that the eigenvector has been cut at the dimensionality of the first field region.
Importantly, these vectors are not eigenvectors of region $A$ alone that would serve to diagonalize the local region CMs.
To construct a symplectic transformation that does not mix $\phi$- and $\pi$-coordinates, zeros are interleaved into each of the half-eigenvectors (denoted by superscript $\vec{0}$), as in the form of Eq.~\eqref{eq:sigmaGH} (with $G, H$ exchanged for $\bar{\mathbf{v}}_{\phi}, \bar{\mathbf{v}}_{\pi}$, respectively).
While there are $2d$ eigenvectors with positive $\lambda_m$ eigenvalue, only the half with the lowest PT symplectic eigenvalues is retained to span the local transformation basis.
For any $\left(n_a+n_b\right)$-mode Gaussian system, the maximum number of negative eigenvalues possibly contributing to the negativity is $\min\left(n_a, n_b\right)$, the minimum number of modes in either region~\cite{Serafini2017}.
As such, the choice of utilizing the lowest $d$ left and right eigenvectors to initialize a local symplectic transformation captures all available negativity.
Note that while entanglement consolidating transformations for pure states can be informed solely through the local CMs~\cite{2003PhRvA..67e2311B,2004PhRvA..70e2329B}, this consolidation for a mixed state depends upon full CM diagonalization, consistent with extensions of entanglement quantifications from pure to mixed states.

While offering a transformation between local basis vectors, $\mathbf{e}$, and the Gaussian modes dominantly contributing to the negativity,
\begin{equation}
\{\mathbf{e}_{\phi,j}, \mathbf{e}_{\pi,j}\} \stackrel{ \bar{S}_A}{\leftrightarrow} \left\{ \left( \mathbf{\bar{v}}_{\phi,j}\right)_A , \left( \mathbf{\bar{v}}_{\pi,j} \right)_A \right\}\ \  \forall \ j \in \{1, \cdots, d\} \ \ \ ,
\end{equation}
the operator as described is not symplectic.
To rectify this property, the symplectic Gram-Schmit (sGS) procedure, discussed in Appendix~\ref{sec:symplecticGS}, is applied to $\bar{S}_A$ to produce a symplectic transformation $S_A$ in the first region.

The associated transformation in the second region, $S_B$, can be determined through symmetry transformation of ${S}_A$, i.e., a reversal of Gaussian mode order.
The local, symplectic operator, $ {S} = {S}_A \oplus  {S}_B$, thus transforms the negativity basis into pairs of basis vectors within each region,
\begin{equation}
   {\sigma}' =  {S}  {\sigma}  {S}^{T} \qquad , \qquad  {S} =  {S}_A \oplus {S}_B \ \ \ ,
\end{equation}
consolidating the extractable entanglement into corresponding pairs of Gaussian modes.
The locally-transformed ${\sigma}'$ exhibits entanglement that manifests the negativity core in the first $n_c = n_-$ modes in either region.

After the above transformation defined by the PT symplectic eigenvectors, the negativity described in Eq.~\eqref{eq:neg} is captured in $(1_A\times 1_B)_{\rm mixed}$ pairs of local modes as governed by the PT symplectic eigenvalues,
\begin{align}
  \mathcal{N}\left[ {\sigma}'(1,2d)\right] &= \mathcal{N}_1  \ \ \ , \nonumber \\ \mathcal{N}\left[ {\sigma}'(2,2d-1)\right] &= \mathcal{N}_2  \ \ \ ,  \nonumber \\ &\vdots \nonumber \\
    \mathcal{N}\left[ {\sigma}'(d,d+1)\right] &= \mathcal{N}_d \ \ \ ,
    \label{eq:negativityadditivity}
\end{align}
where $ {\sigma}'(i, j)$ is the 2-mode reduced CM of $ {\sigma}'$ with the $i^{\text{th}}$ and $j^{\text{th}}$ modes.
Subsequently,  negativity can be extracted through $(1\times 1)_{\rm mixed}$-mode techniques for each of the $n_c = n_-$ core pairs.
See Appendix~\ref{app:consolidationEX} for detailed examples of this entanglement consolidation through local symplectic operation.

In order to describe a physical unitary transformation on the Hilbert space, a symplectic operation ${S}$ has been determined acting upon the CM.
In order for that symplectic operation to generate the hierarchy of negativity contributions observed in the field, the symplectic eigenbasis of the PT CM has been employed in the calculation of $ {S}$.
While this connection is not generically valid---even for local arbitrary operations $ {S}_{arb} =  {S}_{arb,A} \oplus  {S}_{arb,B}$, symplectic operations do not generically commute with the active portion of the partial transposition operator $\left[  {S}_{arb,B}, \tau_z^{\oplus n_B}\right] \neq 0$---the particular choice of local operation in Eq.~\eqref{eq:sAbarGH} that acts as a separate transformation in the local $\phi$ and $\pi$ quadrature bases does allow commutation $\left[  {S}, \Gamma \right] = 0$.  In a system without mixed expectation values ($\langle \phi_i \pi_j\rangle = 0 \ \forall i,j$),  a negativity-informed transformation can thus be determined for application on the physical system.

\subsection{Broader Applicability: D-Type CMs}
\label{sec:broaderapp}

Though the consolidating transformation has not been designed for generic mixed states, 
some aspects of the core structure apply to a broader class of mixed Gaussian states.
For example, consider $D$-type CMs, which may be decomposed as,
\begin{align}
  \sigma_D &= D(n).\Sigma.D^T(n)  \ \ \ , \\
  D(n) &= \begin{pmatrix}
    C & 0 \\
    0 & \left(C^{-1}\right)^T
  \end{pmatrix} \quad , \quad  C \in GL\left(n; \mathbb{R}\right) \ \ \ ,
\end{align}
where $D(n)$ is one generator of the symplectic group written in a basis of $\{ \phi_1, \cdots \phi_n, \pi_1, \cdots \pi_n\}$, and $\Sigma$ is a diagonal matrix with elements $\Sigma_i \geq 1$.
This class of mixed CMs share the property of vanishing $\langle \phi_i \pi_j\rangle$ matrix elements.
This condition, as discussed above, allows the PT operation to commute with the consolidating local transformation, and thus allows the PT symplectic spectrum to inform entanglement organizing unitaries in the physical space.
While not a sufficient condition to organize accesible entanglement into separate $(1\times 1)_{\rm mixed}$ mode pairs as described in Eq.~\eqref{eq:negativityadditivity}, the first $k$ pairs capture the logarithmic negativity associated with the first $k$ symplectic eigenvalues as,
\begin{equation}
  \mathcal{N}\left( \sigma_D' (1, \cdots, k, 2n-k+1, \cdots, 2n)\right) = \sum_{j = 1}^k \mathcal{N}_j \ \ \ ,
\end{equation}
with $\sigma_D'(1, \cdots, k, 2n-k+1, \cdots, 2n)$ the $2k$-mode reduced $D$-type CM after application of the negativity consolidating transformation.
This \emph{accumulative} core structure, relevant beyond the lattice scalar field vacuum, arises as a result of the sGS procedure used to generate a valid symplectic transformation for negativity consolidation from the PT symplectic eigenvectors.

\section{Post-consolidation Entanglement Structure}

One canonical pure-state convex decomposition of a Gaussian CV mixed-state CM is characterized by the identification of an additive deviation from purity,
\begin{equation}
\sigma = \sigma^{(p)} + Y \ \ \ ,
\label{eq:mixedpurenoise}
\end{equation}
where $\sigma^{(p)}$ is a pure CM and $Y$ is a positive semidefinite (PSD) matrix of classical correlations.
Because this structure describes an ensemble of states with first-moment displacements drawn from the multivariate Gaussian distribution governed by $Y$, the PSD addition is also referred to as a noise matrix. 
Operationally, this decomposition conveys one tenable mixed-state preparation strategy, i.e., through unitary operations on the quantum vacuum subsequently modified by classically sampled, correlated displacements in phase space.

When Gaussian mixed states are separable, their CMs may be decomposed in underlying block diagonal form as $\sigma_{\rm sep} = \sigma_a \oplus \sigma_b + Y$~\cite{2001PhRvL..86.3658W,2016arXiv161205215L,Serafini2017}.
This decomposition leads naturally to the following necessary and sufficient condition for Gaussian-state separability~\cite{2001PhRvL..86.3658W},
\begin{equation}
  \sigma_{\rm sep} \geq \sigma_a \oplus \sigma_b \ \ \ .
  \label{eq:sepcondition}
\end{equation}
While efficient determination of separability is in general a challenging problem for QMB states, the separability-preserving flow techniques presented in Ref.~\cite{2001PhRvL..87p7904G} provide a practical numerical approach for not only determining the separability of Gaussian states, but also identifying a viable $\sigma_{a,b}$ underlying tensor product state when available (see Appendix~\ref{app:sepcriterion} for further details).  
In the following, the techniques of Ref.~\cite{2001PhRvL..87p7904G} will be utilized to characterize the simplified entanglement structure between regions of the scalar field vacuum after consolidation.


\subsection{Separability Structure}
\label{sec:separabilitystructure}

After the local negativity-consolidating transformation presented in Section~\ref{sec:consolidation},  disjoint regions of the scalar vacuum are identified to be separable along the core-halo bipartition.
As such, the A-B quantum system is expressible as a tensor-product state immersed in classical correlations as,
\begin{equation}
\sigma' = \sigma_c \oplus \sigma_h + Y  \quad , \quad Y = \begin{pmatrix}
  \alpha & \beta \\
  \beta^T & \delta
\end{pmatrix} \ \ \ .
\label{eq:core-halosep}
\end{equation}  
This bipartition corresponds to the horizontal dashed teal line in the diagram at right of Fig.~\ref{fig:corediagram}.
With all the negativity accounted for in the core, the halo degrees of freedom, when isolated, are subsequently found to be separable along the A-B region divide,
\begin{equation}
\Tr_{\rm c} (\sigma') = \sigma_h + \delta = \sigma_{h,A} \oplus \sigma_{h,B} + Y_h \ \ \ .
\label{eq:halo-halosep}
\end{equation}
This observation corresponds to the vertical dashed teal line in the halo space of the diagram at right of Fig.~\ref{fig:corediagram}.

For pure states, these two observations of sequential separability could be combined to conclude the separability of, for example, $\sigma_{h,A}$ from the rest of the system.
However, this logic does not apply to mixed states, where the classical correlations of $Y$ and $Y_h$ can be incompatible.
As notated by the \emph{exclusive or} between these separable bipartitions in Fig~\ref{fig:corediagram}, the separability observations of Eqs.~\eqref{eq:core-halosep} and~\eqref{eq:halo-halosep} for the consolidated scalar vacuum regions are not compatible.  
In particular, the negativity between the portion of the halo in either region and the rest of the system is found to be non-zero. 
This is a sufficient observation to conclude that $\sigma_{h,A}$ is connected to the remaining many-body system with entanglement, and therefore that this core-halo separated state contains inaccessible entanglement in the halo.

More explicitly, attempts to combine the above separability observations lead to
\begin{align}
  \sigma' &= \sigma_c \oplus  \sigma_{h,A} \oplus \sigma_{h,B} + Y'  \ \ \ , \label{eq:sigprime3} \\
  Y' &\equiv \mathbf{0}_c \oplus Y_h - \mathbf{0}_c \oplus \delta + Y \nonumber \ \ \ .
\end{align}
If the individual separability calculations are satisfied, $Y$, $Y_h$, and $\delta$ are PSD matrices.
Upon simplification,  a necessary and sufficient condition for the three-part simultaneous separability in Eq.~\eqref{eq:sigprime3} becomes the identification of classical correlations such that,
\begin{equation}
  Y' = \begin{pmatrix}
    \alpha & \beta
    \\
    \beta^T & Y_h
  \end{pmatrix} \geq 0 \ \ \ .
  \label{eq:simsepineq}
\end{equation}
Replacing halo contributions identified in the core-halo partition, $\delta$ in Eq.~\eqref{eq:core-halosep}, with that compatible with halo$_A$-halo$_B$ separability, $Y_h$ in Eq.~\eqref{eq:halo-halosep}, must result in a valid classical correlation matrix.
In contrast, if $Y' < 0$, as is presently observed, the partitions are not simultaneously separable.

Focusing next on the structure of the core, the additivity of the negativity among the $(1 \times 1)_{\rm mixed}$ mode pairs as expressed in Eq.~\eqref{eq:negativityadditivity} suggests that these core pairs are all separable from each other.
Furthermore, upon tracing, all bipartitions of $k$-pairs by $\ell$-pairs for $k+\ell \leq n_c$ are identified to be separable.
The remaining question is whether these core-pair separability observations are simultaneously compatible with the core-halo separability, i.e., whether the transformed CM can be decomposed as
\begin{equation}
  \sigma' = \sigma_{c,1} \oplus \sigma_{c,2} \oplus \cdots \oplus \sigma_{c,n_c} \oplus \sigma_{h} + Y \ \ \ ,
  \label{eq:simultaneouscoresep}
\end{equation}
where $\sigma_{c,i}$ is a $(1_A \times 1_B)_{\rm mixed}$ entangled CM.
This question addresses multipartite simultaneous separability of the core pairs.
While progress has been made in extending CV separability criteria to tripartite systems~\cite{Giedke_2001,Adesso_2006}, the $(n_c+1)$-partite nature of this question is beyond current direct techniques.
However, sequential application of the bipartite procedure of Ref.~\cite{2001PhRvL..87p7904G} (see Appendix~\ref{app:sepcriterion}) provides a methodology to pursue numerical identification of viable decompositions.
For modest sized systems, at least one of the $n_c!$ different bipartition orderings is reliably found to yield tensor-product core pairs in the form of Eq.~\eqref{eq:simultaneouscoresep}.
Thus, Eq.~\eqref{eq:simultaneouscoresep} represents the numerically determined final separability structure after entanglement consolidation.

\subsection{Entanglement in Halo}
\label{sec:HaloEnt}

\begin{figure}
  \includegraphics[width = 0.98\columnwidth]{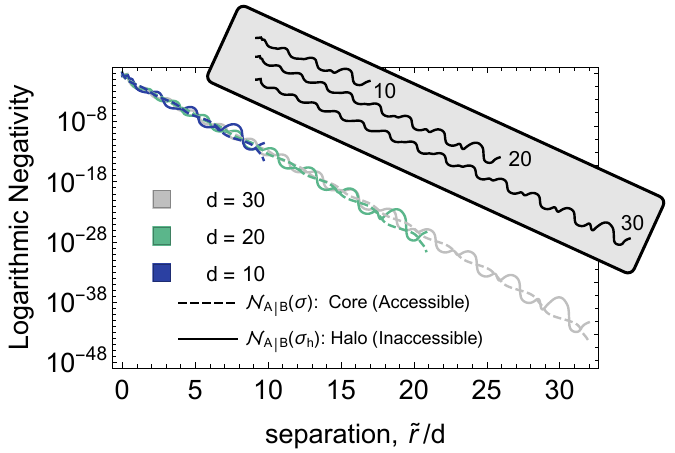}
  \caption{Logarithmic negativity quantifying core (accessible, dashed lines) and halo (inaccessible, solid lines) entanglement between two disjoint regions of the latticized one-dimensional massive ($m = 0.003/d$) scalar field vacuum as a function of dimensionless separation $\tilde{r}/d$, where $\tilde{r}$ is the integer separation in lattice units and $d$ is the pixelated size of each region.  The inset diagrammatically reproduces the solid lines of the main panel, vertically displaced to enhance visibility of emergent structure.
  }
  \label{fig:boundent}
\end{figure}

The incompatibility of $\delta$ and $Y_h$ discussed in Eqs.~\eqref{eq:core-halosep} and~\eqref{eq:halo-halosep} may also be addressed through the entanglement of $\sigma_h$ alone.
In particular, this observable further quantifies the non-simultaneous separability present for the core-halo and halo$_A$-halo$_B$ bipartitions.

The solid lines of Fig.~\ref{fig:boundent} show the negativity of $\sigma_h$  after identification of classical correlations compatible with core-halo separability in Eq.~\eqref{eq:core-halosep} utilizing the separability flow techniques of Ref.~\cite{2001PhRvL..87p7904G}.
The dashed lines show the core entanglement for reference.  By design of the local transformation presented in Sec.~\ref{app:explicitconstruction}, the core entanglement is equal to the full A-B negativity.
For the purpose of demonstration, we have chosen a mass function, $m=3 \times 10^{-3}/d$, such that the correlation length across the regions is held fixed as $d$ increases and reduces the impact of pixelation toward the continuum  limit.

While many solutions in the form of Eq.~\eqref{eq:core-halosep} exist, the decompositions generated by the techniques of Ref.~\cite{2001PhRvL..87p7904G} appear to produce a tight upper bound for the halo entanglement.
The family of decompositions purifying $\sigma_h$ result in the same or larger values of the halo negativity.  Further, numerical searches of $\sigma_h$ perturbations that satisfy the necessary constraints are found to also produce only increases.
Thus, this quantity should be interpreted as an estimate of the entanglement distributed between regions in the halo, with a value of zero incompatible with the implementation of the core-halo separability of Section~\ref{sec:separabilitystructure}.

At fixed region pixelation $d$,
Fig.~\ref{fig:boundent} illustrates that the vanishing of halo entanglement at large separations coincides with the boundary at which regions become separable.
Both the magnitude and exponential decay are consistent with that previously calculated for the full A-B negativity~\cite{Marcovitch:2008sxc}.
The core and halo entanglements fluctuate with amplitudes that diminish toward the continuum limit.
The inset of Fig.~\ref{fig:boundent} shows that the regime from which the continuum limit emerges passes through a phase where this calculation of the halo entanglement exhibits a clear step-wise structure, a feature more commonly found in systems with topological behavior.

\subsection{Inaccessibility of Halo Entanglement}
\label{sec:inaccessibleHaloEnt}

Through consolidation and implementation of core-halo separability, a clear distinction has been generated between the entanglement accessible through $(1 \times 1)_{\rm mixed}$ core modes and the inaccessible entanglement in the halo.  This inaccessibility is generated by classical correlations arising from the isolation of the halo (Eq.~\eqref{eq:halo-halosep}), obscuring the underlying entanglement.  
Operational implications follow for both application directions considered in this work 1.) to prepare the vacuum regions by distributing and locally transforming a series of entangled core pairs governed by PT symplectic eigenvalues and a classically-connected halo state, additional entanglement spanning the regions must be provided within the halo and 2.)  entanglement that can be extracted from regions of a vacuum state through the basic units of $(1_A \times 1_B)_{\rm mixed}$ entangled pairs is set by the core entanglement.

In general, convex decompositions of mixed states are not unique, i.e., there exist multiple sets $\{\sigma^{(p)}, Y\}$ in the decomposition of Eq.~\eqref{eq:mixedpurenoise} compatible with the specific form of a mixed state.   Separability indicates that at least one of these sets has an underlying pure tensor-product CM, and thus there exists a set with vanishing quantum correlations across the bipartition.
The presence of entanglement in the pure CM portion of compatible $\{\sigma^{(p)}, Y\}$ decomposition sets can vary widely, including those characterized by distributed entangled pairs within the classically mixed ensemble of a separable state.

As discussed in Appendix~\ref{app:sepobscent}, such separability-obscured entanglement is a well-known phenomenon, i.e., separable states may contain underlying entanglement visible only through additional classical communication (e.g., for the post-selection of entangled ensemble samples) or restriction/modification of the noise.  
For example, Eqs.~\eqref{eq:mixedpurenoise} and~\eqref{eq:sepcondition} show (along with the positivity of CMs) that the presence of a classical correlation matrix characterized by a tensor product of CMs will result in a mixed state that satisfies the separablility criterion, regardless of the entanglement present in the underlying pure state.
As such, it is possible for the separability designation to obscure the presence of entanglement, e.g.,~\cite{PhysRevLett.80.2493,PhysRevA.58.826,Vidal:1998ch,PhysRevA.63.032306,Yu_2006,Yu_2009}.

In the case of lattice scalar field vacuum regions, classical correlations connecting the entangled core pairs to the halo Hilbert space (Eq.~\eqref{eq:core-halosep}) are found to be incompatible with possible separable $\{\sigma^{(p)}, Y\}$ decomposition sets in the halo.
This can be physically interpreted as the implementation of core-halo separability (rather than tracing of the core) limiting the structure that can be identified as classical correlations in the internal halo-halo system.
The inequality of Eq.~\eqref{eq:simsepineq} expresses how constraints associated with separability observations can propagate through classical correlations to constrain subsequent subsystem evaluations of separability.

\section{Perspectives from Consolidation}
\label{sec:perspectives}

\subsection{Extracting Spacelike Entanglement}
\label{sec:extractingSpacelike}

Though much progress has been made in simulating the distribution of entanglement utilizing vacuum fields as a medium for (quasi-)particle propagation (e.g., Refs.~\cite{Retzker_2005,MartinMartinez2013,Pozas_Kerstjens_2016,Pozas_Kerstjens_2017,Simidzija_2018}) or in boosted reference frames (e.g., Refs.~\cite{Unruh1976,Reznik:2002fz,Reznik:2003mnx,Salton_2015}), techniques capable of reliably extracting spacelike entanglement from a field in the laboratory remain challenging.
Prior to consolidation, the modes associated with the dominant spacelike entanglement are  delocalized within each region.   
To extract entanglement from such a system into a sensing apparatus having basic quantum degrees of freedom requires highly structured interactions strategically distributed across each detection region.
However, by first consolidating the negativity into a core through local unitaries in each region, access to the extractable entanglement can be achieved by simply coupling to $(1_A \times 1_B)$ mode pairs.
It is within this organization of the entanglement that the strategy of entanglement extraction via swapping of a detection mode into the system becomes possible.  
As such, the entanglement consolidation methods we develop are expected to aid in the identification of useful distributed degrees of freedom, and thus the design of future quantum sensors, capable of recovering spacelike entanglement from continuum quantum fields.

For generic Gaussian states, a calculation of negativity provides an upper bound to the entanglement that may be extracted given a bipartite system.
Interestingly, the consolidating transformation developed here shows that the entirety of the accessible entanglement can be reorganized into $(1 \times 1)_{\rm mixed}$ entangled pairs, extending the physical significance of negativity for the free scalar field~\cite{Horodecki:1996nc,Vidal:2002zz,Simon:2000zz,Plenio:2005cwa}.
This reorganization of the negativity into entangled pairs, for which the measure is additive, promotes the negativity from an upper bound to a quantity with physical and operational meaning in terms of extractable two-mode entanglement resources.
There is a one-to-one correspondence between the negativity of a $(1 \times 1)_{\rm mixed}$ core pair and the two-mode squeezing required to prepare it~\cite{2003PhRvL..91j7901G,2004PhRvA..69e2320W}.
Furthermore, such two-mode entangled Gaussian states are known to be distillable~\cite{2000quant.ph..7061G}.

The explicit protocol developed here for extracting spacelike entanglement through $(1 \times 1)_{\rm mixed}$ modes is specifically designed for spatially latticized fields.  
As the continuum limit is approached, the PT symplectic eigenvectors converge, characterizing region-distributed modes that capture the logarithmic negativity.
Rather than pre-processing of the field through local unitaries, a process which may cease to be available when interacting with continuous fields, such modes may be used to inform the design of continuous space interaction profiles for a pair of finite-sized local detectors. 
Due to the vanishing of extractable entanglement at low region pixelation (low UV truncation) where regions become compatible with separable states, design of the detector energy sensitivity is expected to be necessary.

\subsection{Connections to Bound Entanglement}
\label{sec:boundEntConnection}

As discussed in Section~\ref{sec:inaccessibleHaloEnt}, the inaccessibility of halo entanglement indicates that, when utilizing the PT entanglement consolidation strategy, more entanglement must be provided to create the scalar field vacuum regions than could be subsequently extracted.
Such irreversability in the operational treatment of quantum correlations is often expressed within the framework of bound entanglement~\cite{PhysRevLett.80.5239,1999PhRvL..82.5385B,2000PhRvA..61c0301B,PhysRevLett.85.2657,2001PhRvA..63c2306S,2001PhRvL..86.3658W,2006PhRvL..97h0501B,Piani_2007}, which is the difference between the entanglement of formation required to create a state and the distillable entanglement that can be subsequently extracted from it (or asymptotically many samples of it).
In this context, we consider whether the inaccessible entanglement identified to be necessary through the consolidated state preparation procedure  yields insight into the fundamental presence of bound entanglement between regions of the scalar field vacuum, i.e., whether all possible state preparation protocols would similarly require the input of bound entanglement.

While it has become  natural to expect some amount of entanglement to become suppressed or bound upon the introduction of classical noise in the creation of mixed quantum states, it is important to note that mixedness is not a sufficient criteria for establishing operational irreversibility of an entanglement resource.  
For example, mixed density matrices that are transformable via local unitaries, $U = U_A \otimes U_B$, to a form that tags entangled pure states~\cite{VerstraeteQIP2017}, 
\begin{equation}
 \rho = U \left[\sum_{j} \rho \left(|jj\rangle_{AB} \right) \otimes \rho \left(|\psi_j\rangle_{AB} \right)  \right]  U^\dagger  \ \ \ 
\end{equation}
with $\rho\left(|\psi\rangle\right) = |\psi\rangle \langle \psi |$,
can be prepared and extracted reversibly.
Furthermore, relevant to the structure of core pairs after consolidation as in  Eq.~\eqref{eq:simultaneouscoresep}, it has been determined that, for two mode symmetric states, the entanglement of formation over all possible $\sigma^{(p)}$ pure-state decompositions is equal to that of the associated mixed state~\cite{2003PhRvL..91j7901G,
2004PhRvA..69e2320W}. 
Because the core pairs fall into this two-mode symmetric category, 
it is thus possible to identify a pure state decomposition for which $\mathcal{N}\left(\sigma_c^{(p)}\right) = \mathcal{N} \left(\sigma_c\right)$.
Core states of disjoint scalar field vacuum regions thus provide a CV example in which the required mixedness can be generated without suppressing the entanglement, thus requiring no entanglement excess in their creation.

Generic methods do not exist for determining whether a mixed-state density matrix has bound entanglement.
This includes indirect calculations involving the entanglement of formation and distillable entanglement.  
While progress has been reported for non-symmetric two-mode Gaussian states~\cite{Marian:2008zza,2019arXiv190309961T}, performing the necessary constrained optimizations in systems of more than two CV modes remains challenging.  

Though the core-halo entanglement structure has reduced complexity, and an upper-bound excluding zero has been provided for the inaccessible halo entanglement, we have not proven that this reorganization is optimal. 
If the presence of bound entanglement were to persist for all possible state preparation procedures, disjoint pairs of scalar field vacuum regions would be identified as NPT (non-positive PT) states with bound entanglement.
This is distinct from the fully-bound PPT entangled states that are more commonly considered~\cite{2000PhRvA..61f2313D,DiVincenzo_2000,2006quant.ph..8250S,2006quant.ph..9050C,Pankowski_2010}.
Note that, consistent with the core organization of tensor-product entangled pairs, the presence of distillable entanglement has been established in all NPT Gaussian states~\cite{giedke2001distillability}.


\section{Conclusion and Outlook}

We have studied the entanglement structure between two disjoint finite regions of infinite-volume, one-dimensional, non-interacting 
lattice scalar field theory vacuua.
The motivations for this analysis include designing protocols of laboratory preparations for quantum simulation, as well as informing quantum sensor design through identification of the distributed modes within each region that are dominantly responsible for the accessible entanglement.

To illuminate the entanglement structure within the mixed states of disjoint vacuum regions, local transformations are created that consolidate the entanglement  into a core-halo system. 
The core is comprised of  $(1\times 1)_{\rm mixed}$ CV pairs that carry all the accessible entanglement between the regions, with an exponential hierarchy of contributions following the PT symplectic eigenvalues. 
The consolidated system is separable along the core-halo partition, and the isolated halo is separable between the two regions.
Generally, however, when a separable mixed state is a subset of a larger QMB system, 
entanglement may still be required to prepare it, even when the subset is connected to the QMB system only via classical correlations.
Analysis of the consolidated separability structure reveals that initialization performed through the distribution of entangled core pairs requires the presence of inaccessible entanglement in the halo portion of the state.

Through iterated application of this protocol, transformations may be applied to the halo subsystem for further consolidation of this bound entanglement, allowing a complete pairwise protocol for preparing the quantum system.
Despite the clarity of the consolidated entanglement organization, 
the possibility that alternative protocols may exist that require less (or even no) inaccessible entanglement has not been precluded by the current analysis.

The consolidation of accessible entanglement in the field utilizes local operations to transform region volume-to-volume entanglement calculations plagued by a sign problem~\cite{Klco:2020rga,Klco:2021biu} into pairs of additively contributing entangled modes.
This suggests that both computational and conceptual advantages are possible through a  delocalization of lattice degrees of freedom, likely with analogy to the holographic techniques connecting entanglement in conformal fields to geometries in bulk gravitational duals~\cite{Ryu:2006bv,Nishioka:2009un,Pastawski:2015qua}.
In particular, the exponential hierarchy of entanglement present in the core-mode pairs may provide a non-local scheme for entanglement truncation with systematically quantifiable uncertainties.

While aspects of these entanglement reorganizations and subsequent conclusions apply more broadly~(\ref{sec:broaderapp}), it should be stressed that the results do not apply to generic mixed quantum states.
However, in the Standard Model, analogous consolidations are expected to be useful in regimes where the effective degrees of freedom are perturbatively close to Gaussian.
The asymptotic freedom of quantum chromodynamics (QCD) in the UV, which provides a perturbative vacuum of quarks and gluons, and chiral symmetry breaking in the IR, which provides a perturbative vacuum of pions, suggests the possibility of non-trivial evolution in the analogous QCD negativity core and halo through the confinement and chiral symmetry breaking scale(s).
Hints of interesting behaviors in such symmetry transition regimes, where low-energy effective interactions become increasingly non-local, may be found in the connection between emergent global symmetries and suppressed fluctuations of entanglement, e.g., as displayed in nuclear forces in low-energy scattering~\cite{Beane:2018oxh,Wagman:2017tmp,Beane:2020wjl,low2021symmetry,Beane:2021zvo}.


\begin{acknowledgments}
We would like to thank Silas Beane, Roland Farrell, Neda Hosseinidehaj, Aidan Murran, and John Preskill for valuable discussions.
We would also like to thank the Center for Experimental Nuclear Physics and Astrophysics (CENPA) for providing a creative environment for developmental periods of this work.
We have made extensive use of Wolfram Mathematica~\cite{Mathematica}.  DB is supported in part by NSF Nuclear Physics grant PHY-2111046.  NK is supported in part by the Walter Burke Institute for Theoretical Physics, and by the U.S. Department of Energy Office of Science, Office of Advanced Scientific Computing Research, (DE-SC0020290), and
Office of High Energy Physics DE-ACO2-07CH11359.
MJS is supported in part by the U.S. Department of Energy,
Office of Science, Office of Nuclear Physics, InQubator for Quantum Simulation (IQuS) under Award Number DOE
(NP) Award DE-SC0020970, and was supported in part by the Institute for Nuclear Theory with DOE grant No. DE-FG02-00ER41132.
\end{acknowledgments}

\bibliography{hqbib}

\begin{thebibliography}{172}%
\makeatletter
\providecommand \@ifxundefined [1]{%
 \@ifx{#1\undefined}
}%
\providecommand \@ifnum [1]{%
 \ifnum #1\expandafter \@firstoftwo
 \else \expandafter \@secondoftwo
 \fi
}%
\providecommand \@ifx [1]{%
 \ifx #1\expandafter \@firstoftwo
 \else \expandafter \@secondoftwo
 \fi
}%
\providecommand \natexlab [1]{#1}%
\providecommand \enquote  [1]{``#1''}%
\providecommand \bibnamefont  [1]{#1}%
\providecommand \bibfnamefont [1]{#1}%
\providecommand \citenamefont [1]{#1}%
\providecommand \href@noop [0]{\@secondoftwo}%
\providecommand \href [0]{\begingroup \@sanitize@url \@href}%
\providecommand \@href[1]{\@@startlink{#1}\@@href}%
\providecommand \@@href[1]{\endgroup#1\@@endlink}%
\providecommand \@sanitize@url [0]{\catcode `\\12\catcode `\$12\catcode
  `\&12\catcode `\#12\catcode `\^12\catcode `\_12\catcode `\%12\relax}%
\providecommand \@@startlink[1]{}%
\providecommand \@@endlink[0]{}%
\providecommand \url  [0]{\begingroup\@sanitize@url \@url }%
\providecommand \@url [1]{\endgroup\@href {#1}{\urlprefix }}%
\providecommand \urlprefix  [0]{URL }%
\providecommand \Eprint [0]{\href }%
\providecommand \doibase [0]{https://doi.org/}%
\providecommand \selectlanguage [0]{\@gobble}%
\providecommand \bibinfo  [0]{\@secondoftwo}%
\providecommand \bibfield  [0]{\@secondoftwo}%
\providecommand \translation [1]{[#1]}%
\providecommand \BibitemOpen [0]{}%
\providecommand \bibitemStop [0]{}%
\providecommand \bibitemNoStop [0]{.\EOS\space}%
\providecommand \EOS [0]{\spacefactor3000\relax}%
\providecommand \BibitemShut  [1]{\csname bibitem#1\endcsname}%
\let\auto@bib@innerbib\@empty
\bibitem [{\citenamefont {Feynman}(1982)}]{Feynman1982}%
  \BibitemOpen
  \bibfield  {author} {\bibinfo {author} {\bibfnamefont {R.~P.}\ \bibnamefont
  {Feynman}},\ }\bibfield  {title} {\bibinfo {title} {Simulating physics with
  computers},\ }\href {https://doi.org/10.1007/BF02650179} {\bibfield
  {journal} {\bibinfo  {journal} {International Journal of Theoretical
  Physics}\ }\textbf {\bibinfo {volume} {21}},\ \bibinfo {pages} {467}
  (\bibinfo {year} {1982})}\BibitemShut {NoStop}%
\bibitem [{\citenamefont {Ho}\ and\ \citenamefont {Hsu}(2016)}]{Ho:2015rga}%
  \BibitemOpen
  \bibfield  {author} {\bibinfo {author} {\bibfnamefont {C.~M.}\ \bibnamefont
  {Ho}}\ and\ \bibinfo {author} {\bibfnamefont {S.~D.~H.}\ \bibnamefont
  {Hsu}},\ }\bibfield  {title} {\bibinfo {title} {{Entanglement and Fast
  Quantum Thermalization in Heavy Ion Collisions}},\ }\href
  {https://doi.org/10.1142/S0217732316501108} {\bibfield  {journal} {\bibinfo
  {journal} {Mod. Phys. Lett. A}\ }\textbf {\bibinfo {volume} {31}},\ \bibinfo
  {pages} {1650110} (\bibinfo {year} {2016})},\ \Eprint
  {https://arxiv.org/abs/1506.03696} {arXiv:1506.03696 [hep-th]} \BibitemShut
  {NoStop}%
\bibitem [{\citenamefont {Kharzeev}\ and\ \citenamefont
  {Levin}(2017)}]{Kharzeev:2017qzs}%
  \BibitemOpen
  \bibfield  {author} {\bibinfo {author} {\bibfnamefont {D.~E.}\ \bibnamefont
  {Kharzeev}}\ and\ \bibinfo {author} {\bibfnamefont {E.~M.}\ \bibnamefont
  {Levin}},\ }\bibfield  {title} {\bibinfo {title} {{Deep inelastic scattering
  as a probe of entanglement}},\ }\href
  {https://doi.org/10.1103/PhysRevD.95.114008} {\bibfield  {journal} {\bibinfo
  {journal} {Phys. Rev. D}\ }\textbf {\bibinfo {volume} {95}},\ \bibinfo
  {pages} {114008} (\bibinfo {year} {2017})},\ \Eprint
  {https://arxiv.org/abs/1702.03489} {arXiv:1702.03489 [hep-ph]} \BibitemShut
  {NoStop}%
\bibitem [{\citenamefont {Baker}\ and\ \citenamefont
  {Kharzeev}(2018)}]{Baker:2017wtt}%
  \BibitemOpen
  \bibfield  {author} {\bibinfo {author} {\bibfnamefont {O.~K.}\ \bibnamefont
  {Baker}}\ and\ \bibinfo {author} {\bibfnamefont {D.~E.}\ \bibnamefont
  {Kharzeev}},\ }\bibfield  {title} {\bibinfo {title} {{Thermal radiation and
  entanglement in proton-proton collisions at energies available at the CERN
  Large Hadron Collider}},\ }\href {https://doi.org/10.1103/PhysRevD.98.054007}
  {\bibfield  {journal} {\bibinfo  {journal} {Phys. Rev. D}\ }\textbf {\bibinfo
  {volume} {98}},\ \bibinfo {pages} {054007} (\bibinfo {year} {2018})},\
  \Eprint {https://arxiv.org/abs/1712.04558} {arXiv:1712.04558 [hep-ph]}
  \BibitemShut {NoStop}%
\bibitem [{\citenamefont {Cervera-Lierta}\ \emph {et~al.}(2017)\citenamefont
  {Cervera-Lierta}, \citenamefont {Latorre}, \citenamefont {Rojo},\ and\
  \citenamefont {Rottoli}}]{Cervera-Lierta:2017tdt}%
  \BibitemOpen
  \bibfield  {author} {\bibinfo {author} {\bibfnamefont {A.}~\bibnamefont
  {Cervera-Lierta}}, \bibinfo {author} {\bibfnamefont {J.~I.}\ \bibnamefont
  {Latorre}}, \bibinfo {author} {\bibfnamefont {J.}~\bibnamefont {Rojo}},\ and\
  \bibinfo {author} {\bibfnamefont {L.}~\bibnamefont {Rottoli}},\ }\bibfield
  {title} {\bibinfo {title} {{Maximal Entanglement in High Energy Physics}},\
  }\href {https://doi.org/10.21468/SciPostPhys.3.5.036} {\bibfield  {journal}
  {\bibinfo  {journal} {SciPost Phys.}\ }\textbf {\bibinfo {volume} {3}},\
  \bibinfo {pages} {036} (\bibinfo {year} {2017})},\ \Eprint
  {https://arxiv.org/abs/1703.02989} {arXiv:1703.02989 [hep-th]} \BibitemShut
  {NoStop}%
\bibitem [{\citenamefont {Berges}\ \emph {et~al.}(2019)\citenamefont {Berges},
  \citenamefont {Floerchinger},\ and\ \citenamefont
  {Venugopalan}}]{Berges:2018cny}%
  \BibitemOpen
  \bibfield  {author} {\bibinfo {author} {\bibfnamefont {J.}~\bibnamefont
  {Berges}}, \bibinfo {author} {\bibfnamefont {S.}~\bibnamefont
  {Floerchinger}},\ and\ \bibinfo {author} {\bibfnamefont {R.}~\bibnamefont
  {Venugopalan}},\ }\bibfield  {title} {\bibinfo {title} {{Entanglement and
  thermalization}},\ }\href {https://doi.org/10.1016/j.nuclphysa.2018.12.008}
  {\bibfield  {journal} {\bibinfo  {journal} {Nucl. Phys. A}\ }\textbf
  {\bibinfo {volume} {982}},\ \bibinfo {pages} {819} (\bibinfo {year}
  {2019})},\ \Eprint {https://arxiv.org/abs/1812.08120} {arXiv:1812.08120
  [hep-th]} \BibitemShut {NoStop}%
\bibitem [{\citenamefont {Beane}\ \emph {et~al.}(2019)\citenamefont {Beane},
  \citenamefont {Kaplan}, \citenamefont {Klco},\ and\ \citenamefont
  {Savage}}]{Beane:2018oxh}%
  \BibitemOpen
  \bibfield  {author} {\bibinfo {author} {\bibfnamefont {S.~R.}\ \bibnamefont
  {Beane}}, \bibinfo {author} {\bibfnamefont {D.~B.}\ \bibnamefont {Kaplan}},
  \bibinfo {author} {\bibfnamefont {N.}~\bibnamefont {Klco}},\ and\ \bibinfo
  {author} {\bibfnamefont {M.~J.}\ \bibnamefont {Savage}},\ }\bibfield  {title}
  {\bibinfo {title} {{Entanglement Suppression and Emergent Symmetries of
  Strong Interactions}},\ }\href
  {https://doi.org/10.1103/PhysRevLett.122.102001} {\bibfield  {journal}
  {\bibinfo  {journal} {Phys. Rev. Lett.}\ }\textbf {\bibinfo {volume} {122}},\
  \bibinfo {pages} {102001} (\bibinfo {year} {2019})},\ \Eprint
  {https://arxiv.org/abs/1812.03138} {arXiv:1812.03138 [nucl-th]} \BibitemShut
  {NoStop}%
\bibitem [{\citenamefont {Gorton}\ and\ \citenamefont
  {Johnson}(2019)}]{GortonJohnson2019a}%
  \BibitemOpen
  \bibfield  {author} {\bibinfo {author} {\bibfnamefont {O.}~\bibnamefont
  {Gorton}}\ and\ \bibinfo {author} {\bibfnamefont {C.~W.}\ \bibnamefont
  {Johnson}},\ }\bibfield  {title} {\bibinfo {title} {Entanglement entropy and
  proton-neutron interactions}} (\bibinfo {year} {2019}),\ \bibinfo {note}
  {{ESNT} workshop on proton-neutron pairing,
  http://esnt.cea.fr/Phocea/Page/index.php?id=84}\BibitemShut {NoStop}%
\bibitem [{\citenamefont {Beane}\ and\ \citenamefont
  {Ehlers}(2019)}]{Beane:2019loz}%
  \BibitemOpen
  \bibfield  {author} {\bibinfo {author} {\bibfnamefont {S.~R.}\ \bibnamefont
  {Beane}}\ and\ \bibinfo {author} {\bibfnamefont {P.}~\bibnamefont {Ehlers}},\
  }\bibfield  {title} {\bibinfo {title} {{Chiral symmetry breaking,
  entanglement, and the nucleon spin decomposition}},\ }\href
  {https://doi.org/10.1142/S0217732320500480} {\bibfield  {journal} {\bibinfo
  {journal} {Mod. Phys. Lett. A}\ }\textbf {\bibinfo {volume} {35}},\ \bibinfo
  {pages} {2050048} (\bibinfo {year} {2019})},\ \Eprint
  {https://arxiv.org/abs/1905.03295} {arXiv:1905.03295 [hep-ph]} \BibitemShut
  {NoStop}%
\bibitem [{\citenamefont {Tu}\ \emph {et~al.}(2020)\citenamefont {Tu},
  \citenamefont {Kharzeev},\ and\ \citenamefont {Ullrich}}]{Tu:2019ouv}%
  \BibitemOpen
  \bibfield  {author} {\bibinfo {author} {\bibfnamefont {Z.}~\bibnamefont
  {Tu}}, \bibinfo {author} {\bibfnamefont {D.~E.}\ \bibnamefont {Kharzeev}},\
  and\ \bibinfo {author} {\bibfnamefont {T.}~\bibnamefont {Ullrich}},\
  }\bibfield  {title} {\bibinfo {title} {{Einstein-Podolsky-Rosen Paradox and
  Quantum Entanglement at Subnucleonic Scales}},\ }\href
  {https://doi.org/10.1103/PhysRevLett.124.062001} {\bibfield  {journal}
  {\bibinfo  {journal} {Phys. Rev. Lett.}\ }\textbf {\bibinfo {volume} {124}},\
  \bibinfo {pages} {062001} (\bibinfo {year} {2020})},\ \Eprint
  {https://arxiv.org/abs/1904.11974} {arXiv:1904.11974 [hep-ph]} \BibitemShut
  {NoStop}%
\bibitem [{\citenamefont {Beane}\ and\ \citenamefont
  {Farrell}(2021)}]{Beane:2020wjl}%
  \BibitemOpen
  \bibfield  {author} {\bibinfo {author} {\bibfnamefont {S.~R.}\ \bibnamefont
  {Beane}}\ and\ \bibinfo {author} {\bibfnamefont {R.~C.}\ \bibnamefont
  {Farrell}},\ }\bibfield  {title} {\bibinfo {title} {{Geometry and
  entanglement in the scattering matrix}},\ }\href
  {https://doi.org/10.1016/j.aop.2021.168581} {\bibfield  {journal} {\bibinfo
  {journal} {Annals Phys.}\ }\textbf {\bibinfo {volume} {433}},\ \bibinfo
  {pages} {168581} (\bibinfo {year} {2021})},\ \Eprint
  {https://arxiv.org/abs/2011.01278} {arXiv:2011.01278 [hep-th]} \BibitemShut
  {NoStop}%
\bibitem [{\citenamefont {Beane}\ \emph {et~al.}(2021)\citenamefont {Beane},
  \citenamefont {Farrell},\ and\ \citenamefont {Varma}}]{Beane:2021zvo}%
  \BibitemOpen
  \bibfield  {author} {\bibinfo {author} {\bibfnamefont {S.~R.}\ \bibnamefont
  {Beane}}, \bibinfo {author} {\bibfnamefont {R.~C.}\ \bibnamefont {Farrell}},\
  and\ \bibinfo {author} {\bibfnamefont {M.}~\bibnamefont {Varma}},\ }\bibfield
   {title} {\bibinfo {title} {{Entanglement minimization in hadronic scattering
  with pions}},\ }\href {https://doi.org/10.1142/S0217751X21502055} {\bibfield
  {journal} {\bibinfo  {journal} {Int. J. Mod. Phys. A}\ }\textbf {\bibinfo
  {volume} {36}},\ \bibinfo {pages} {2150205} (\bibinfo {year} {2021})},\
  \Eprint {https://arxiv.org/abs/2108.00646} {arXiv:2108.00646 [hep-ph]}
  \BibitemShut {NoStop}%
\bibitem [{\citenamefont {Kharzeev}\ and\ \citenamefont
  {Levin}(2021)}]{Kharzeev:2021yyf}%
  \BibitemOpen
  \bibfield  {author} {\bibinfo {author} {\bibfnamefont {D.~E.}\ \bibnamefont
  {Kharzeev}}\ and\ \bibinfo {author} {\bibfnamefont {E.}~\bibnamefont
  {Levin}},\ }\bibfield  {title} {\bibinfo {title} {{Deep inelastic scattering
  as a probe of entanglement: Confronting experimental data}},\ }\href
  {https://doi.org/10.1103/PhysRevD.104.L031503} {\bibfield  {journal}
  {\bibinfo  {journal} {Phys. Rev. D}\ }\textbf {\bibinfo {volume} {104}},\
  \bibinfo {pages} {L031503} (\bibinfo {year} {2021})},\ \Eprint
  {https://arxiv.org/abs/2102.09773} {arXiv:2102.09773 [hep-ph]} \BibitemShut
  {NoStop}%
\bibitem [{\citenamefont {Robin}\ \emph {et~al.}(2021)\citenamefont {Robin},
  \citenamefont {Savage},\ and\ \citenamefont {Pillet}}]{Robin:2020aeh}%
  \BibitemOpen
  \bibfield  {author} {\bibinfo {author} {\bibfnamefont {C.}~\bibnamefont
  {Robin}}, \bibinfo {author} {\bibfnamefont {M.~J.}\ \bibnamefont {Savage}},\
  and\ \bibinfo {author} {\bibfnamefont {N.}~\bibnamefont {Pillet}},\
  }\bibfield  {title} {\bibinfo {title} {{Entanglement Rearrangement in
  Self-Consistent Nuclear Structure Calculations}},\ }\href
  {https://doi.org/10.1103/PhysRevC.103.034325} {\bibfield  {journal} {\bibinfo
   {journal} {Phys. Rev. C}\ }\textbf {\bibinfo {volume} {103}},\ \bibinfo
  {pages} {034325} (\bibinfo {year} {2021})},\ \Eprint
  {https://arxiv.org/abs/2007.09157} {arXiv:2007.09157 [nucl-th]} \BibitemShut
  {NoStop}%
\bibitem [{\citenamefont {Low}\ and\ \citenamefont
  {Mehen}(2021)}]{low2021symmetry}%
  \BibitemOpen
  \bibfield  {author} {\bibinfo {author} {\bibfnamefont {I.}~\bibnamefont
  {Low}}\ and\ \bibinfo {author} {\bibfnamefont {T.}~\bibnamefont {Mehen}},\
  }\bibfield  {title} {\bibinfo {title} {Symmetry from entanglement
  suppression},\ }\href@noop {} {\  (\bibinfo {year} {2021})},\ \Eprint
  {https://arxiv.org/abs/2104.10835} {arXiv:2104.10835 [hep-th]} \BibitemShut
  {NoStop}%
\bibitem [{\citenamefont {Gong}\ \emph {et~al.}(2021)\citenamefont {Gong},
  \citenamefont {Parida}, \citenamefont {Tu},\ and\ \citenamefont
  {Venugopalan}}]{Gong:2021bcp}%
  \BibitemOpen
  \bibfield  {author} {\bibinfo {author} {\bibfnamefont {W.}~\bibnamefont
  {Gong}}, \bibinfo {author} {\bibfnamefont {G.}~\bibnamefont {Parida}},
  \bibinfo {author} {\bibfnamefont {Z.}~\bibnamefont {Tu}},\ and\ \bibinfo
  {author} {\bibfnamefont {R.}~\bibnamefont {Venugopalan}},\ }\bibfield
  {title} {\bibinfo {title} {{Bell-type inequality tests and quantum
  entanglement from $\Lambda$-hyperon spin correlations at high energy
  colliders}},\ }\href@noop {} {\  (\bibinfo {year} {2021})},\ \Eprint
  {https://arxiv.org/abs/2107.13007} {arXiv:2107.13007 [hep-ph]} \BibitemShut
  {NoStop}%
\bibitem [{\citenamefont {Roggero}(2021)}]{Roggero:2021asb}%
  \BibitemOpen
  \bibfield  {author} {\bibinfo {author} {\bibfnamefont {A.}~\bibnamefont
  {Roggero}},\ }\bibfield  {title} {\bibinfo {title} {{Entanglement and
  many-body effects in collective neutrino oscillations}},\ }\href
  {https://doi.org/10.1103/PhysRevD.104.103016} {\bibfield  {journal} {\bibinfo
   {journal} {Phys. Rev. D}\ }\textbf {\bibinfo {volume} {104}},\ \bibinfo
  {pages} {103016} (\bibinfo {year} {2021})},\ \Eprint
  {https://arxiv.org/abs/2102.10188} {arXiv:2102.10188 [hep-ph]} \BibitemShut
  {NoStop}%
\bibitem [{\citenamefont {Mueller}\ \emph {et~al.}(2021)\citenamefont
  {Mueller}, \citenamefont {Zache},\ and\ \citenamefont
  {Ott}}]{Mueller:2021gxd}%
  \BibitemOpen
  \bibfield  {author} {\bibinfo {author} {\bibfnamefont {N.}~\bibnamefont
  {Mueller}}, \bibinfo {author} {\bibfnamefont {T.~V.}\ \bibnamefont {Zache}},\
  and\ \bibinfo {author} {\bibfnamefont {R.}~\bibnamefont {Ott}},\ }\bibfield
  {title} {\bibinfo {title} {{Thermalization of Gauge Theories from their
  Entanglement Spectrum}},\ }\href@noop {} {\  (\bibinfo {year} {2021})},\
  \Eprint {https://arxiv.org/abs/2107.11416} {arXiv:2107.11416 [quant-ph]}
  \BibitemShut {NoStop}%
\bibitem [{\citenamefont {Ba\~nuls}\ \emph {et~al.}(2020)\citenamefont
  {Ba\~nuls} \emph {et~al.}}]{Banuls:2019bmf}%
  \BibitemOpen
  \bibfield  {author} {\bibinfo {author} {\bibfnamefont {M.~C.}\ \bibnamefont
  {Ba\~nuls}} \emph {et~al.},\ }\bibfield  {title} {\bibinfo {title}
  {{Simulating Lattice Gauge Theories within Quantum Technologies}},\ }\href
  {https://doi.org/10.1140/epjd/e2020-100571-8} {\bibfield  {journal} {\bibinfo
   {journal} {Eur. Phys. J. D}\ }\textbf {\bibinfo {volume} {74}},\ \bibinfo
  {pages} {165} (\bibinfo {year} {2020})},\ \Eprint
  {https://arxiv.org/abs/1911.00003} {arXiv:1911.00003 [quant-ph]} \BibitemShut
  {NoStop}%
\bibitem [{\citenamefont {Alexeev}\ \emph {et~al.}(2021)\citenamefont {Alexeev}
  \emph {et~al.}}]{Alexeev:2020xrq}%
  \BibitemOpen
  \bibfield  {author} {\bibinfo {author} {\bibfnamefont {Y.}~\bibnamefont
  {Alexeev}} \emph {et~al.},\ }\bibfield  {title} {\bibinfo {title} {{Quantum
  Computer Systems for Scientific Discovery}},\ }\href
  {https://doi.org/10.1103/PRXQuantum.2.017001} {\bibfield  {journal} {\bibinfo
   {journal} {P. R. X. Quantum.}\ }\textbf {\bibinfo {volume} {2}},\ \bibinfo
  {pages} {017001} (\bibinfo {year} {2021})},\ \Eprint
  {https://arxiv.org/abs/1912.07577} {arXiv:1912.07577 [quant-ph]} \BibitemShut
  {NoStop}%
\bibitem [{\citenamefont {Klco}\ \emph {et~al.}(2021)\citenamefont {Klco},
  \citenamefont {Roggero},\ and\ \citenamefont {Savage}}]{Klco:2021lap}%
  \BibitemOpen
  \bibfield  {author} {\bibinfo {author} {\bibfnamefont {N.}~\bibnamefont
  {Klco}}, \bibinfo {author} {\bibfnamefont {A.}~\bibnamefont {Roggero}},\ and\
  \bibinfo {author} {\bibfnamefont {M.~J.}\ \bibnamefont {Savage}},\ }\bibfield
   {title} {\bibinfo {title} {{Standard Model Physics and the Digital Quantum
  Revolution: Thoughts about the Interface}},\ }\href@noop {} {\  (\bibinfo
  {year} {2021})},\ \Eprint {https://arxiv.org/abs/2107.04769}
  {arXiv:2107.04769 [quant-ph]} \BibitemShut {NoStop}%
\bibitem [{\citenamefont {Zohar}(2021)}]{Zohar:2021nyc}%
  \BibitemOpen
  \bibfield  {author} {\bibinfo {author} {\bibfnamefont {E.}~\bibnamefont
  {Zohar}},\ }\bibfield  {title} {\bibinfo {title} {{Quantum simulation of
  lattice gauge theories in more than one space
  dimension\textemdash{}requirements, challenges and methods}},\ }\href
  {https://doi.org/10.1098/rsta.2021.0069} {\bibfield  {journal} {\bibinfo
  {journal} {Phil. Trans. A. Math. Phys. Eng. Sci.}\ }\textbf {\bibinfo
  {volume} {380}},\ \bibinfo {pages} {20210069} (\bibinfo {year} {2021})},\
  \Eprint {https://arxiv.org/abs/2106.04609} {arXiv:2106.04609 [quant-ph]}
  \BibitemShut {NoStop}%
\bibitem [{\citenamefont {Halvorson}\ and\ \citenamefont
  {Clifton}(2000)}]{Halvorson:1999pz}%
  \BibitemOpen
  \bibfield  {author} {\bibinfo {author} {\bibfnamefont {H.}~\bibnamefont
  {Halvorson}}\ and\ \bibinfo {author} {\bibfnamefont {R.}~\bibnamefont
  {Clifton}},\ }\bibfield  {title} {\bibinfo {title} {{Generic Bell correlation
  between arbitrary local algebras in quantum field theory}},\ }\href
  {https://doi.org/10.1063/1.533253} {\bibfield  {journal} {\bibinfo  {journal}
  {J. Math. Phys.}\ }\textbf {\bibinfo {volume} {41}},\ \bibinfo {pages} {1711}
  (\bibinfo {year} {2000})},\ \Eprint {https://arxiv.org/abs/math-ph/9909013}
  {arXiv:math-ph/9909013} \BibitemShut {NoStop}%
\bibitem [{\citenamefont {Reeh}\ and\ \citenamefont
  {Schlieder}(1961)}]{Reeh1961}%
  \BibitemOpen
  \bibfield  {author} {\bibinfo {author} {\bibfnamefont {H.}~\bibnamefont
  {Reeh}}\ and\ \bibinfo {author} {\bibfnamefont {S.}~\bibnamefont
  {Schlieder}},\ }\bibfield  {title} {\bibinfo {title} {Bemerkungen zur
  unit{\"a}r{\"a}quivalenz von lorentzinvarianten feldern},\ }\href
  {https://doi.org/10.1007/BF02787889} {\bibfield  {journal} {\bibinfo
  {journal} {Il Nuovo Cimento (1955-1965)}\ }\textbf {\bibinfo {volume} {22}},\
  \bibinfo {pages} {1051} (\bibinfo {year} {1961})}\BibitemShut {NoStop}%
\bibitem [{\citenamefont {Summers}\ and\ \citenamefont
  {Werner}(1985)}]{summers1985vacuum}%
  \BibitemOpen
  \bibfield  {author} {\bibinfo {author} {\bibfnamefont {S.~J.}\ \bibnamefont
  {Summers}}\ and\ \bibinfo {author} {\bibfnamefont {R.}~\bibnamefont
  {Werner}},\ }\bibfield  {title} {\bibinfo {title} {The vacuum violates bell's
  inequalities},\ }\href
  {https://doi.org/https://doi.org/10.1016/0375-9601(85)90093-3} {\bibfield
  {journal} {\bibinfo  {journal} {Physics Letters A}\ }\textbf {\bibinfo
  {volume} {110}},\ \bibinfo {pages} {257} (\bibinfo {year}
  {1985})}\BibitemShut {NoStop}%
\bibitem [{\citenamefont {Summers}\ and\ \citenamefont
  {Werner}(1987{\natexlab{a}})}]{summers1987bell1}%
  \BibitemOpen
  \bibfield  {author} {\bibinfo {author} {\bibfnamefont {S.~J.}\ \bibnamefont
  {Summers}}\ and\ \bibinfo {author} {\bibfnamefont {R.}~\bibnamefont
  {Werner}},\ }\bibfield  {title} {\bibinfo {title} {Bell's inequalities and
  quantum field theory. i. general setting},\ }\href
  {https://doi.org/10.1063/1.527733} {\bibfield  {journal} {\bibinfo  {journal}
  {Journal of Mathematical Physics}\ }\textbf {\bibinfo {volume} {28}},\
  \bibinfo {pages} {2440} (\bibinfo {year} {1987}{\natexlab{a}})},\ \Eprint
  {https://arxiv.org/abs/https://doi.org/10.1063/1.527733}
  {https://doi.org/10.1063/1.527733} \BibitemShut {NoStop}%
\bibitem [{\citenamefont {Summers}\ and\ \citenamefont
  {Werner}(1987{\natexlab{b}})}]{summers1987bell2}%
  \BibitemOpen
  \bibfield  {author} {\bibinfo {author} {\bibfnamefont {S.~J.}\ \bibnamefont
  {Summers}}\ and\ \bibinfo {author} {\bibfnamefont {R.}~\bibnamefont
  {Werner}},\ }\bibfield  {title} {\bibinfo {title} {Bell's inequalities and
  quantum field theory. ii. bell's inequalities are maximally violated in the
  vacuum},\ }\href {https://doi.org/10.1063/1.527734} {\bibfield  {journal}
  {\bibinfo  {journal} {Journal of Mathematical Physics}\ }\textbf {\bibinfo
  {volume} {28}},\ \bibinfo {pages} {2448} (\bibinfo {year}
  {1987}{\natexlab{b}})},\ \Eprint
  {https://arxiv.org/abs/https://doi.org/10.1063/1.527734}
  {https://doi.org/10.1063/1.527734} \BibitemShut {NoStop}%
\bibitem [{\citenamefont {Valentini}(1991)}]{VALENTINI1991321}%
  \BibitemOpen
  \bibfield  {author} {\bibinfo {author} {\bibfnamefont {A.}~\bibnamefont
  {Valentini}},\ }\bibfield  {title} {\bibinfo {title} {Non-local correlations
  in quantum electrodynamics},\ }\href
  {https://doi.org/https://doi.org/10.1016/0375-9601(91)90952-5} {\bibfield
  {journal} {\bibinfo  {journal} {Physics Letters A}\ }\textbf {\bibinfo
  {volume} {153}},\ \bibinfo {pages} {321 } (\bibinfo {year}
  {1991})}\BibitemShut {NoStop}%
\bibitem [{\citenamefont {Reznik}(2003)}]{Reznik:2002fz}%
  \BibitemOpen
  \bibfield  {author} {\bibinfo {author} {\bibfnamefont {B.}~\bibnamefont
  {Reznik}},\ }\bibfield  {title} {\bibinfo {title} {{Entanglement from the
  vacuum}},\ }\href {https://doi.org/10.1023/A:1022875910744} {\bibfield
  {journal} {\bibinfo  {journal} {Found. Phys.}\ }\textbf {\bibinfo {volume}
  {33}},\ \bibinfo {pages} {167} (\bibinfo {year} {2003})},\ \Eprint
  {https://arxiv.org/abs/quant-ph/0212044} {arXiv:quant-ph/0212044}
  \BibitemShut {NoStop}%
\bibitem [{\citenamefont {Reznik}\ \emph {et~al.}(2005)\citenamefont {Reznik},
  \citenamefont {Retzker},\ and\ \citenamefont {Silman}}]{Reznik:2003mnx}%
  \BibitemOpen
  \bibfield  {author} {\bibinfo {author} {\bibfnamefont {B.}~\bibnamefont
  {Reznik}}, \bibinfo {author} {\bibfnamefont {A.}~\bibnamefont {Retzker}},\
  and\ \bibinfo {author} {\bibfnamefont {J.}~\bibnamefont {Silman}},\
  }\bibfield  {title} {\bibinfo {title} {{Violating Bell's inequalities in the
  vacuum}},\ }\href {https://doi.org/10.1103/PhysRevA.71.042104} {\bibfield
  {journal} {\bibinfo  {journal} {Phys. Rev. A}\ }\textbf {\bibinfo {volume}
  {71}},\ \bibinfo {pages} {042104} (\bibinfo {year} {2005})},\ \Eprint
  {https://arxiv.org/abs/quant-ph/0310058} {arXiv:quant-ph/0310058}
  \BibitemShut {NoStop}%
\bibitem [{\citenamefont {Srednicki}(1993)}]{Srednicki:1993im}%
  \BibitemOpen
  \bibfield  {author} {\bibinfo {author} {\bibfnamefont {M.}~\bibnamefont
  {Srednicki}},\ }\bibfield  {title} {\bibinfo {title} {{Entropy and area}},\
  }\href {https://doi.org/10.1103/PhysRevLett.71.666} {\bibfield  {journal}
  {\bibinfo  {journal} {Phys. Rev. Lett.}\ }\textbf {\bibinfo {volume} {71}},\
  \bibinfo {pages} {666} (\bibinfo {year} {1993})},\ \Eprint
  {https://arxiv.org/abs/hep-th/9303048} {arXiv:hep-th/9303048 [hep-th]}
  \BibitemShut {NoStop}%
\bibitem [{\citenamefont {Holzhey}\ \emph {et~al.}(1994)\citenamefont
  {Holzhey}, \citenamefont {Larsen},\ and\ \citenamefont
  {Wilczek}}]{Holzhey:1994we}%
  \BibitemOpen
  \bibfield  {author} {\bibinfo {author} {\bibfnamefont {C.}~\bibnamefont
  {Holzhey}}, \bibinfo {author} {\bibfnamefont {F.}~\bibnamefont {Larsen}},\
  and\ \bibinfo {author} {\bibfnamefont {F.}~\bibnamefont {Wilczek}},\
  }\bibfield  {title} {\bibinfo {title} {{Geometric and renormalized entropy in
  conformal field theory}},\ }\href
  {https://doi.org/10.1016/0550-3213(94)90402-2} {\bibfield  {journal}
  {\bibinfo  {journal} {Nucl. Phys. B}\ }\textbf {\bibinfo {volume} {424}},\
  \bibinfo {pages} {443} (\bibinfo {year} {1994})},\ \Eprint
  {https://arxiv.org/abs/hep-th/9403108} {arXiv:hep-th/9403108} \BibitemShut
  {NoStop}%
\bibitem [{\citenamefont {Callan}\ and\ \citenamefont
  {Wilczek}(1994)}]{Callan:1994py}%
  \BibitemOpen
  \bibfield  {author} {\bibinfo {author} {\bibfnamefont {C.~G.}\ \bibnamefont
  {Callan}, \bibfnamefont {Jr.}}\ and\ \bibinfo {author} {\bibfnamefont
  {F.}~\bibnamefont {Wilczek}},\ }\bibfield  {title} {\bibinfo {title} {{On
  geometric entropy}},\ }\href {https://doi.org/10.1016/0370-2693(94)91007-3}
  {\bibfield  {journal} {\bibinfo  {journal} {Phys. Lett. B}\ }\textbf
  {\bibinfo {volume} {333}},\ \bibinfo {pages} {55} (\bibinfo {year} {1994})},\
  \Eprint {https://arxiv.org/abs/hep-th/9401072} {arXiv:hep-th/9401072}
  \BibitemShut {NoStop}%
\bibitem [{\citenamefont {Audenaert}\ \emph {et~al.}(2002)\citenamefont
  {Audenaert}, \citenamefont {Eisert}, \citenamefont {Plenio},\ and\
  \citenamefont {Werner}}]{Audenaert:2002xfl}%
  \BibitemOpen
  \bibfield  {author} {\bibinfo {author} {\bibfnamefont {K.}~\bibnamefont
  {Audenaert}}, \bibinfo {author} {\bibfnamefont {J.}~\bibnamefont {Eisert}},
  \bibinfo {author} {\bibfnamefont {M.}~\bibnamefont {Plenio}},\ and\ \bibinfo
  {author} {\bibfnamefont {R.}~\bibnamefont {Werner}},\ }\bibfield  {title}
  {\bibinfo {title} {{Entanglement Properties of the Harmonic Chain}},\ }\href
  {https://doi.org/10.1103/PhysRevA.66.042327} {\bibfield  {journal} {\bibinfo
  {journal} {Phys. Rev. A}\ }\textbf {\bibinfo {volume} {66}},\ \bibinfo
  {pages} {042327} (\bibinfo {year} {2002})},\ \Eprint
  {https://arxiv.org/abs/quant-ph/0205025} {arXiv:quant-ph/0205025}
  \BibitemShut {NoStop}%
\bibitem [{\citenamefont {{Botero}}\ and\ \citenamefont
  {{Reznik}}(2004)}]{2004PhRvA..70e2329B}%
  \BibitemOpen
  \bibfield  {author} {\bibinfo {author} {\bibfnamefont {A.}~\bibnamefont
  {{Botero}}}\ and\ \bibinfo {author} {\bibfnamefont {B.}~\bibnamefont
  {{Reznik}}},\ }\bibfield  {title} {\bibinfo {title} {{Spatial structures and
  localization of vacuum entanglement in the linear harmonic chain}},\ }\href
  {https://doi.org/10.1103/PhysRevA.70.052329} {\bibfield  {journal} {\bibinfo
  {journal} {\pra}\ }\textbf {\bibinfo {volume} {70}},\ \bibinfo {eid} {052329}
  (\bibinfo {year} {2004})},\ \Eprint {https://arxiv.org/abs/quant-ph/0403233}
  {arXiv:quant-ph/0403233 [quant-ph]} \BibitemShut {NoStop}%
\bibitem [{\citenamefont {Calabrese}\ and\ \citenamefont
  {Cardy}(2004)}]{Calabrese:2004eu}%
  \BibitemOpen
  \bibfield  {author} {\bibinfo {author} {\bibfnamefont {P.}~\bibnamefont
  {Calabrese}}\ and\ \bibinfo {author} {\bibfnamefont {J.~L.}\ \bibnamefont
  {Cardy}},\ }\bibfield  {title} {\bibinfo {title} {{Entanglement entropy and
  quantum field theory}},\ }\href
  {https://doi.org/10.1088/1742-5468/2004/06/P06002} {\bibfield  {journal}
  {\bibinfo  {journal} {J. Stat. Mech.}\ }\textbf {\bibinfo {volume} {0406}},\
  \bibinfo {pages} {P06002} (\bibinfo {year} {2004})},\ \Eprint
  {https://arxiv.org/abs/hep-th/0405152} {arXiv:hep-th/0405152} \BibitemShut
  {NoStop}%
\bibitem [{\citenamefont {Ryu}\ and\ \citenamefont
  {Takayanagi}(2006)}]{Ryu:2006bv}%
  \BibitemOpen
  \bibfield  {author} {\bibinfo {author} {\bibfnamefont {S.}~\bibnamefont
  {Ryu}}\ and\ \bibinfo {author} {\bibfnamefont {T.}~\bibnamefont
  {Takayanagi}},\ }\bibfield  {title} {\bibinfo {title} {{Holographic
  derivation of entanglement entropy from AdS/CFT}},\ }\href
  {https://doi.org/10.1103/PhysRevLett.96.181602} {\bibfield  {journal}
  {\bibinfo  {journal} {Phys. Rev. Lett.}\ }\textbf {\bibinfo {volume} {96}},\
  \bibinfo {pages} {181602} (\bibinfo {year} {2006})},\ \Eprint
  {https://arxiv.org/abs/hep-th/0603001} {arXiv:hep-th/0603001} \BibitemShut
  {NoStop}%
\bibitem [{\citenamefont {Kofler}\ \emph {et~al.}(2006)\citenamefont {Kofler},
  \citenamefont {Vedral}, \citenamefont {Kim},\ and\ \citenamefont
  {Brukner}}]{kofler2006entanglement}%
  \BibitemOpen
  \bibfield  {author} {\bibinfo {author} {\bibfnamefont {J.}~\bibnamefont
  {Kofler}}, \bibinfo {author} {\bibfnamefont {V.}~\bibnamefont {Vedral}},
  \bibinfo {author} {\bibfnamefont {M.~S.}\ \bibnamefont {Kim}},\ and\ \bibinfo
  {author} {\bibfnamefont {{\v{C}}.}~\bibnamefont {Brukner}},\ }\bibfield
  {title} {\bibinfo {title} {Entanglement between collective operators in a
  linear harmonic chain},\ }\href {https://doi.org/10.1103/physreva.73.052107}
  {\bibfield  {journal} {\bibinfo  {journal} {Physical Review A}\ }\textbf
  {\bibinfo {volume} {73}},\ \bibinfo {pages} {052107} (\bibinfo {year}
  {2006})}\BibitemShut {NoStop}%
\bibitem [{\citenamefont {Marcovitch}\ \emph {et~al.}(2009)\citenamefont
  {Marcovitch}, \citenamefont {Retzker}, \citenamefont {Plenio},\ and\
  \citenamefont {Reznik}}]{Marcovitch:2008sxc}%
  \BibitemOpen
  \bibfield  {author} {\bibinfo {author} {\bibfnamefont {S.}~\bibnamefont
  {Marcovitch}}, \bibinfo {author} {\bibfnamefont {A.}~\bibnamefont {Retzker}},
  \bibinfo {author} {\bibfnamefont {M.}~\bibnamefont {Plenio}},\ and\ \bibinfo
  {author} {\bibfnamefont {B.}~\bibnamefont {Reznik}},\ }\bibfield  {title}
  {\bibinfo {title} {{Critical and noncritical long-range entanglement in
  Klein-Gordon fields}},\ }\href {https://doi.org/10.1103/PhysRevA.80.012325}
  {\bibfield  {journal} {\bibinfo  {journal} {Phys. Rev. A}\ }\textbf {\bibinfo
  {volume} {80}},\ \bibinfo {pages} {012325} (\bibinfo {year} {2009})},\
  \Eprint {https://arxiv.org/abs/0811.1288} {arXiv:0811.1288 [quant-ph]}
  \BibitemShut {NoStop}%
\bibitem [{\citenamefont {Casini}\ and\ \citenamefont
  {Huerta}(2009)}]{Casini:2009sr}%
  \BibitemOpen
  \bibfield  {author} {\bibinfo {author} {\bibfnamefont {H.}~\bibnamefont
  {Casini}}\ and\ \bibinfo {author} {\bibfnamefont {M.}~\bibnamefont
  {Huerta}},\ }\bibfield  {title} {\bibinfo {title} {{Entanglement entropy in
  free quantum field theory}},\ }\href
  {https://doi.org/10.1088/1751-8113/42/50/504007} {\bibfield  {journal}
  {\bibinfo  {journal} {J. Phys. A}\ }\textbf {\bibinfo {volume} {42}},\
  \bibinfo {pages} {504007} (\bibinfo {year} {2009})},\ \Eprint
  {https://arxiv.org/abs/0905.2562} {arXiv:0905.2562 [hep-th]} \BibitemShut
  {NoStop}%
\bibitem [{\citenamefont {Nishioka}\ \emph {et~al.}(2009)\citenamefont
  {Nishioka}, \citenamefont {Ryu},\ and\ \citenamefont
  {Takayanagi}}]{Nishioka:2009un}%
  \BibitemOpen
  \bibfield  {author} {\bibinfo {author} {\bibfnamefont {T.}~\bibnamefont
  {Nishioka}}, \bibinfo {author} {\bibfnamefont {S.}~\bibnamefont {Ryu}},\ and\
  \bibinfo {author} {\bibfnamefont {T.}~\bibnamefont {Takayanagi}},\ }\bibfield
   {title} {\bibinfo {title} {{Holographic Entanglement Entropy: An
  Overview}},\ }\href {https://doi.org/10.1088/1751-8113/42/50/504008}
  {\bibfield  {journal} {\bibinfo  {journal} {J. Phys. A}\ }\textbf {\bibinfo
  {volume} {42}},\ \bibinfo {pages} {504008} (\bibinfo {year} {2009})},\
  \Eprint {https://arxiv.org/abs/0905.0932} {arXiv:0905.0932 [hep-th]}
  \BibitemShut {NoStop}%
\bibitem [{\citenamefont {Calabrese}\ \emph {et~al.}(2009)\citenamefont
  {Calabrese}, \citenamefont {Cardy},\ and\ \citenamefont
  {Tonni}}]{Calabrese:2009ez}%
  \BibitemOpen
  \bibfield  {author} {\bibinfo {author} {\bibfnamefont {P.}~\bibnamefont
  {Calabrese}}, \bibinfo {author} {\bibfnamefont {J.}~\bibnamefont {Cardy}},\
  and\ \bibinfo {author} {\bibfnamefont {E.}~\bibnamefont {Tonni}},\ }\bibfield
   {title} {\bibinfo {title} {{Entanglement entropy of two disjoint intervals
  in conformal field theory}},\ }\href
  {https://doi.org/10.1088/1742-5468/2009/11/P11001} {\bibfield  {journal}
  {\bibinfo  {journal} {J. Stat. Mech.}\ }\textbf {\bibinfo {volume} {0911}},\
  \bibinfo {pages} {P11001} (\bibinfo {year} {2009})},\ \Eprint
  {https://arxiv.org/abs/0905.2069} {arXiv:0905.2069 [hep-th]} \BibitemShut
  {NoStop}%
\bibitem [{\citenamefont {Calabrese}\ and\ \citenamefont
  {Cardy}(2009)}]{Calabrese:2009qy}%
  \BibitemOpen
  \bibfield  {author} {\bibinfo {author} {\bibfnamefont {P.}~\bibnamefont
  {Calabrese}}\ and\ \bibinfo {author} {\bibfnamefont {J.}~\bibnamefont
  {Cardy}},\ }\bibfield  {title} {\bibinfo {title} {{Entanglement entropy and
  conformal field theory}},\ }\href
  {https://doi.org/10.1088/1751-8113/42/50/504005} {\bibfield  {journal}
  {\bibinfo  {journal} {J. Phys. A}\ }\textbf {\bibinfo {volume} {42}},\
  \bibinfo {pages} {504005} (\bibinfo {year} {2009})},\ \Eprint
  {https://arxiv.org/abs/0905.4013} {arXiv:0905.4013 [cond-mat.stat-mech]}
  \BibitemShut {NoStop}%
\bibitem [{\citenamefont {Casini}\ \emph {et~al.}(2011)\citenamefont {Casini},
  \citenamefont {Huerta},\ and\ \citenamefont {Myers}}]{Casini:2011kv}%
  \BibitemOpen
  \bibfield  {author} {\bibinfo {author} {\bibfnamefont {H.}~\bibnamefont
  {Casini}}, \bibinfo {author} {\bibfnamefont {M.}~\bibnamefont {Huerta}},\
  and\ \bibinfo {author} {\bibfnamefont {R.~C.}\ \bibnamefont {Myers}},\
  }\bibfield  {title} {\bibinfo {title} {{Towards a derivation of holographic
  entanglement entropy}},\ }\href {https://doi.org/10.1007/JHEP05(2011)036}
  {\bibfield  {journal} {\bibinfo  {journal} {JHEP}\ }\textbf {\bibinfo
  {volume} {05}},\ \bibinfo {pages} {036}},\ \Eprint
  {https://arxiv.org/abs/1102.0440} {arXiv:1102.0440 [hep-th]} \BibitemShut
  {NoStop}%
\bibitem [{\citenamefont {Calabrese}\ \emph {et~al.}(2012)\citenamefont
  {Calabrese}, \citenamefont {Cardy},\ and\ \citenamefont
  {Tonni}}]{Calabrese:2012ew}%
  \BibitemOpen
  \bibfield  {author} {\bibinfo {author} {\bibfnamefont {P.}~\bibnamefont
  {Calabrese}}, \bibinfo {author} {\bibfnamefont {J.}~\bibnamefont {Cardy}},\
  and\ \bibinfo {author} {\bibfnamefont {E.}~\bibnamefont {Tonni}},\ }\bibfield
   {title} {\bibinfo {title} {{Entanglement negativity in quantum field
  theory}},\ }\href {https://doi.org/10.1103/PhysRevLett.109.130502} {\bibfield
   {journal} {\bibinfo  {journal} {Phys. Rev. Lett.}\ }\textbf {\bibinfo
  {volume} {109}},\ \bibinfo {pages} {130502} (\bibinfo {year} {2012})},\
  \Eprint {https://arxiv.org/abs/1206.3092} {arXiv:1206.3092
  [cond-mat.stat-mech]} \BibitemShut {NoStop}%
\bibitem [{\citenamefont {Calabrese}\ \emph {et~al.}(2013)\citenamefont
  {Calabrese}, \citenamefont {Cardy},\ and\ \citenamefont
  {Tonni}}]{Calabrese:2012nk}%
  \BibitemOpen
  \bibfield  {author} {\bibinfo {author} {\bibfnamefont {P.}~\bibnamefont
  {Calabrese}}, \bibinfo {author} {\bibfnamefont {J.}~\bibnamefont {Cardy}},\
  and\ \bibinfo {author} {\bibfnamefont {E.}~\bibnamefont {Tonni}},\ }\bibfield
   {title} {\bibinfo {title} {{Entanglement negativity in extended systems: A
  field theoretical approach}},\ }\href
  {https://doi.org/10.1088/1742-5468/2013/02/P02008} {\bibfield  {journal}
  {\bibinfo  {journal} {J. Stat. Mech.}\ }\textbf {\bibinfo {volume} {1302}},\
  \bibinfo {pages} {P02008} (\bibinfo {year} {2013})},\ \Eprint
  {https://arxiv.org/abs/1210.5359} {arXiv:1210.5359 [cond-mat.stat-mech]}
  \BibitemShut {NoStop}%
\bibitem [{\citenamefont {Mohammadi~Mozaffar}\ and\ \citenamefont
  {Mollabashi}(2017)}]{MohammadiMozaffar:2017nri}%
  \BibitemOpen
  \bibfield  {author} {\bibinfo {author} {\bibfnamefont {M.~R.}\ \bibnamefont
  {Mohammadi~Mozaffar}}\ and\ \bibinfo {author} {\bibfnamefont
  {A.}~\bibnamefont {Mollabashi}},\ }\bibfield  {title} {\bibinfo {title}
  {{Entanglement in Lifshitz-type Quantum Field Theories}},\ }\href
  {https://doi.org/10.1007/JHEP07(2017)120} {\bibfield  {journal} {\bibinfo
  {journal} {JHEP}\ }\textbf {\bibinfo {volume} {07}},\ \bibinfo {pages}
  {120}},\ \Eprint {https://arxiv.org/abs/1705.00483} {arXiv:1705.00483
  [hep-th]} \BibitemShut {NoStop}%
\bibitem [{\citenamefont {Coser}\ \emph {et~al.}(2017)\citenamefont {Coser},
  \citenamefont {De~Nobili},\ and\ \citenamefont {Tonni}}]{Coser_2017}%
  \BibitemOpen
  \bibfield  {author} {\bibinfo {author} {\bibfnamefont {A.}~\bibnamefont
  {Coser}}, \bibinfo {author} {\bibfnamefont {C.}~\bibnamefont {De~Nobili}},\
  and\ \bibinfo {author} {\bibfnamefont {E.}~\bibnamefont {Tonni}},\ }\bibfield
   {title} {\bibinfo {title} {A contour for the entanglement entropies in
  harmonic lattices},\ }\href {https://doi.org/10.1088/1751-8121/aa7902}
  {\bibfield  {journal} {\bibinfo  {journal} {Journal of Physics A:
  Mathematical and Theoretical}\ }\textbf {\bibinfo {volume} {50}},\ \bibinfo
  {pages} {314001} (\bibinfo {year} {2017})}\BibitemShut {NoStop}%
\bibitem [{\citenamefont {Ruggiero}\ \emph {et~al.}(2018)\citenamefont
  {Ruggiero}, \citenamefont {Tonni},\ and\ \citenamefont
  {Calabrese}}]{Ruggiero:2018hyl}%
  \BibitemOpen
  \bibfield  {author} {\bibinfo {author} {\bibfnamefont {P.}~\bibnamefont
  {Ruggiero}}, \bibinfo {author} {\bibfnamefont {E.}~\bibnamefont {Tonni}},\
  and\ \bibinfo {author} {\bibfnamefont {P.}~\bibnamefont {Calabrese}},\
  }\bibfield  {title} {\bibinfo {title} {{Entanglement entropy of two disjoint
  intervals and the recursion formula for conformal blocks}},\ }\href
  {https://doi.org/10.1088/1742-5468/aae5a8} {\bibfield  {journal} {\bibinfo
  {journal} {J. Stat. Mech.}\ }\textbf {\bibinfo {volume} {1811}},\ \bibinfo
  {pages} {113101} (\bibinfo {year} {2018})},\ \Eprint
  {https://arxiv.org/abs/1805.05975} {arXiv:1805.05975 [cond-mat.stat-mech]}
  \BibitemShut {NoStop}%
\bibitem [{\citenamefont {Di~Giulio}\ and\ \citenamefont
  {Tonni}(2020)}]{DiGiulio:2019cxv}%
  \BibitemOpen
  \bibfield  {author} {\bibinfo {author} {\bibfnamefont {G.}~\bibnamefont
  {Di~Giulio}}\ and\ \bibinfo {author} {\bibfnamefont {E.}~\bibnamefont
  {Tonni}},\ }\bibfield  {title} {\bibinfo {title} {{On entanglement
  hamiltonians of an interval in massless harmonic chains}},\ }\href
  {https://doi.org/10.1088/1742-5468/ab7129} {\bibfield  {journal} {\bibinfo
  {journal} {J. Stat. Mech.}\ }\textbf {\bibinfo {volume} {2003}},\ \bibinfo
  {pages} {033102} (\bibinfo {year} {2020})},\ \Eprint
  {https://arxiv.org/abs/1911.07188} {arXiv:1911.07188 [cond-mat.stat-mech]}
  \BibitemShut {NoStop}%
\bibitem [{\citenamefont {Klco}\ and\ \citenamefont
  {Savage}(2021{\natexlab{a}})}]{Klco:2020rga}%
  \BibitemOpen
  \bibfield  {author} {\bibinfo {author} {\bibfnamefont {N.}~\bibnamefont
  {Klco}}\ and\ \bibinfo {author} {\bibfnamefont {M.~J.}\ \bibnamefont
  {Savage}},\ }\bibfield  {title} {\bibinfo {title} {{Geometric quantum
  information structure in quantum fields and their lattice simulation}},\
  }\href {https://doi.org/10.1103/PhysRevD.103.065007} {\bibfield  {journal}
  {\bibinfo  {journal} {Phys. Rev. D}\ }\textbf {\bibinfo {volume} {103}},\
  \bibinfo {pages} {065007} (\bibinfo {year} {2021}{\natexlab{a}})},\ \Eprint
  {https://arxiv.org/abs/2008.03647} {arXiv:2008.03647 [quant-ph]} \BibitemShut
  {NoStop}%
\bibitem [{\citenamefont {Roy}\ \emph {et~al.}(2020)\citenamefont {Roy},
  \citenamefont {Pollmann},\ and\ \citenamefont {Saleur}}]{Roy:2020frd}%
  \BibitemOpen
  \bibfield  {author} {\bibinfo {author} {\bibfnamefont {A.}~\bibnamefont
  {Roy}}, \bibinfo {author} {\bibfnamefont {F.}~\bibnamefont {Pollmann}},\ and\
  \bibinfo {author} {\bibfnamefont {H.}~\bibnamefont {Saleur}},\ }\bibfield
  {title} {\bibinfo {title} {{Entanglement Hamiltonian of the 1+1-dimensional
  free, compactified boson conformal field theory}},\ }\href
  {https://doi.org/10.1088/1742-5468/aba498} {\bibfield  {journal} {\bibinfo
  {journal} {J. Stat. Mech.}\ }\textbf {\bibinfo {volume} {2008}},\ \bibinfo
  {pages} {083104} (\bibinfo {year} {2020})},\ \Eprint
  {https://arxiv.org/abs/2004.14370} {arXiv:2004.14370 [cond-mat.stat-mech]}
  \BibitemShut {NoStop}%
\bibitem [{\citenamefont {Klco}\ and\ \citenamefont
  {Savage}(2021{\natexlab{b}})}]{Klco:2021biu}%
  \BibitemOpen
  \bibfield  {author} {\bibinfo {author} {\bibfnamefont {N.}~\bibnamefont
  {Klco}}\ and\ \bibinfo {author} {\bibfnamefont {M.~J.}\ \bibnamefont
  {Savage}},\ }\bibfield  {title} {\bibinfo {title} {{Entanglement Spheres and
  a UV-IR Connection in Effective Field Theories}},\ }\href
  {https://doi.org/10.1103/PhysRevLett.127.211602} {\bibfield  {journal}
  {\bibinfo  {journal} {Phys. Rev. Lett.}\ }\textbf {\bibinfo {volume} {127}},\
  \bibinfo {pages} {211602} (\bibinfo {year} {2021}{\natexlab{b}})},\ \Eprint
  {https://arxiv.org/abs/2103.14999} {arXiv:2103.14999 [hep-th]} \BibitemShut
  {NoStop}%
\bibitem [{\citenamefont {Wang}\ \emph {et~al.}(2019)\citenamefont {Wang},
  \citenamefont {Blencowe}, \citenamefont {Wilson},\ and\ \citenamefont
  {Rimberg}}]{Wang:2018kla}%
  \BibitemOpen
  \bibfield  {author} {\bibinfo {author} {\bibfnamefont {H.}~\bibnamefont
  {Wang}}, \bibinfo {author} {\bibfnamefont {M.~P.}\ \bibnamefont {Blencowe}},
  \bibinfo {author} {\bibfnamefont {C.~M.}\ \bibnamefont {Wilson}},\ and\
  \bibinfo {author} {\bibfnamefont {A.~J.}\ \bibnamefont {Rimberg}},\
  }\bibfield  {title} {\bibinfo {title} {{Mechanically generating entangled
  photons from the vacuum: A microwave circuit-acoustic resonator analog of the
  oscillatory Unruh effect}},\ }\href
  {https://doi.org/10.1103/PhysRevA.99.053833} {\bibfield  {journal} {\bibinfo
  {journal} {Phys. Rev. A}\ }\textbf {\bibinfo {volume} {99}},\ \bibinfo
  {pages} {053833} (\bibinfo {year} {2019})},\ \Eprint
  {https://arxiv.org/abs/1811.10065} {arXiv:1811.10065 [quant-ph]} \BibitemShut
  {NoStop}%
\bibitem [{\citenamefont {Roy}\ \emph {et~al.}(2021)\citenamefont {Roy},
  \citenamefont {Schuricht}, \citenamefont {Hauschild}, \citenamefont
  {Pollmann},\ and\ \citenamefont {Saleur}}]{Roy:2020ppa}%
  \BibitemOpen
  \bibfield  {author} {\bibinfo {author} {\bibfnamefont {A.}~\bibnamefont
  {Roy}}, \bibinfo {author} {\bibfnamefont {D.}~\bibnamefont {Schuricht}},
  \bibinfo {author} {\bibfnamefont {J.}~\bibnamefont {Hauschild}}, \bibinfo
  {author} {\bibfnamefont {F.}~\bibnamefont {Pollmann}},\ and\ \bibinfo
  {author} {\bibfnamefont {H.}~\bibnamefont {Saleur}},\ }\bibfield  {title}
  {\bibinfo {title} {{The quantum sine-Gordon model with quantum circuits}},\
  }\href {https://doi.org/10.1016/j.nuclphysb.2021.115445} {\bibfield
  {journal} {\bibinfo  {journal} {Nucl. Phys. B}\ }\textbf {\bibinfo {volume}
  {968}},\ \bibinfo {pages} {115445} (\bibinfo {year} {2021})},\ \Eprint
  {https://arxiv.org/abs/2007.06874} {arXiv:2007.06874 [quant-ph]} \BibitemShut
  {NoStop}%
\bibitem [{\citenamefont {Hung}\ \emph {et~al.}(2021)\citenamefont {Hung},
  \citenamefont {Busnaina}, \citenamefont {Chang}, \citenamefont {Vadiraj},
  \citenamefont {Nsanzineza}, \citenamefont {Solano}, \citenamefont {Alaeian},
  \citenamefont {Rico},\ and\ \citenamefont {Wilson}}]{hung2021quantum}%
  \BibitemOpen
  \bibfield  {author} {\bibinfo {author} {\bibfnamefont {J.~S.~C.}\
  \bibnamefont {Hung}}, \bibinfo {author} {\bibfnamefont {J.~H.}\ \bibnamefont
  {Busnaina}}, \bibinfo {author} {\bibfnamefont {C.~W.~S.}\ \bibnamefont
  {Chang}}, \bibinfo {author} {\bibfnamefont {A.~M.}\ \bibnamefont {Vadiraj}},
  \bibinfo {author} {\bibfnamefont {I.}~\bibnamefont {Nsanzineza}}, \bibinfo
  {author} {\bibfnamefont {E.}~\bibnamefont {Solano}}, \bibinfo {author}
  {\bibfnamefont {H.}~\bibnamefont {Alaeian}}, \bibinfo {author} {\bibfnamefont
  {E.}~\bibnamefont {Rico}},\ and\ \bibinfo {author} {\bibfnamefont {C.~M.}\
  \bibnamefont {Wilson}},\ }\bibfield  {title} {\bibinfo {title} {Quantum
  simulation of the bosonic creutz ladder with a parametric cavity},\
  }\bibfield  {journal} {\bibinfo  {journal} {Physical Review Letters}\
  }\textbf {\bibinfo {volume} {127}},\ \href
  {https://doi.org/10.1103/physrevlett.127.100503}
  {10.1103/physrevlett.127.100503} (\bibinfo {year} {2021})\BibitemShut
  {NoStop}%
\bibitem [{\citenamefont {Kitaev}\ and\ \citenamefont
  {Webb}(2009)}]{kitaev2009wavefunction}%
  \BibitemOpen
  \bibfield  {author} {\bibinfo {author} {\bibfnamefont {A.}~\bibnamefont
  {Kitaev}}\ and\ \bibinfo {author} {\bibfnamefont {W.~A.}\ \bibnamefont
  {Webb}},\ }\bibfield  {title} {\bibinfo {title} {Wavefunction preparation and
  resampling using a quantum computer},\ }\href@noop {} {\  (\bibinfo {year}
  {2009})},\ \Eprint {https://arxiv.org/abs/0801.0342} {arXiv:0801.0342
  [quant-ph]} \BibitemShut {NoStop}%
\bibitem [{\citenamefont {Jordan}\ \emph {et~al.}(2014)\citenamefont {Jordan},
  \citenamefont {Lee},\ and\ \citenamefont {Preskill}}]{Jordan:2011ci}%
  \BibitemOpen
  \bibfield  {author} {\bibinfo {author} {\bibfnamefont {S.~P.}\ \bibnamefont
  {Jordan}}, \bibinfo {author} {\bibfnamefont {K.~S.~M.}\ \bibnamefont {Lee}},\
  and\ \bibinfo {author} {\bibfnamefont {J.}~\bibnamefont {Preskill}},\
  }\bibfield  {title} {\bibinfo {title} {{Quantum Computation of Scattering in
  Scalar Quantum Field Theories}},\ }\href@noop {} {\bibfield  {journal}
  {\bibinfo  {journal} {Quant. Inf. Comput.}\ }\textbf {\bibinfo {volume}
  {14}},\ \bibinfo {pages} {1014} (\bibinfo {year} {2014})},\ \Eprint
  {https://arxiv.org/abs/1112.4833} {arXiv:1112.4833 [hep-th]} \BibitemShut
  {NoStop}%
\bibitem [{\citenamefont {Jordan}\ \emph {et~al.}(2012)\citenamefont {Jordan},
  \citenamefont {Lee},\ and\ \citenamefont {Preskill}}]{Jordan:2011ne}%
  \BibitemOpen
  \bibfield  {author} {\bibinfo {author} {\bibfnamefont {S.~P.}\ \bibnamefont
  {Jordan}}, \bibinfo {author} {\bibfnamefont {K.~S.~M.}\ \bibnamefont {Lee}},\
  and\ \bibinfo {author} {\bibfnamefont {J.}~\bibnamefont {Preskill}},\
  }\bibfield  {title} {\bibinfo {title} {{Quantum Algorithms for Quantum Field
  Theories}},\ }\href {https://doi.org/10.1126/science.1217069} {\bibfield
  {journal} {\bibinfo  {journal} {Science}\ }\textbf {\bibinfo {volume}
  {336}},\ \bibinfo {pages} {1130} (\bibinfo {year} {2012})},\ \Eprint
  {https://arxiv.org/abs/1111.3633} {arXiv:1111.3633 [quant-ph]} \BibitemShut
  {NoStop}%
\bibitem [{\citenamefont {Marshall}\ \emph {et~al.}(2015)\citenamefont
  {Marshall}, \citenamefont {Pooser}, \citenamefont {Siopsis},\ and\
  \citenamefont {Weedbrook}}]{Marshall:2015mna}%
  \BibitemOpen
  \bibfield  {author} {\bibinfo {author} {\bibfnamefont {K.}~\bibnamefont
  {Marshall}}, \bibinfo {author} {\bibfnamefont {R.}~\bibnamefont {Pooser}},
  \bibinfo {author} {\bibfnamefont {G.}~\bibnamefont {Siopsis}},\ and\ \bibinfo
  {author} {\bibfnamefont {C.}~\bibnamefont {Weedbrook}},\ }\bibfield  {title}
  {\bibinfo {title} {{Quantum simulation of quantum field theory using
  continuous variables}},\ }\href {https://doi.org/10.1103/PhysRevA.92.063825}
  {\bibfield  {journal} {\bibinfo  {journal} {Phys. Rev. A}\ }\textbf {\bibinfo
  {volume} {92}},\ \bibinfo {pages} {063825} (\bibinfo {year} {2015})},\
  \Eprint {https://arxiv.org/abs/1503.08121} {arXiv:1503.08121 [quant-ph]}
  \BibitemShut {NoStop}%
\bibitem [{\citenamefont {Bermudez}\ \emph {et~al.}(2017)\citenamefont
  {Bermudez}, \citenamefont {Aarts},\ and\ \citenamefont
  {M\"uller}}]{Bermudez:2017yrq}%
  \BibitemOpen
  \bibfield  {author} {\bibinfo {author} {\bibfnamefont {A.}~\bibnamefont
  {Bermudez}}, \bibinfo {author} {\bibfnamefont {G.}~\bibnamefont {Aarts}},\
  and\ \bibinfo {author} {\bibfnamefont {M.}~\bibnamefont {M\"uller}},\
  }\bibfield  {title} {\bibinfo {title} {{Quantum sensors for the generating
  functional of interacting quantum field theories}},\ }\href
  {https://doi.org/10.1103/PhysRevX.7.041012} {\bibfield  {journal} {\bibinfo
  {journal} {Phys. Rev. X}\ }\textbf {\bibinfo {volume} {7}},\ \bibinfo {pages}
  {041012} (\bibinfo {year} {2017})},\ \Eprint
  {https://arxiv.org/abs/1704.02877} {arXiv:1704.02877 [quant-ph]} \BibitemShut
  {NoStop}%
\bibitem [{\citenamefont {Yeter-Aydeniz}\ and\ \citenamefont
  {Siopsis}(2018)}]{Yeter-Aydeniz:2017ubh}%
  \BibitemOpen
  \bibfield  {author} {\bibinfo {author} {\bibfnamefont {K.}~\bibnamefont
  {Yeter-Aydeniz}}\ and\ \bibinfo {author} {\bibfnamefont {G.}~\bibnamefont
  {Siopsis}},\ }\bibfield  {title} {\bibinfo {title} {{Quantum Computation of
  Scattering Amplitudes in Scalar Quantum Electrodynamics}},\ }\href
  {https://doi.org/10.1103/PhysRevD.97.036004} {\bibfield  {journal} {\bibinfo
  {journal} {Phys. Rev.}\ }\textbf {\bibinfo {volume} {D97}},\ \bibinfo {pages}
  {036004} (\bibinfo {year} {2018})},\ \Eprint
  {https://arxiv.org/abs/1709.02355} {arXiv:1709.02355 [quant-ph]} \BibitemShut
  {NoStop}%
\bibitem [{\citenamefont {Klco}\ and\ \citenamefont
  {Savage}(2019)}]{Klco:2018zqz}%
  \BibitemOpen
  \bibfield  {author} {\bibinfo {author} {\bibfnamefont {N.}~\bibnamefont
  {Klco}}\ and\ \bibinfo {author} {\bibfnamefont {M.~J.}\ \bibnamefont
  {Savage}},\ }\bibfield  {title} {\bibinfo {title} {{Digitization of scalar
  fields for quantum computing}},\ }\href
  {https://doi.org/10.1103/PhysRevA.99.052335} {\bibfield  {journal} {\bibinfo
  {journal} {Phys. Rev.}\ }\textbf {\bibinfo {volume} {A99}},\ \bibinfo {pages}
  {052335} (\bibinfo {year} {2019})},\ \Eprint
  {https://arxiv.org/abs/1808.10378} {arXiv:1808.10378 [quant-ph]} \BibitemShut
  {NoStop}%
\bibitem [{\citenamefont {Macridin}\ \emph {et~al.}(2018)\citenamefont
  {Macridin}, \citenamefont {Spentzouris}, \citenamefont {Amundson},\ and\
  \citenamefont {Harnik}}]{Macridin:2018oli}%
  \BibitemOpen
  \bibfield  {author} {\bibinfo {author} {\bibfnamefont {A.}~\bibnamefont
  {Macridin}}, \bibinfo {author} {\bibfnamefont {P.}~\bibnamefont
  {Spentzouris}}, \bibinfo {author} {\bibfnamefont {J.}~\bibnamefont
  {Amundson}},\ and\ \bibinfo {author} {\bibfnamefont {R.}~\bibnamefont
  {Harnik}},\ }\bibfield  {title} {\bibinfo {title} {{Digital quantum
  computation of fermion-boson interacting systems}},\ }\href
  {https://doi.org/10.1103/PhysRevA.98.042312} {\bibfield  {journal} {\bibinfo
  {journal} {Phys. Rev. A}\ }\textbf {\bibinfo {volume} {98}},\ \bibinfo
  {pages} {042312} (\bibinfo {year} {2018})},\ \Eprint
  {https://arxiv.org/abs/1805.09928} {arXiv:1805.09928 [quant-ph]} \BibitemShut
  {NoStop}%
\bibitem [{\citenamefont {Yeter-Aydeniz}\ \emph {et~al.}(2019)\citenamefont
  {Yeter-Aydeniz}, \citenamefont {Dumitrescu}, \citenamefont {McCaskey},
  \citenamefont {Bennink}, \citenamefont {Pooser},\ and\ \citenamefont
  {Siopsis}}]{Yeter-Aydeniz:2019scalar}%
  \BibitemOpen
  \bibfield  {author} {\bibinfo {author} {\bibfnamefont {K.}~\bibnamefont
  {Yeter-Aydeniz}}, \bibinfo {author} {\bibfnamefont {E.~F.}\ \bibnamefont
  {Dumitrescu}}, \bibinfo {author} {\bibfnamefont {A.~J.}\ \bibnamefont
  {McCaskey}}, \bibinfo {author} {\bibfnamefont {R.~S.}\ \bibnamefont
  {Bennink}}, \bibinfo {author} {\bibfnamefont {R.~C.}\ \bibnamefont
  {Pooser}},\ and\ \bibinfo {author} {\bibfnamefont {G.}~\bibnamefont
  {Siopsis}},\ }\bibfield  {title} {\bibinfo {title} {Scalar quantum field
  theories as a benchmark for near-term quantum computers},\ }\href
  {https://doi.org/10.1103/PhysRevA.99.032306} {\bibfield  {journal} {\bibinfo
  {journal} {Phys. Rev. A}\ }\textbf {\bibinfo {volume} {99}},\ \bibinfo
  {pages} {032306} (\bibinfo {year} {2019})}\BibitemShut {NoStop}%
\bibitem [{\citenamefont {Klco}\ and\ \citenamefont
  {Savage}(2020{\natexlab{a}})}]{Klco:2019yrb}%
  \BibitemOpen
  \bibfield  {author} {\bibinfo {author} {\bibfnamefont {N.}~\bibnamefont
  {Klco}}\ and\ \bibinfo {author} {\bibfnamefont {M.~J.}\ \bibnamefont
  {Savage}},\ }\bibfield  {title} {\bibinfo {title} {{Systematically
  Localizable Operators for Quantum Simulations of Quantum Field Theories}},\
  }\href {https://doi.org/10.1103/PhysRevA.102.012619} {\bibfield  {journal}
  {\bibinfo  {journal} {Phys. Rev. A}\ }\textbf {\bibinfo {volume} {102}},\
  \bibinfo {pages} {012619} (\bibinfo {year} {2020}{\natexlab{a}})},\ \Eprint
  {https://arxiv.org/abs/1912.03577} {arXiv:1912.03577 [quant-ph]} \BibitemShut
  {NoStop}%
\bibitem [{\citenamefont {Klco}\ and\ \citenamefont
  {Savage}(2020{\natexlab{b}})}]{Klco:2020aud}%
  \BibitemOpen
  \bibfield  {author} {\bibinfo {author} {\bibfnamefont {N.}~\bibnamefont
  {Klco}}\ and\ \bibinfo {author} {\bibfnamefont {M.~J.}\ \bibnamefont
  {Savage}},\ }\bibfield  {title} {\bibinfo {title} {{Fixed-point quantum
  circuits for quantum field theories}},\ }\href
  {https://doi.org/10.1103/PhysRevA.102.052422} {\bibfield  {journal} {\bibinfo
   {journal} {Phys. Rev. A}\ }\textbf {\bibinfo {volume} {102}},\ \bibinfo
  {pages} {052422} (\bibinfo {year} {2020}{\natexlab{b}})},\ \Eprint
  {https://arxiv.org/abs/2002.02018} {arXiv:2002.02018 [quant-ph]} \BibitemShut
  {NoStop}%
\bibitem [{\citenamefont {Barata}\ \emph {et~al.}(2021)\citenamefont {Barata},
  \citenamefont {Mueller}, \citenamefont {Tarasov},\ and\ \citenamefont
  {Venugopalan}}]{Barata:2020jtq}%
  \BibitemOpen
  \bibfield  {author} {\bibinfo {author} {\bibfnamefont {J.~a.}\ \bibnamefont
  {Barata}}, \bibinfo {author} {\bibfnamefont {N.}~\bibnamefont {Mueller}},
  \bibinfo {author} {\bibfnamefont {A.}~\bibnamefont {Tarasov}},\ and\ \bibinfo
  {author} {\bibfnamefont {R.}~\bibnamefont {Venugopalan}},\ }\bibfield
  {title} {\bibinfo {title} {{Single-particle digitization strategy for quantum
  computation of a $\phi^4$ scalar field theory}},\ }\href
  {https://doi.org/10.1103/PhysRevA.103.042410} {\bibfield  {journal} {\bibinfo
   {journal} {Phys. Rev. A}\ }\textbf {\bibinfo {volume} {103}},\ \bibinfo
  {pages} {042410} (\bibinfo {year} {2021})},\ \Eprint
  {https://arxiv.org/abs/2012.00020} {arXiv:2012.00020 [hep-th]} \BibitemShut
  {NoStop}%
\bibitem [{\citenamefont {Macridin}\ \emph {et~al.}(2021)\citenamefont
  {Macridin}, \citenamefont {Li}, \citenamefont {Mrenna},\ and\ \citenamefont
  {Spentzouris}}]{Macridin:2021uwn}%
  \BibitemOpen
  \bibfield  {author} {\bibinfo {author} {\bibfnamefont {A.}~\bibnamefont
  {Macridin}}, \bibinfo {author} {\bibfnamefont {A.~C.~Y.}\ \bibnamefont {Li}},
  \bibinfo {author} {\bibfnamefont {S.}~\bibnamefont {Mrenna}},\ and\ \bibinfo
  {author} {\bibfnamefont {P.}~\bibnamefont {Spentzouris}},\ }\bibfield
  {title} {\bibinfo {title} {{Bosonic field digitization for quantum
  computers}},\ }\href@noop {} {\  (\bibinfo {year} {2021})},\ \Eprint
  {https://arxiv.org/abs/2108.10793} {arXiv:2108.10793 [quant-ph]} \BibitemShut
  {NoStop}%
\bibitem [{\citenamefont {Kurkcuoglu}\ \emph {et~al.}(2021)\citenamefont
  {Kurkcuoglu}, \citenamefont {Alam}, \citenamefont {Li}, \citenamefont
  {Macridin},\ and\ \citenamefont {Perdue}}]{Kurkcuoglu:2021dnw}%
  \BibitemOpen
  \bibfield  {author} {\bibinfo {author} {\bibfnamefont {D.~M.}\ \bibnamefont
  {Kurkcuoglu}}, \bibinfo {author} {\bibfnamefont {M.~S.}\ \bibnamefont
  {Alam}}, \bibinfo {author} {\bibfnamefont {A.~C.~Y.}\ \bibnamefont {Li}},
  \bibinfo {author} {\bibfnamefont {A.}~\bibnamefont {Macridin}},\ and\
  \bibinfo {author} {\bibfnamefont {G.~N.}\ \bibnamefont {Perdue}},\ }\bibfield
   {title} {\bibinfo {title} {{Quantum simulation of $\phi^4$ theories in qudit
  systems}},\ }\href@noop {} {\  (\bibinfo {year} {2021})},\ \Eprint
  {https://arxiv.org/abs/2108.13357} {arXiv:2108.13357 [quant-ph]} \BibitemShut
  {NoStop}%
\bibitem [{\citenamefont {Deliyannis}\ \emph {et~al.}(2021)\citenamefont
  {Deliyannis}, \citenamefont {Freytsis}, \citenamefont {Nachman},\ and\
  \citenamefont {Bauer}}]{Deliyannis:2021che}%
  \BibitemOpen
  \bibfield  {author} {\bibinfo {author} {\bibfnamefont {P.}~\bibnamefont
  {Deliyannis}}, \bibinfo {author} {\bibfnamefont {M.}~\bibnamefont
  {Freytsis}}, \bibinfo {author} {\bibfnamefont {B.}~\bibnamefont {Nachman}},\
  and\ \bibinfo {author} {\bibfnamefont {C.~W.}\ \bibnamefont {Bauer}},\
  }\bibfield  {title} {\bibinfo {title} {{Practical considerations for the
  preparation of multivariate Gaussian states on quantum computers}},\
  }\href@noop {} {\  (\bibinfo {year} {2021})},\ \Eprint
  {https://arxiv.org/abs/2109.10918} {arXiv:2109.10918 [quant-ph]} \BibitemShut
  {NoStop}%
\bibitem [{\citenamefont {Byrnes}\ and\ \citenamefont
  {Yamamoto}(2006)}]{PhysRevA.73.022328}%
  \BibitemOpen
  \bibfield  {author} {\bibinfo {author} {\bibfnamefont {T.}~\bibnamefont
  {Byrnes}}\ and\ \bibinfo {author} {\bibfnamefont {Y.}~\bibnamefont
  {Yamamoto}},\ }\bibfield  {title} {\bibinfo {title} {Simulating lattice gauge
  theories on a quantum computer},\ }\href
  {https://doi.org/10.1103/PhysRevA.73.022328} {\bibfield  {journal} {\bibinfo
  {journal} {Phys. Rev. A}\ }\textbf {\bibinfo {volume} {73}},\ \bibinfo
  {pages} {022328} (\bibinfo {year} {2006})}\BibitemShut {NoStop}%
\bibitem [{\citenamefont {Zohar}\ \emph
  {et~al.}(2013{\natexlab{a}})\citenamefont {Zohar}, \citenamefont {Cirac},\
  and\ \citenamefont {Reznik}}]{Zohar:2012xf}%
  \BibitemOpen
  \bibfield  {author} {\bibinfo {author} {\bibfnamefont {E.}~\bibnamefont
  {Zohar}}, \bibinfo {author} {\bibfnamefont {J.~I.}\ \bibnamefont {Cirac}},\
  and\ \bibinfo {author} {\bibfnamefont {B.}~\bibnamefont {Reznik}},\
  }\bibfield  {title} {\bibinfo {title} {{Cold-Atom Quantum Simulator for SU(2)
  Yang-Mills Lattice Gauge Theory}},\ }\href
  {https://doi.org/10.1103/PhysRevLett.110.125304} {\bibfield  {journal}
  {\bibinfo  {journal} {Phys. Rev. Lett.}\ }\textbf {\bibinfo {volume} {110}},\
  \bibinfo {pages} {125304} (\bibinfo {year} {2013}{\natexlab{a}})},\ \Eprint
  {https://arxiv.org/abs/1211.2241} {arXiv:1211.2241 [quant-ph]} \BibitemShut
  {NoStop}%
\bibitem [{\citenamefont {Banerjee}\ \emph {et~al.}(2013)\citenamefont
  {Banerjee}, \citenamefont {B\"ogli}, \citenamefont {Dalmonte}, \citenamefont
  {Rico}, \citenamefont {Stebler}, \citenamefont {Wiese},\ and\ \citenamefont
  {Zoller}}]{Banerjee:2012xg}%
  \BibitemOpen
  \bibfield  {author} {\bibinfo {author} {\bibfnamefont {D.}~\bibnamefont
  {Banerjee}}, \bibinfo {author} {\bibfnamefont {M.}~\bibnamefont {B\"ogli}},
  \bibinfo {author} {\bibfnamefont {M.}~\bibnamefont {Dalmonte}}, \bibinfo
  {author} {\bibfnamefont {E.}~\bibnamefont {Rico}}, \bibinfo {author}
  {\bibfnamefont {P.}~\bibnamefont {Stebler}}, \bibinfo {author} {\bibfnamefont
  {U.~J.}\ \bibnamefont {Wiese}},\ and\ \bibinfo {author} {\bibfnamefont
  {P.}~\bibnamefont {Zoller}},\ }\bibfield  {title} {\bibinfo {title} {{Atomic
  Quantum Simulation of U(N) and SU(N) Non-Abelian Lattice Gauge Theories}},\
  }\href {https://doi.org/10.1103/PhysRevLett.110.125303} {\bibfield  {journal}
  {\bibinfo  {journal} {Phys. Rev. Lett.}\ }\textbf {\bibinfo {volume} {110}},\
  \bibinfo {pages} {125303} (\bibinfo {year} {2013})},\ \Eprint
  {https://arxiv.org/abs/1211.2242} {arXiv:1211.2242 [cond-mat.quant-gas]}
  \BibitemShut {NoStop}%
\bibitem [{\citenamefont {Tagliacozzo}\ \emph {et~al.}(2013)\citenamefont
  {Tagliacozzo}, \citenamefont {Celi}, \citenamefont {Orland},\ and\
  \citenamefont {Lewenstein}}]{Tagliacozzo:2012df}%
  \BibitemOpen
  \bibfield  {author} {\bibinfo {author} {\bibfnamefont {L.}~\bibnamefont
  {Tagliacozzo}}, \bibinfo {author} {\bibfnamefont {A.}~\bibnamefont {Celi}},
  \bibinfo {author} {\bibfnamefont {P.}~\bibnamefont {Orland}},\ and\ \bibinfo
  {author} {\bibfnamefont {M.}~\bibnamefont {Lewenstein}},\ }\bibfield  {title}
  {\bibinfo {title} {{Simulations of non-Abelian gauge theories with optical
  lattices}},\ }\href {https://doi.org/10.1038/ncomms3615} {\bibfield
  {journal} {\bibinfo  {journal} {Nature Commun.}\ }\textbf {\bibinfo {volume}
  {4}},\ \bibinfo {pages} {2615} (\bibinfo {year} {2013})},\ \Eprint
  {https://arxiv.org/abs/1211.2704} {arXiv:1211.2704 [cond-mat.quant-gas]}
  \BibitemShut {NoStop}%
\bibitem [{\citenamefont {Stannigel}\ \emph {et~al.}(2014)\citenamefont
  {Stannigel}, \citenamefont {Hauke}, \citenamefont {Marcos}, \citenamefont
  {Hafezi}, \citenamefont {Diehl}, \citenamefont {Dalmonte},\ and\
  \citenamefont {Zoller}}]{Stannigel:2013zka}%
  \BibitemOpen
  \bibfield  {author} {\bibinfo {author} {\bibfnamefont {K.}~\bibnamefont
  {Stannigel}}, \bibinfo {author} {\bibfnamefont {P.}~\bibnamefont {Hauke}},
  \bibinfo {author} {\bibfnamefont {D.}~\bibnamefont {Marcos}}, \bibinfo
  {author} {\bibfnamefont {M.}~\bibnamefont {Hafezi}}, \bibinfo {author}
  {\bibfnamefont {S.}~\bibnamefont {Diehl}}, \bibinfo {author} {\bibfnamefont
  {M.}~\bibnamefont {Dalmonte}},\ and\ \bibinfo {author} {\bibfnamefont
  {P.}~\bibnamefont {Zoller}},\ }\bibfield  {title} {\bibinfo {title}
  {{Constrained dynamics via the Zeno effect in quantum simulation:
  Implementing non-Abelian lattice gauge theories with cold atoms}},\ }\href
  {https://doi.org/10.1103/PhysRevLett.112.120406} {\bibfield  {journal}
  {\bibinfo  {journal} {Phys. Rev. Lett.}\ }\textbf {\bibinfo {volume} {112}},\
  \bibinfo {pages} {120406} (\bibinfo {year} {2014})},\ \Eprint
  {https://arxiv.org/abs/1308.0528} {arXiv:1308.0528 [quant-ph]} \BibitemShut
  {NoStop}%
\bibitem [{\citenamefont {Zohar}\ \emph
  {et~al.}(2013{\natexlab{b}})\citenamefont {Zohar}, \citenamefont {Cirac},\
  and\ \citenamefont {Reznik}}]{Zohar:2013zla}%
  \BibitemOpen
  \bibfield  {author} {\bibinfo {author} {\bibfnamefont {E.}~\bibnamefont
  {Zohar}}, \bibinfo {author} {\bibfnamefont {J.~I.}\ \bibnamefont {Cirac}},\
  and\ \bibinfo {author} {\bibfnamefont {B.}~\bibnamefont {Reznik}},\
  }\bibfield  {title} {\bibinfo {title} {{Quantum simulations of gauge theories
  with ultracold atoms: local gauge invariance from angular momentum
  conservation}},\ }\href {https://doi.org/10.1103/PhysRevA.88.023617}
  {\bibfield  {journal} {\bibinfo  {journal} {Phys. Rev. A}\ }\textbf {\bibinfo
  {volume} {88}},\ \bibinfo {pages} {023617} (\bibinfo {year}
  {2013}{\natexlab{b}})},\ \Eprint {https://arxiv.org/abs/1303.5040}
  {arXiv:1303.5040 [quant-ph]} \BibitemShut {NoStop}%
\bibitem [{\citenamefont {Zohar}\ and\ \citenamefont
  {Burrello}(2015)}]{Zohar:2014qma}%
  \BibitemOpen
  \bibfield  {author} {\bibinfo {author} {\bibfnamefont {E.}~\bibnamefont
  {Zohar}}\ and\ \bibinfo {author} {\bibfnamefont {M.}~\bibnamefont
  {Burrello}},\ }\bibfield  {title} {\bibinfo {title} {{Formulation of lattice
  gauge theories for quantum simulations}},\ }\href
  {https://doi.org/10.1103/PhysRevD.91.054506} {\bibfield  {journal} {\bibinfo
  {journal} {Phys. Rev. D}\ }\textbf {\bibinfo {volume} {91}},\ \bibinfo
  {pages} {054506} (\bibinfo {year} {2015})},\ \Eprint
  {https://arxiv.org/abs/1409.3085} {arXiv:1409.3085 [quant-ph]} \BibitemShut
  {NoStop}%
\bibitem [{\citenamefont {Kasper}\ \emph {et~al.}(2016)\citenamefont {Kasper},
  \citenamefont {Hebenstreit}, \citenamefont {Oberthaler},\ and\ \citenamefont
  {Berges}}]{Kasper:2015cca}%
  \BibitemOpen
  \bibfield  {author} {\bibinfo {author} {\bibfnamefont {V.}~\bibnamefont
  {Kasper}}, \bibinfo {author} {\bibfnamefont {F.}~\bibnamefont {Hebenstreit}},
  \bibinfo {author} {\bibfnamefont {M.}~\bibnamefont {Oberthaler}},\ and\
  \bibinfo {author} {\bibfnamefont {J.}~\bibnamefont {Berges}},\ }\bibfield
  {title} {\bibinfo {title} {{Schwinger pair production with ultracold
  atoms}},\ }\href {https://doi.org/10.1016/j.physletb.2016.07.036} {\bibfield
  {journal} {\bibinfo  {journal} {Phys. Lett. B}\ }\textbf {\bibinfo {volume}
  {760}},\ \bibinfo {pages} {742} (\bibinfo {year} {2016})},\ \Eprint
  {https://arxiv.org/abs/1506.01238} {arXiv:1506.01238 [cond-mat.quant-gas]}
  \BibitemShut {NoStop}%
\bibitem [{\citenamefont {Mezzacapo}\ \emph {et~al.}(2015)\citenamefont
  {Mezzacapo}, \citenamefont {Rico}, \citenamefont {Sab\'\i{}n}, \citenamefont
  {Egusquiza}, \citenamefont {Lamata},\ and\ \citenamefont
  {Solano}}]{Mezzacapo:2015bra}%
  \BibitemOpen
  \bibfield  {author} {\bibinfo {author} {\bibfnamefont {A.}~\bibnamefont
  {Mezzacapo}}, \bibinfo {author} {\bibfnamefont {E.}~\bibnamefont {Rico}},
  \bibinfo {author} {\bibfnamefont {C.}~\bibnamefont {Sab\'\i{}n}}, \bibinfo
  {author} {\bibfnamefont {I.~L.}\ \bibnamefont {Egusquiza}}, \bibinfo {author}
  {\bibfnamefont {L.}~\bibnamefont {Lamata}},\ and\ \bibinfo {author}
  {\bibfnamefont {E.}~\bibnamefont {Solano}},\ }\bibfield  {title} {\bibinfo
  {title} {{Non-Abelian $SU(2)$ Lattice Gauge Theories in Superconducting
  Circuits}},\ }\href {https://doi.org/10.1103/PhysRevLett.115.240502}
  {\bibfield  {journal} {\bibinfo  {journal} {Phys. Rev. Lett.}\ }\textbf
  {\bibinfo {volume} {115}},\ \bibinfo {pages} {240502} (\bibinfo {year}
  {2015})},\ \Eprint {https://arxiv.org/abs/1505.04720} {arXiv:1505.04720
  [quant-ph]} \BibitemShut {NoStop}%
\bibitem [{\citenamefont {Bazavov}\ \emph {et~al.}(2015)\citenamefont
  {Bazavov}, \citenamefont {Meurice}, \citenamefont {Tsai}, \citenamefont
  {Unmuth-Yockey},\ and\ \citenamefont {Zhang}}]{Bazavov:2015kka}%
  \BibitemOpen
  \bibfield  {author} {\bibinfo {author} {\bibfnamefont {A.}~\bibnamefont
  {Bazavov}}, \bibinfo {author} {\bibfnamefont {Y.}~\bibnamefont {Meurice}},
  \bibinfo {author} {\bibfnamefont {S.-W.}\ \bibnamefont {Tsai}}, \bibinfo
  {author} {\bibfnamefont {J.}~\bibnamefont {Unmuth-Yockey}},\ and\ \bibinfo
  {author} {\bibfnamefont {J.}~\bibnamefont {Zhang}},\ }\bibfield  {title}
  {\bibinfo {title} {{Gauge-invariant implementation of the Abelian Higgs model
  on optical lattices}},\ }\href {https://doi.org/10.1103/PhysRevD.92.076003}
  {\bibfield  {journal} {\bibinfo  {journal} {Phys. Rev. D}\ }\textbf {\bibinfo
  {volume} {92}},\ \bibinfo {pages} {076003} (\bibinfo {year} {2015})},\
  \Eprint {https://arxiv.org/abs/1503.08354} {arXiv:1503.08354 [hep-lat]}
  \BibitemShut {NoStop}%
\bibitem [{\citenamefont {Martinez}\ \emph {et~al.}(2016)\citenamefont
  {Martinez} \emph {et~al.}}]{Martinez:2016yna}%
  \BibitemOpen
  \bibfield  {author} {\bibinfo {author} {\bibfnamefont {E.~A.}\ \bibnamefont
  {Martinez}} \emph {et~al.},\ }\bibfield  {title} {\bibinfo {title}
  {{Real-time dynamics of lattice gauge theories with a few-qubit quantum
  computer}},\ }\href {https://doi.org/10.1038/nature18318} {\bibfield
  {journal} {\bibinfo  {journal} {Nature}\ }\textbf {\bibinfo {volume} {534}},\
  \bibinfo {pages} {516} (\bibinfo {year} {2016})},\ \Eprint
  {https://arxiv.org/abs/1605.04570} {arXiv:1605.04570 [quant-ph]} \BibitemShut
  {NoStop}%
\bibitem [{\citenamefont {Gonz\'alez-Cuadra}\ \emph {et~al.}(2017)\citenamefont
  {Gonz\'alez-Cuadra}, \citenamefont {Zohar},\ and\ \citenamefont
  {Cirac}}]{Gonzalez-Cuadra:2017lvz}%
  \BibitemOpen
  \bibfield  {author} {\bibinfo {author} {\bibfnamefont {D.}~\bibnamefont
  {Gonz\'alez-Cuadra}}, \bibinfo {author} {\bibfnamefont {E.}~\bibnamefont
  {Zohar}},\ and\ \bibinfo {author} {\bibfnamefont {J.~I.}\ \bibnamefont
  {Cirac}},\ }\bibfield  {title} {\bibinfo {title} {{Quantum Simulation of the
  Abelian-Higgs Lattice Gauge Theory with Ultracold Atoms}},\ }\href
  {https://doi.org/10.1088/1367-2630/aa6f37} {\bibfield  {journal} {\bibinfo
  {journal} {New J. Phys.}\ }\textbf {\bibinfo {volume} {19}},\ \bibinfo
  {pages} {063038} (\bibinfo {year} {2017})},\ \Eprint
  {https://arxiv.org/abs/1702.05492} {arXiv:1702.05492 [quant-ph]} \BibitemShut
  {NoStop}%
\bibitem [{\citenamefont {Zohar}\ and\ \citenamefont
  {Cirac}(2018)}]{Zohar:2018cwb}%
  \BibitemOpen
  \bibfield  {author} {\bibinfo {author} {\bibfnamefont {E.}~\bibnamefont
  {Zohar}}\ and\ \bibinfo {author} {\bibfnamefont {J.~I.}\ \bibnamefont
  {Cirac}},\ }\bibfield  {title} {\bibinfo {title} {{Eliminating fermionic
  matter fields in lattice gauge theories}},\ }\href
  {https://doi.org/10.1103/PhysRevB.98.075119} {\bibfield  {journal} {\bibinfo
  {journal} {Phys. Rev. B}\ }\textbf {\bibinfo {volume} {98}},\ \bibinfo
  {pages} {075119} (\bibinfo {year} {2018})},\ \Eprint
  {https://arxiv.org/abs/1805.05347} {arXiv:1805.05347 [quant-ph]} \BibitemShut
  {NoStop}%
\bibitem [{\citenamefont {Raychowdhury}\ and\ \citenamefont
  {Stryker}(2020)}]{Raychowdhury:2018osk}%
  \BibitemOpen
  \bibfield  {author} {\bibinfo {author} {\bibfnamefont {I.}~\bibnamefont
  {Raychowdhury}}\ and\ \bibinfo {author} {\bibfnamefont {J.~R.}\ \bibnamefont
  {Stryker}},\ }\bibfield  {title} {\bibinfo {title} {{Solving Gauss's Law on
  Digital Quantum Computers with Loop-String-Hadron Digitization}},\ }\href
  {https://doi.org/10.1103/PhysRevResearch.2.033039} {\bibfield  {journal}
  {\bibinfo  {journal} {Phys. Rev. Res.}\ }\textbf {\bibinfo {volume} {2}},\
  \bibinfo {pages} {033039} (\bibinfo {year} {2020})},\ \Eprint
  {https://arxiv.org/abs/1812.07554} {arXiv:1812.07554 [hep-lat]} \BibitemShut
  {NoStop}%
\bibitem [{\citenamefont {Stryker}(2019)}]{Stryker:2018efp}%
  \BibitemOpen
  \bibfield  {author} {\bibinfo {author} {\bibfnamefont {J.~R.}\ \bibnamefont
  {Stryker}},\ }\bibfield  {title} {\bibinfo {title} {{Oracles for Gauss's law
  on digital quantum computers}},\ }\href
  {https://doi.org/10.1103/PhysRevA.99.042301} {\bibfield  {journal} {\bibinfo
  {journal} {Phys. Rev. A}\ }\textbf {\bibinfo {volume} {99}},\ \bibinfo
  {pages} {042301} (\bibinfo {year} {2019})},\ \Eprint
  {https://arxiv.org/abs/1812.01617} {arXiv:1812.01617 [quant-ph]} \BibitemShut
  {NoStop}%
\bibitem [{\citenamefont {Zohar}\ and\ \citenamefont
  {Cirac}(2019)}]{Zohar:2019ygc}%
  \BibitemOpen
  \bibfield  {author} {\bibinfo {author} {\bibfnamefont {E.}~\bibnamefont
  {Zohar}}\ and\ \bibinfo {author} {\bibfnamefont {J.~I.}\ \bibnamefont
  {Cirac}},\ }\bibfield  {title} {\bibinfo {title} {{Removing Staggered
  Fermionic Matter in $U(N)$ and $SU(N)$ Lattice Gauge Theories}},\ }\href
  {https://doi.org/10.1103/PhysRevD.99.114511} {\bibfield  {journal} {\bibinfo
  {journal} {Phys. Rev. D}\ }\textbf {\bibinfo {volume} {99}},\ \bibinfo
  {pages} {114511} (\bibinfo {year} {2019})},\ \Eprint
  {https://arxiv.org/abs/1905.00652} {arXiv:1905.00652 [quant-ph]} \BibitemShut
  {NoStop}%
\bibitem [{\citenamefont {Zache}\ \emph {et~al.}(2020)\citenamefont {Zache},
  \citenamefont {Schweigler}, \citenamefont {Erne}, \citenamefont
  {Schmiedmayer},\ and\ \citenamefont {Berges}}]{Zache:2019xkx}%
  \BibitemOpen
  \bibfield  {author} {\bibinfo {author} {\bibfnamefont {T.~V.}\ \bibnamefont
  {Zache}}, \bibinfo {author} {\bibfnamefont {T.}~\bibnamefont {Schweigler}},
  \bibinfo {author} {\bibfnamefont {S.}~\bibnamefont {Erne}}, \bibinfo {author}
  {\bibfnamefont {J.}~\bibnamefont {Schmiedmayer}},\ and\ \bibinfo {author}
  {\bibfnamefont {J.}~\bibnamefont {Berges}},\ }\bibfield  {title} {\bibinfo
  {title} {{Extracting the field theory description of a quantum many-body
  system from experimental data}},\ }\href
  {https://doi.org/10.1103/PhysRevX.10.011020} {\bibfield  {journal} {\bibinfo
  {journal} {Phys. Rev. X}\ }\textbf {\bibinfo {volume} {10}},\ \bibinfo
  {pages} {011020} (\bibinfo {year} {2020})},\ \Eprint
  {https://arxiv.org/abs/1909.12815} {arXiv:1909.12815 [cond-mat.quant-gas]}
  \BibitemShut {NoStop}%
\bibitem [{\citenamefont {Alexandru}\ \emph {et~al.}(2019)\citenamefont
  {Alexandru}, \citenamefont {Bedaque}, \citenamefont {Harmalkar},
  \citenamefont {Lamm}, \citenamefont {Lawrence},\ and\ \citenamefont
  {Warrington}}]{Alexandru:2019nsa}%
  \BibitemOpen
  \bibfield  {author} {\bibinfo {author} {\bibfnamefont {A.}~\bibnamefont
  {Alexandru}}, \bibinfo {author} {\bibfnamefont {P.~F.}\ \bibnamefont
  {Bedaque}}, \bibinfo {author} {\bibfnamefont {S.}~\bibnamefont {Harmalkar}},
  \bibinfo {author} {\bibfnamefont {H.}~\bibnamefont {Lamm}}, \bibinfo {author}
  {\bibfnamefont {S.}~\bibnamefont {Lawrence}},\ and\ \bibinfo {author}
  {\bibfnamefont {N.~C.}\ \bibnamefont {Warrington}} (\bibinfo {collaboration}
  {NuQS}),\ }\bibfield  {title} {\bibinfo {title} {{Gluon Field Digitization
  for Quantum Computers}},\ }\href
  {https://doi.org/10.1103/PhysRevD.100.114501} {\bibfield  {journal} {\bibinfo
   {journal} {Phys. Rev. D}\ }\textbf {\bibinfo {volume} {100}},\ \bibinfo
  {pages} {114501} (\bibinfo {year} {2019})},\ \Eprint
  {https://arxiv.org/abs/1906.11213} {arXiv:1906.11213 [hep-lat]} \BibitemShut
  {NoStop}%
\bibitem [{\citenamefont {Singh}\ and\ \citenamefont
  {Chandrasekharan}(2019)}]{Singh:2019uwd}%
  \BibitemOpen
  \bibfield  {author} {\bibinfo {author} {\bibfnamefont {H.}~\bibnamefont
  {Singh}}\ and\ \bibinfo {author} {\bibfnamefont {S.}~\bibnamefont
  {Chandrasekharan}},\ }\bibfield  {title} {\bibinfo {title} {{Qubit
  regularization of the $O(3)$ sigma model}},\ }\href
  {https://doi.org/10.1103/PhysRevD.100.054505} {\bibfield  {journal} {\bibinfo
   {journal} {Phys. Rev. D}\ }\textbf {\bibinfo {volume} {100}},\ \bibinfo
  {pages} {054505} (\bibinfo {year} {2019})},\ \Eprint
  {https://arxiv.org/abs/1905.13204} {arXiv:1905.13204 [hep-lat]} \BibitemShut
  {NoStop}%
\bibitem [{\citenamefont {Luo}\ \emph {et~al.}(2020)\citenamefont {Luo},
  \citenamefont {Shen}, \citenamefont {Highman}, \citenamefont {Clark},
  \citenamefont {DeMarco}, \citenamefont {El-Khadra},\ and\ \citenamefont
  {Gadway}}]{Luo:2019vmi}%
  \BibitemOpen
  \bibfield  {author} {\bibinfo {author} {\bibfnamefont {D.}~\bibnamefont
  {Luo}}, \bibinfo {author} {\bibfnamefont {J.}~\bibnamefont {Shen}}, \bibinfo
  {author} {\bibfnamefont {M.}~\bibnamefont {Highman}}, \bibinfo {author}
  {\bibfnamefont {B.~K.}\ \bibnamefont {Clark}}, \bibinfo {author}
  {\bibfnamefont {B.}~\bibnamefont {DeMarco}}, \bibinfo {author} {\bibfnamefont
  {A.~X.}\ \bibnamefont {El-Khadra}},\ and\ \bibinfo {author} {\bibfnamefont
  {B.}~\bibnamefont {Gadway}},\ }\bibfield  {title} {\bibinfo {title}
  {{Framework for simulating gauge theories with dipolar spin systems}},\
  }\href {https://doi.org/10.1103/PhysRevA.102.032617} {\bibfield  {journal}
  {\bibinfo  {journal} {Phys. Rev. A}\ }\textbf {\bibinfo {volume} {102}},\
  \bibinfo {pages} {032617} (\bibinfo {year} {2020})},\ \Eprint
  {https://arxiv.org/abs/1912.11488} {arXiv:1912.11488 [quant-ph]} \BibitemShut
  {NoStop}%
\bibitem [{\citenamefont {Surace}\ \emph
  {et~al.}(2020{\natexlab{a}})\citenamefont {Surace}, \citenamefont {Mazza},
  \citenamefont {Giudici}, \citenamefont {Lerose}, \citenamefont {Gambassi},\
  and\ \citenamefont {Dalmonte}}]{Surace:2019dtp}%
  \BibitemOpen
  \bibfield  {author} {\bibinfo {author} {\bibfnamefont {F.~M.}\ \bibnamefont
  {Surace}}, \bibinfo {author} {\bibfnamefont {P.~P.}\ \bibnamefont {Mazza}},
  \bibinfo {author} {\bibfnamefont {G.}~\bibnamefont {Giudici}}, \bibinfo
  {author} {\bibfnamefont {A.}~\bibnamefont {Lerose}}, \bibinfo {author}
  {\bibfnamefont {A.}~\bibnamefont {Gambassi}},\ and\ \bibinfo {author}
  {\bibfnamefont {M.}~\bibnamefont {Dalmonte}},\ }\bibfield  {title} {\bibinfo
  {title} {{Lattice gauge theories and string dynamics in Rydberg atom quantum
  simulators}},\ }\href {https://doi.org/10.1103/PhysRevX.10.021041} {\bibfield
   {journal} {\bibinfo  {journal} {Phys. Rev. X}\ }\textbf {\bibinfo {volume}
  {10}},\ \bibinfo {pages} {021041} (\bibinfo {year} {2020}{\natexlab{a}})},\
  \Eprint {https://arxiv.org/abs/1902.09551} {arXiv:1902.09551
  [cond-mat.quant-gas]} \BibitemShut {NoStop}%
\bibitem [{\citenamefont {Klco}\ \emph {et~al.}(2020)\citenamefont {Klco},
  \citenamefont {Stryker},\ and\ \citenamefont {Savage}}]{Klco:2019evd}%
  \BibitemOpen
  \bibfield  {author} {\bibinfo {author} {\bibfnamefont {N.}~\bibnamefont
  {Klco}}, \bibinfo {author} {\bibfnamefont {J.~R.}\ \bibnamefont {Stryker}},\
  and\ \bibinfo {author} {\bibfnamefont {M.~J.}\ \bibnamefont {Savage}},\
  }\bibfield  {title} {\bibinfo {title} {{SU(2) non-Abelian gauge field theory
  in one dimension on digital quantum computers}},\ }\href
  {https://doi.org/10.1103/PhysRevD.101.074512} {\bibfield  {journal} {\bibinfo
   {journal} {Phys. Rev. D}\ }\textbf {\bibinfo {volume} {101}},\ \bibinfo
  {pages} {074512} (\bibinfo {year} {2020})},\ \Eprint
  {https://arxiv.org/abs/1908.06935} {arXiv:1908.06935 [quant-ph]} \BibitemShut
  {NoStop}%
\bibitem [{\citenamefont {Mil}\ \emph {et~al.}(2020)\citenamefont {Mil},
  \citenamefont {Zache}, \citenamefont {Hegde}, \citenamefont {Xia},
  \citenamefont {Bhatt}, \citenamefont {Oberthaler}, \citenamefont {Hauke},
  \citenamefont {Berges},\ and\ \citenamefont {Jendrzejewski}}]{Mil:2019pbt}%
  \BibitemOpen
  \bibfield  {author} {\bibinfo {author} {\bibfnamefont {A.}~\bibnamefont
  {Mil}}, \bibinfo {author} {\bibfnamefont {T.~V.}\ \bibnamefont {Zache}},
  \bibinfo {author} {\bibfnamefont {A.}~\bibnamefont {Hegde}}, \bibinfo
  {author} {\bibfnamefont {A.}~\bibnamefont {Xia}}, \bibinfo {author}
  {\bibfnamefont {R.~P.}\ \bibnamefont {Bhatt}}, \bibinfo {author}
  {\bibfnamefont {M.~K.}\ \bibnamefont {Oberthaler}}, \bibinfo {author}
  {\bibfnamefont {P.}~\bibnamefont {Hauke}}, \bibinfo {author} {\bibfnamefont
  {J.}~\bibnamefont {Berges}},\ and\ \bibinfo {author} {\bibfnamefont
  {F.}~\bibnamefont {Jendrzejewski}},\ }\bibfield  {title} {\bibinfo {title}
  {{A scalable realization of local U(1) gauge invariance in cold atomic
  mixtures}},\ }\href {https://doi.org/10.1126/science.aaz5312} {\bibfield
  {journal} {\bibinfo  {journal} {Science}\ }\textbf {\bibinfo {volume}
  {367}},\ \bibinfo {pages} {1128} (\bibinfo {year} {2020})},\ \Eprint
  {https://arxiv.org/abs/1909.07641} {arXiv:1909.07641 [cond-mat.quant-gas]}
  \BibitemShut {NoStop}%
\bibitem [{\citenamefont {Ott}\ \emph {et~al.}(2021)\citenamefont {Ott},
  \citenamefont {Zache}, \citenamefont {Jendrzejewski},\ and\ \citenamefont
  {Berges}}]{Ott:2020ycj}%
  \BibitemOpen
  \bibfield  {author} {\bibinfo {author} {\bibfnamefont {R.}~\bibnamefont
  {Ott}}, \bibinfo {author} {\bibfnamefont {T.~V.}\ \bibnamefont {Zache}},
  \bibinfo {author} {\bibfnamefont {F.}~\bibnamefont {Jendrzejewski}},\ and\
  \bibinfo {author} {\bibfnamefont {J.}~\bibnamefont {Berges}},\ }\bibfield
  {title} {\bibinfo {title} {{Scalable Cold-Atom Quantum Simulator for
  Two-Dimensional QED}},\ }\href
  {https://doi.org/10.1103/PhysRevLett.127.130504} {\bibfield  {journal}
  {\bibinfo  {journal} {Phys. Rev. Lett.}\ }\textbf {\bibinfo {volume} {127}},\
  \bibinfo {pages} {130504} (\bibinfo {year} {2021})},\ \Eprint
  {https://arxiv.org/abs/2012.10432} {arXiv:2012.10432 [cond-mat.quant-gas]}
  \BibitemShut {NoStop}%
\bibitem [{\citenamefont {Haase}\ \emph {et~al.}(2021)\citenamefont {Haase},
  \citenamefont {Dellantonio}, \citenamefont {Celi}, \citenamefont {Paulson},
  \citenamefont {Kan}, \citenamefont {Jansen},\ and\ \citenamefont
  {Muschik}}]{Haase:2020kaj}%
  \BibitemOpen
  \bibfield  {author} {\bibinfo {author} {\bibfnamefont {J.~F.}\ \bibnamefont
  {Haase}}, \bibinfo {author} {\bibfnamefont {L.}~\bibnamefont {Dellantonio}},
  \bibinfo {author} {\bibfnamefont {A.}~\bibnamefont {Celi}}, \bibinfo {author}
  {\bibfnamefont {D.}~\bibnamefont {Paulson}}, \bibinfo {author} {\bibfnamefont
  {A.}~\bibnamefont {Kan}}, \bibinfo {author} {\bibfnamefont {K.}~\bibnamefont
  {Jansen}},\ and\ \bibinfo {author} {\bibfnamefont {C.~A.}\ \bibnamefont
  {Muschik}},\ }\bibfield  {title} {\bibinfo {title} {{A resource efficient
  approach for quantum and classical simulations of gauge theories in particle
  physics}},\ }\href {https://doi.org/10.22331/q-2021-02-04-393} {\bibfield
  {journal} {\bibinfo  {journal} {Quantum}\ }\textbf {\bibinfo {volume} {5}},\
  \bibinfo {pages} {393} (\bibinfo {year} {2021})},\ \Eprint
  {https://arxiv.org/abs/2006.14160} {arXiv:2006.14160 [quant-ph]} \BibitemShut
  {NoStop}%
\bibitem [{\citenamefont {Dasgupta}\ and\ \citenamefont
  {Raychowdhury}(2020)}]{Dasgupta:2020itb}%
  \BibitemOpen
  \bibfield  {author} {\bibinfo {author} {\bibfnamefont {R.}~\bibnamefont
  {Dasgupta}}\ and\ \bibinfo {author} {\bibfnamefont {I.}~\bibnamefont
  {Raychowdhury}},\ }\bibfield  {title} {\bibinfo {title} {{Cold Atom Quantum
  Simulator for String and Hadron Dynamics in Non-Abelian Lattice Gauge
  Theory}},\ }\href@noop {} {\  (\bibinfo {year} {2020})},\ \Eprint
  {https://arxiv.org/abs/2009.13969} {arXiv:2009.13969 [hep-lat]} \BibitemShut
  {NoStop}%
\bibitem [{\citenamefont {Shaw}\ \emph {et~al.}(2020)\citenamefont {Shaw},
  \citenamefont {Lougovski}, \citenamefont {Stryker},\ and\ \citenamefont
  {Wiebe}}]{Shaw2020quantumalgorithms}%
  \BibitemOpen
  \bibfield  {author} {\bibinfo {author} {\bibfnamefont {A.~F.}\ \bibnamefont
  {Shaw}}, \bibinfo {author} {\bibfnamefont {P.}~\bibnamefont {Lougovski}},
  \bibinfo {author} {\bibfnamefont {J.~R.}\ \bibnamefont {Stryker}},\ and\
  \bibinfo {author} {\bibfnamefont {N.}~\bibnamefont {Wiebe}},\ }\bibfield
  {title} {\bibinfo {title} {Quantum {A}lgorithms for {S}imulating the
  {L}attice {S}chwinger {M}odel},\ }\href
  {https://doi.org/10.22331/q-2020-08-10-306} {\bibfield  {journal} {\bibinfo
  {journal} {{Quantum}}\ }\textbf {\bibinfo {volume} {4}},\ \bibinfo {pages}
  {306} (\bibinfo {year} {2020})}\BibitemShut {NoStop}%
\bibitem [{\citenamefont {Buser}\ \emph {et~al.}(2021)\citenamefont {Buser},
  \citenamefont {Gharibyan}, \citenamefont {Hanada}, \citenamefont {Honda},\
  and\ \citenamefont {Liu}}]{Buser:2020cvn}%
  \BibitemOpen
  \bibfield  {author} {\bibinfo {author} {\bibfnamefont {A.~J.}\ \bibnamefont
  {Buser}}, \bibinfo {author} {\bibfnamefont {H.}~\bibnamefont {Gharibyan}},
  \bibinfo {author} {\bibfnamefont {M.}~\bibnamefont {Hanada}}, \bibinfo
  {author} {\bibfnamefont {M.}~\bibnamefont {Honda}},\ and\ \bibinfo {author}
  {\bibfnamefont {J.}~\bibnamefont {Liu}},\ }\bibfield  {title} {\bibinfo
  {title} {{Quantum simulation of gauge theory via orbifold lattice}},\ }\href
  {https://doi.org/10.1007/JHEP09(2021)034} {\bibfield  {journal} {\bibinfo
  {journal} {JHEP}\ }\textbf {\bibinfo {volume} {09}},\ \bibinfo {pages}
  {034}},\ \Eprint {https://arxiv.org/abs/2011.06576} {arXiv:2011.06576
  [hep-th]} \BibitemShut {NoStop}%
\bibitem [{\citenamefont {Shen}\ \emph {et~al.}(2020)\citenamefont {Shen},
  \citenamefont {Luo}, \citenamefont {Highman}, \citenamefont {Clark},
  \citenamefont {DeMarco}, \citenamefont {El-Khadra},\ and\ \citenamefont
  {Gadway}}]{Shen:2020coq}%
  \BibitemOpen
  \bibfield  {author} {\bibinfo {author} {\bibfnamefont {J.}~\bibnamefont
  {Shen}}, \bibinfo {author} {\bibfnamefont {D.}~\bibnamefont {Luo}}, \bibinfo
  {author} {\bibfnamefont {M.}~\bibnamefont {Highman}}, \bibinfo {author}
  {\bibfnamefont {B.~K.}\ \bibnamefont {Clark}}, \bibinfo {author}
  {\bibfnamefont {B.}~\bibnamefont {DeMarco}}, \bibinfo {author} {\bibfnamefont
  {A.~X.}\ \bibnamefont {El-Khadra}},\ and\ \bibinfo {author} {\bibfnamefont
  {B.}~\bibnamefont {Gadway}},\ }\bibfield  {title} {\bibinfo {title} {{(1+1)-d
  U(1) Quantum link models from effective Hamiltonians of dipolar molecules}},\
  }\href {https://doi.org/10.22323/1.363.0125} {\bibfield  {journal} {\bibinfo
  {journal} {PoS}\ }\textbf {\bibinfo {volume} {LATTICE2019}},\ \bibinfo
  {pages} {125} (\bibinfo {year} {2020})},\ \Eprint
  {https://arxiv.org/abs/2001.10002} {arXiv:2001.10002 [hep-lat]} \BibitemShut
  {NoStop}%
\bibitem [{\citenamefont {Surace}\ \emph
  {et~al.}(2020{\natexlab{b}})\citenamefont {Surace}, \citenamefont {Mazza},
  \citenamefont {Giudici}, \citenamefont {Lerose}, \citenamefont {Gambassi},\
  and\ \citenamefont {Dalmonte}}]{Surace_2020}%
  \BibitemOpen
  \bibfield  {author} {\bibinfo {author} {\bibfnamefont {F.~M.}\ \bibnamefont
  {Surace}}, \bibinfo {author} {\bibfnamefont {P.~P.}\ \bibnamefont {Mazza}},
  \bibinfo {author} {\bibfnamefont {G.}~\bibnamefont {Giudici}}, \bibinfo
  {author} {\bibfnamefont {A.}~\bibnamefont {Lerose}}, \bibinfo {author}
  {\bibfnamefont {A.}~\bibnamefont {Gambassi}},\ and\ \bibinfo {author}
  {\bibfnamefont {M.}~\bibnamefont {Dalmonte}},\ }\bibfield  {title} {\bibinfo
  {title} {Lattice gauge theories and string dynamics in rydberg atom quantum
  simulators},\ }\bibfield  {journal} {\bibinfo  {journal} {Physical Review X}\
  }\textbf {\bibinfo {volume} {10}},\ \href
  {https://doi.org/10.1103/physrevx.10.021041} {10.1103/physrevx.10.021041}
  (\bibinfo {year} {2020}{\natexlab{b}})\BibitemShut {NoStop}%
\bibitem [{\citenamefont {Van~Damme}\ \emph {et~al.}(2020)\citenamefont
  {Van~Damme}, \citenamefont {Halimeh},\ and\ \citenamefont
  {Hauke}}]{VanDamme:2020rur}%
  \BibitemOpen
  \bibfield  {author} {\bibinfo {author} {\bibfnamefont {M.}~\bibnamefont
  {Van~Damme}}, \bibinfo {author} {\bibfnamefont {J.~C.}\ \bibnamefont
  {Halimeh}},\ and\ \bibinfo {author} {\bibfnamefont {P.}~\bibnamefont
  {Hauke}},\ }\bibfield  {title} {\bibinfo {title} {{Gauge-Symmetry Violation
  Quantum Phase Transition in Lattice Gauge Theories}},\ }\href@noop {} {\
  (\bibinfo {year} {2020})},\ \Eprint {https://arxiv.org/abs/2010.07338}
  {arXiv:2010.07338 [cond-mat.quant-gas]} \BibitemShut {NoStop}%
\bibitem [{\citenamefont {Yang}\ \emph {et~al.}(2020)\citenamefont {Yang},
  \citenamefont {Sun}, \citenamefont {Ott}, \citenamefont {Wang}, \citenamefont
  {Zache}, \citenamefont {Halimeh}, \citenamefont {Yuan}, \citenamefont
  {Hauke},\ and\ \citenamefont {Pan}}]{Yang_2020}%
  \BibitemOpen
  \bibfield  {author} {\bibinfo {author} {\bibfnamefont {B.}~\bibnamefont
  {Yang}}, \bibinfo {author} {\bibfnamefont {H.}~\bibnamefont {Sun}}, \bibinfo
  {author} {\bibfnamefont {R.}~\bibnamefont {Ott}}, \bibinfo {author}
  {\bibfnamefont {H.-Y.}\ \bibnamefont {Wang}}, \bibinfo {author}
  {\bibfnamefont {T.~V.}\ \bibnamefont {Zache}}, \bibinfo {author}
  {\bibfnamefont {J.~C.}\ \bibnamefont {Halimeh}}, \bibinfo {author}
  {\bibfnamefont {Z.-S.}\ \bibnamefont {Yuan}}, \bibinfo {author}
  {\bibfnamefont {P.}~\bibnamefont {Hauke}},\ and\ \bibinfo {author}
  {\bibfnamefont {J.-W.}\ \bibnamefont {Pan}},\ }\bibfield  {title} {\bibinfo
  {title} {Observation of gauge invariance in a 71-site bose–hubbard quantum
  simulator},\ }\href {https://doi.org/10.1038/s41586-020-2910-8} {\bibfield
  {journal} {\bibinfo  {journal} {Nature}\ }\textbf {\bibinfo {volume} {587}},\
  \bibinfo {pages} {392–396} (\bibinfo {year} {2020})}\BibitemShut {NoStop}%
\bibitem [{\citenamefont {Kreshchuk}\ \emph {et~al.}(2020)\citenamefont
  {Kreshchuk}, \citenamefont {Kirby}, \citenamefont {Goldstein}, \citenamefont
  {Beauchemin},\ and\ \citenamefont {Love}}]{Kreshchuk:2020dla}%
  \BibitemOpen
  \bibfield  {author} {\bibinfo {author} {\bibfnamefont {M.}~\bibnamefont
  {Kreshchuk}}, \bibinfo {author} {\bibfnamefont {W.~M.}\ \bibnamefont
  {Kirby}}, \bibinfo {author} {\bibfnamefont {G.}~\bibnamefont {Goldstein}},
  \bibinfo {author} {\bibfnamefont {H.}~\bibnamefont {Beauchemin}},\ and\
  \bibinfo {author} {\bibfnamefont {P.~J.}\ \bibnamefont {Love}},\ }\bibfield
  {title} {\bibinfo {title} {{Quantum Simulation of Quantum Field Theory in the
  Light-Front Formulation}},\ }\href@noop {} {\  (\bibinfo {year} {2020})},\
  \Eprint {https://arxiv.org/abs/2002.04016} {arXiv:2002.04016 [quant-ph]}
  \BibitemShut {NoStop}%
\bibitem [{\citenamefont {Kasper}\ \emph
  {et~al.}(2020{\natexlab{a}})\citenamefont {Kasper}, \citenamefont
  {Juzeliunas}, \citenamefont {Lewenstein}, \citenamefont {Jendrzejewski},\
  and\ \citenamefont {Zohar}}]{Kasper:2020akk}%
  \BibitemOpen
  \bibfield  {author} {\bibinfo {author} {\bibfnamefont {V.}~\bibnamefont
  {Kasper}}, \bibinfo {author} {\bibfnamefont {G.}~\bibnamefont {Juzeliunas}},
  \bibinfo {author} {\bibfnamefont {M.}~\bibnamefont {Lewenstein}}, \bibinfo
  {author} {\bibfnamefont {F.}~\bibnamefont {Jendrzejewski}},\ and\ \bibinfo
  {author} {\bibfnamefont {E.}~\bibnamefont {Zohar}},\ }\bibfield  {title}
  {\bibinfo {title} {{From the Jaynes\textendash{}Cummings model to non-abelian
  gauge theories: a guided tour for the quantum engineer}},\ }\href
  {https://doi.org/10.1088/1367-2630/abb961} {\bibfield  {journal} {\bibinfo
  {journal} {New J. Phys.}\ }\textbf {\bibinfo {volume} {22}},\ \bibinfo
  {pages} {103027} (\bibinfo {year} {2020}{\natexlab{a}})},\ \Eprint
  {https://arxiv.org/abs/2006.01258} {arXiv:2006.01258 [quant-ph]} \BibitemShut
  {NoStop}%
\bibitem [{\citenamefont {Kreshchuk}\ \emph {et~al.}(2021)\citenamefont
  {Kreshchuk}, \citenamefont {Jia}, \citenamefont {Kirby}, \citenamefont
  {Goldstein}, \citenamefont {Vary},\ and\ \citenamefont
  {Love}}]{Kreshchuk:2020aiq}%
  \BibitemOpen
  \bibfield  {author} {\bibinfo {author} {\bibfnamefont {M.}~\bibnamefont
  {Kreshchuk}}, \bibinfo {author} {\bibfnamefont {S.}~\bibnamefont {Jia}},
  \bibinfo {author} {\bibfnamefont {W.~M.}\ \bibnamefont {Kirby}}, \bibinfo
  {author} {\bibfnamefont {G.}~\bibnamefont {Goldstein}}, \bibinfo {author}
  {\bibfnamefont {J.~P.}\ \bibnamefont {Vary}},\ and\ \bibinfo {author}
  {\bibfnamefont {P.~J.}\ \bibnamefont {Love}},\ }\bibfield  {title} {\bibinfo
  {title} {{Simulating Hadronic Physics on NISQ devices using Basis Light-Front
  Quantization}},\ }\href {https://doi.org/10.1103/PhysRevA.103.062601}
  {\bibfield  {journal} {\bibinfo  {journal} {Phys. Rev. A}\ }\textbf {\bibinfo
  {volume} {103}},\ \bibinfo {pages} {062601} (\bibinfo {year} {2021})},\
  \Eprint {https://arxiv.org/abs/2011.13443} {arXiv:2011.13443 [quant-ph]}
  \BibitemShut {NoStop}%
\bibitem [{\citenamefont {Kasper}\ \emph
  {et~al.}(2020{\natexlab{b}})\citenamefont {Kasper}, \citenamefont {Zache},
  \citenamefont {Jendrzejewski}, \citenamefont {Lewenstein},\ and\
  \citenamefont {Zohar}}]{Kasper:2020owz}%
  \BibitemOpen
  \bibfield  {author} {\bibinfo {author} {\bibfnamefont {V.}~\bibnamefont
  {Kasper}}, \bibinfo {author} {\bibfnamefont {T.~V.}\ \bibnamefont {Zache}},
  \bibinfo {author} {\bibfnamefont {F.}~\bibnamefont {Jendrzejewski}}, \bibinfo
  {author} {\bibfnamefont {M.}~\bibnamefont {Lewenstein}},\ and\ \bibinfo
  {author} {\bibfnamefont {E.}~\bibnamefont {Zohar}},\ }\bibfield  {title}
  {\bibinfo {title} {{Non-Abelian gauge invariance from dynamical
  decoupling}},\ }\href@noop {} {\  (\bibinfo {year} {2020}{\natexlab{b}})},\
  \Eprint {https://arxiv.org/abs/2012.08620} {arXiv:2012.08620 [quant-ph]}
  \BibitemShut {NoStop}%
\bibitem [{\citenamefont {Wiese}(2020)}]{Uwe:CERN_2020}%
  \BibitemOpen
  \bibfield  {author} {\bibinfo {author} {\bibfnamefont {U.-J.}\ \bibnamefont
  {Wiese}},\ }\bibfield  {title} {\bibinfo {title} {{Quantum Link Models for
  the Quantum Simulation of Gauge Theories}},\ }\href@noop {} {\  (\bibinfo
  {year} {2020})},\ \bibinfo {note} {accessed: 2021-05-31}\BibitemShut
  {NoStop}%
\bibitem [{\citenamefont {Davoudi}\ \emph
  {et~al.}(2021{\natexlab{a}})\citenamefont {Davoudi}, \citenamefont
  {Raychowdhury},\ and\ \citenamefont {Shaw}}]{Davoudi:2020yln}%
  \BibitemOpen
  \bibfield  {author} {\bibinfo {author} {\bibfnamefont {Z.}~\bibnamefont
  {Davoudi}}, \bibinfo {author} {\bibfnamefont {I.}~\bibnamefont
  {Raychowdhury}},\ and\ \bibinfo {author} {\bibfnamefont {A.}~\bibnamefont
  {Shaw}},\ }\bibfield  {title} {\bibinfo {title} {{Search for efficient
  formulations for Hamiltonian simulation of non-Abelian lattice gauge
  theories}},\ }\href {https://doi.org/10.1103/PhysRevD.104.074505} {\bibfield
  {journal} {\bibinfo  {journal} {Phys. Rev. D}\ }\textbf {\bibinfo {volume}
  {104}},\ \bibinfo {pages} {074505} (\bibinfo {year} {2021}{\natexlab{a}})},\
  \Eprint {https://arxiv.org/abs/2009.11802} {arXiv:2009.11802 [hep-lat]}
  \BibitemShut {NoStop}%
\bibitem [{\citenamefont {Ciavarella}\ \emph {et~al.}(2021)\citenamefont
  {Ciavarella}, \citenamefont {Klco},\ and\ \citenamefont
  {Savage}}]{Ciavarella:2021nmj}%
  \BibitemOpen
  \bibfield  {author} {\bibinfo {author} {\bibfnamefont {A.}~\bibnamefont
  {Ciavarella}}, \bibinfo {author} {\bibfnamefont {N.}~\bibnamefont {Klco}},\
  and\ \bibinfo {author} {\bibfnamefont {M.~J.}\ \bibnamefont {Savage}},\
  }\bibfield  {title} {\bibinfo {title} {{Trailhead for quantum simulation of
  SU(3) Yang-Mills lattice gauge theory in the local multiplet basis}},\ }\href
  {https://doi.org/10.1103/PhysRevD.103.094501} {\bibfield  {journal} {\bibinfo
   {journal} {Phys. Rev. D}\ }\textbf {\bibinfo {volume} {103}},\ \bibinfo
  {pages} {094501} (\bibinfo {year} {2021})},\ \Eprint
  {https://arxiv.org/abs/2101.10227} {arXiv:2101.10227 [quant-ph]} \BibitemShut
  {NoStop}%
\bibitem [{\citenamefont {Bauer}\ \emph {et~al.}(2021)\citenamefont {Bauer},
  \citenamefont {Freytsis},\ and\ \citenamefont {Nachman}}]{Bauer:2021gup}%
  \BibitemOpen
  \bibfield  {author} {\bibinfo {author} {\bibfnamefont {C.~W.}\ \bibnamefont
  {Bauer}}, \bibinfo {author} {\bibfnamefont {M.}~\bibnamefont {Freytsis}},\
  and\ \bibinfo {author} {\bibfnamefont {B.}~\bibnamefont {Nachman}},\
  }\bibfield  {title} {\bibinfo {title} {{Simulating Collider Physics on
  Quantum Computers Using Effective Field Theories}},\ }\href
  {https://doi.org/10.1103/PhysRevLett.127.212001} {\bibfield  {journal}
  {\bibinfo  {journal} {Phys. Rev. Lett.}\ }\textbf {\bibinfo {volume} {127}},\
  \bibinfo {pages} {212001} (\bibinfo {year} {2021})},\ \Eprint
  {https://arxiv.org/abs/2102.05044} {arXiv:2102.05044 [hep-ph]} \BibitemShut
  {NoStop}%
\bibitem [{\citenamefont {Klco}\ and\ \citenamefont
  {Savage}(2021{\natexlab{c}})}]{Klco:2021jxl}%
  \BibitemOpen
  \bibfield  {author} {\bibinfo {author} {\bibfnamefont {N.}~\bibnamefont
  {Klco}}\ and\ \bibinfo {author} {\bibfnamefont {M.~J.}\ \bibnamefont
  {Savage}},\ }\bibfield  {title} {\bibinfo {title} {{Hierarchical qubit maps
  and hierarchically implemented quantum error correction}},\ }\href
  {https://doi.org/10.1103/PhysRevA.104.062425} {\bibfield  {journal} {\bibinfo
   {journal} {Phys. Rev. A}\ }\textbf {\bibinfo {volume} {104}},\ \bibinfo
  {pages} {062425} (\bibinfo {year} {2021}{\natexlab{c}})},\ \Eprint
  {https://arxiv.org/abs/2109.01953} {arXiv:2109.01953 [quant-ph]} \BibitemShut
  {NoStop}%
\bibitem [{\citenamefont {Stryker}(2021)}]{Stryker:2021asy}%
  \BibitemOpen
  \bibfield  {author} {\bibinfo {author} {\bibfnamefont {J.~R.}\ \bibnamefont
  {Stryker}},\ }\bibfield  {title} {\bibinfo {title} {{Shearing approach to
  gauge invariant Trotterization}},\ }\href@noop {} {\  (\bibinfo {year}
  {2021})},\ \Eprint {https://arxiv.org/abs/2105.11548} {arXiv:2105.11548
  [hep-lat]} \BibitemShut {NoStop}%
\bibitem [{\citenamefont {Davoudi}\ \emph
  {et~al.}(2021{\natexlab{b}})\citenamefont {Davoudi}, \citenamefont {Linke},\
  and\ \citenamefont {Pagano}}]{Davoudi:2021ney}%
  \BibitemOpen
  \bibfield  {author} {\bibinfo {author} {\bibfnamefont {Z.}~\bibnamefont
  {Davoudi}}, \bibinfo {author} {\bibfnamefont {N.~M.}\ \bibnamefont {Linke}},\
  and\ \bibinfo {author} {\bibfnamefont {G.}~\bibnamefont {Pagano}},\
  }\bibfield  {title} {\bibinfo {title} {{Toward simulating quantum field
  theories with controlled phonon-ion dynamics: A hybrid analog-digital
  approach}},\ }\href {https://doi.org/10.1103/PhysRevResearch.3.043072}
  {\bibfield  {journal} {\bibinfo  {journal} {Phys. Rev. Res.}\ }\textbf
  {\bibinfo {volume} {3}},\ \bibinfo {pages} {043072} (\bibinfo {year}
  {2021}{\natexlab{b}})},\ \Eprint {https://arxiv.org/abs/2104.09346}
  {arXiv:2104.09346 [quant-ph]} \BibitemShut {NoStop}%
\bibitem [{\citenamefont {Atas}\ \emph {et~al.}(2021)\citenamefont {Atas},
  \citenamefont {Zhang}, \citenamefont {Lewis}, \citenamefont {Jahanpour},
  \citenamefont {Haase},\ and\ \citenamefont {Muschik}}]{Atas:2021ext}%
  \BibitemOpen
  \bibfield  {author} {\bibinfo {author} {\bibfnamefont {Y.}~\bibnamefont
  {Atas}}, \bibinfo {author} {\bibfnamefont {J.}~\bibnamefont {Zhang}},
  \bibinfo {author} {\bibfnamefont {R.}~\bibnamefont {Lewis}}, \bibinfo
  {author} {\bibfnamefont {A.}~\bibnamefont {Jahanpour}}, \bibinfo {author}
  {\bibfnamefont {J.~F.}\ \bibnamefont {Haase}},\ and\ \bibinfo {author}
  {\bibfnamefont {C.~A.}\ \bibnamefont {Muschik}},\ }\bibfield  {title}
  {\bibinfo {title} {{SU(2) hadrons on a quantum computer}},\ }\href@noop {} {\
   (\bibinfo {year} {2021})},\ \Eprint {https://arxiv.org/abs/2102.08920}
  {arXiv:2102.08920 [quant-ph]} \BibitemShut {NoStop}%
\bibitem [{\citenamefont {A~Rahman}\ \emph {et~al.}(2021)\citenamefont
  {A~Rahman}, \citenamefont {Lewis}, \citenamefont {Mendicelli},\ and\
  \citenamefont {Powell}}]{ARahman:2021ktn}%
  \BibitemOpen
  \bibfield  {author} {\bibinfo {author} {\bibfnamefont {S.}~\bibnamefont
  {A~Rahman}}, \bibinfo {author} {\bibfnamefont {R.}~\bibnamefont {Lewis}},
  \bibinfo {author} {\bibfnamefont {E.}~\bibnamefont {Mendicelli}},\ and\
  \bibinfo {author} {\bibfnamefont {S.}~\bibnamefont {Powell}},\ }\bibfield
  {title} {\bibinfo {title} {{SU(2) lattice gauge theory on a quantum
  annealer}},\ }\href {https://doi.org/10.1103/PhysRevD.104.034501} {\bibfield
  {journal} {\bibinfo  {journal} {Phys. Rev. D}\ }\textbf {\bibinfo {volume}
  {104}},\ \bibinfo {pages} {034501} (\bibinfo {year} {2021})},\ \Eprint
  {https://arxiv.org/abs/2103.08661} {arXiv:2103.08661 [hep-lat]} \BibitemShut
  {NoStop}%
\bibitem [{\citenamefont {Aidelsburger}\ \emph {et~al.}(2021)\citenamefont
  {Aidelsburger}, \citenamefont {Barbiero}, \citenamefont {Bermudez},
  \citenamefont {Chanda}, \citenamefont {Dauphin}, \citenamefont
  {González-Cuadra}, \citenamefont {Grzybowski}, \citenamefont {Hands},
  \citenamefont {Jendrzejewski}, \citenamefont {Jünemann}, \citenamefont
  {Juzeliunas}, \citenamefont {Kasper}, \citenamefont {Piga}, \citenamefont
  {Ran}, \citenamefont {Rizzi}, \citenamefont {Sierra}, \citenamefont
  {Tagliacozzo}, \citenamefont {Tirrito}, \citenamefont {Zache}, \citenamefont
  {Zakrzewski}, \citenamefont {Zohar},\ and\ \citenamefont
  {Lewenstein}}]{aidelsburger2021cold}%
  \BibitemOpen
  \bibfield  {author} {\bibinfo {author} {\bibfnamefont {M.}~\bibnamefont
  {Aidelsburger}}, \bibinfo {author} {\bibfnamefont {L.}~\bibnamefont
  {Barbiero}}, \bibinfo {author} {\bibfnamefont {A.}~\bibnamefont {Bermudez}},
  \bibinfo {author} {\bibfnamefont {T.}~\bibnamefont {Chanda}}, \bibinfo
  {author} {\bibfnamefont {A.}~\bibnamefont {Dauphin}}, \bibinfo {author}
  {\bibfnamefont {D.}~\bibnamefont {González-Cuadra}}, \bibinfo {author}
  {\bibfnamefont {P.~R.}\ \bibnamefont {Grzybowski}}, \bibinfo {author}
  {\bibfnamefont {S.}~\bibnamefont {Hands}}, \bibinfo {author} {\bibfnamefont
  {F.}~\bibnamefont {Jendrzejewski}}, \bibinfo {author} {\bibfnamefont
  {J.}~\bibnamefont {Jünemann}}, \bibinfo {author} {\bibfnamefont
  {G.}~\bibnamefont {Juzeliunas}}, \bibinfo {author} {\bibfnamefont
  {V.}~\bibnamefont {Kasper}}, \bibinfo {author} {\bibfnamefont
  {A.}~\bibnamefont {Piga}}, \bibinfo {author} {\bibfnamefont {S.-J.}\
  \bibnamefont {Ran}}, \bibinfo {author} {\bibfnamefont {M.}~\bibnamefont
  {Rizzi}}, \bibinfo {author} {\bibfnamefont {G.}~\bibnamefont {Sierra}},
  \bibinfo {author} {\bibfnamefont {L.}~\bibnamefont {Tagliacozzo}}, \bibinfo
  {author} {\bibfnamefont {E.}~\bibnamefont {Tirrito}}, \bibinfo {author}
  {\bibfnamefont {T.~V.}\ \bibnamefont {Zache}}, \bibinfo {author}
  {\bibfnamefont {J.}~\bibnamefont {Zakrzewski}}, \bibinfo {author}
  {\bibfnamefont {E.}~\bibnamefont {Zohar}},\ and\ \bibinfo {author}
  {\bibfnamefont {M.}~\bibnamefont {Lewenstein}},\ }\bibfield  {title}
  {\bibinfo {title} {Cold atoms meet lattice gauge theory},\ }\href@noop {}
  {\bibfield  {journal} {\bibinfo  {journal} {arXiv preprint arXiv:2106.03063}\
  } (\bibinfo {year} {2021})}\BibitemShut {NoStop}%
\bibitem [{\citenamefont {{Giedke}}\ \emph {et~al.}(2001)\citenamefont
  {{Giedke}}, \citenamefont {{Kraus}}, \citenamefont {{Lewenstein}},\ and\
  \citenamefont {{Cirac}}}]{2001PhRvL..87p7904G}%
  \BibitemOpen
  \bibfield  {author} {\bibinfo {author} {\bibfnamefont {G.}~\bibnamefont
  {{Giedke}}}, \bibinfo {author} {\bibfnamefont {B.}~\bibnamefont {{Kraus}}},
  \bibinfo {author} {\bibfnamefont {M.}~\bibnamefont {{Lewenstein}}},\ and\
  \bibinfo {author} {\bibfnamefont {J.~I.}\ \bibnamefont {{Cirac}}},\
  }\bibfield  {title} {\bibinfo {title} {{Entanglement Criteria for All
  Bipartite Gaussian States}},\ }\href
  {https://doi.org/10.1103/PhysRevLett.87.167904} {\bibfield  {journal}
  {\bibinfo  {journal} {\prl}\ }\textbf {\bibinfo {volume} {87}},\ \bibinfo
  {eid} {quant-ph/0104050} (\bibinfo {year} {2001})},\ \Eprint
  {https://arxiv.org/abs/quant-ph/0104050} {arXiv:quant-ph/0104050 [quant-ph]}
  \BibitemShut {NoStop}%
\bibitem [{\citenamefont {Williamson}(1936)}]{Williamson1936}%
  \BibitemOpen
  \bibfield  {author} {\bibinfo {author} {\bibfnamefont {J.}~\bibnamefont
  {Williamson}},\ }\bibfield  {title} {\bibinfo {title} {On the algebraic
  problem concerning the normal forms of linear dynamical systems},\ }\href
  {https://doi.org/10.2307/2371062} {\bibfield  {journal} {\bibinfo  {journal}
  {American Journal of Mathematics}\ }\textbf {\bibinfo {volume} {58}},\
  \bibinfo {pages} {141} (\bibinfo {year} {1936})},\ \bibinfo {note} {full
  publication date: Jan., 1936}\BibitemShut {NoStop}%
\bibitem [{\citenamefont {Horodecki}\ \emph {et~al.}(1996)\citenamefont
  {Horodecki}, \citenamefont {Horodecki},\ and\ \citenamefont
  {Horodecki}}]{Horodecki:1996nc}%
  \BibitemOpen
  \bibfield  {author} {\bibinfo {author} {\bibfnamefont {M.}~\bibnamefont
  {Horodecki}}, \bibinfo {author} {\bibfnamefont {P.}~\bibnamefont
  {Horodecki}},\ and\ \bibinfo {author} {\bibfnamefont {R.}~\bibnamefont
  {Horodecki}},\ }\bibfield  {title} {\bibinfo {title} {{On the necessary and
  sufficient conditions for separability of mixed quantum states}},\ }\href
  {https://doi.org/10.1016/S0375-9601(96)00706-2} {\bibfield  {journal}
  {\bibinfo  {journal} {Phys. Lett. A}\ }\textbf {\bibinfo {volume} {223}},\
  \bibinfo {pages} {1} (\bibinfo {year} {1996})},\ \Eprint
  {https://arxiv.org/abs/quant-ph/9605038} {arXiv:quant-ph/9605038}
  \BibitemShut {NoStop}%
\bibitem [{\citenamefont {Vidal}\ and\ \citenamefont
  {Werner}(2002)}]{Vidal:2002zz}%
  \BibitemOpen
  \bibfield  {author} {\bibinfo {author} {\bibfnamefont {G.}~\bibnamefont
  {Vidal}}\ and\ \bibinfo {author} {\bibfnamefont {R.}~\bibnamefont {Werner}},\
  }\bibfield  {title} {\bibinfo {title} {{Computable measure of
  entanglement}},\ }\href {https://doi.org/10.1103/PhysRevA.65.032314}
  {\bibfield  {journal} {\bibinfo  {journal} {Phys. Rev. A}\ }\textbf {\bibinfo
  {volume} {65}},\ \bibinfo {pages} {032314} (\bibinfo {year} {2002})},\
  \Eprint {https://arxiv.org/abs/quant-ph/0102117} {arXiv:quant-ph/0102117}
  \BibitemShut {NoStop}%
\bibitem [{\citenamefont {Simon}(2000)}]{Simon:2000zz}%
  \BibitemOpen
  \bibfield  {author} {\bibinfo {author} {\bibfnamefont {R.}~\bibnamefont
  {Simon}},\ }\bibfield  {title} {\bibinfo {title} {{Peres-Horodecki
  Separability Criterion for Continuous Variable Systems}},\ }\href
  {https://doi.org/10.1103/PhysRevLett.84.2726} {\bibfield  {journal} {\bibinfo
   {journal} {Phys. Rev. Lett.}\ }\textbf {\bibinfo {volume} {84}},\ \bibinfo
  {pages} {2726} (\bibinfo {year} {2000})},\ \Eprint
  {https://arxiv.org/abs/quant-ph/9909044} {arXiv:quant-ph/9909044}
  \BibitemShut {NoStop}%
\bibitem [{\citenamefont {Plenio}(2005)}]{Plenio:2005cwa}%
  \BibitemOpen
  \bibfield  {author} {\bibinfo {author} {\bibfnamefont {M.~B.}\ \bibnamefont
  {Plenio}},\ }\bibfield  {title} {\bibinfo {title} {{Logarithmic Negativity: A
  Full Entanglement Monotone That is not Convex}},\ }\href
  {https://doi.org/10.1103/PhysRevLett.95.090503} {\bibfield  {journal}
  {\bibinfo  {journal} {Phys. Rev. Lett.}\ }\textbf {\bibinfo {volume} {95}},\
  \bibinfo {pages} {090503} (\bibinfo {year} {2005})},\ \Eprint
  {https://arxiv.org/abs/quant-ph/0505071} {arXiv:quant-ph/0505071}
  \BibitemShut {NoStop}%
\bibitem [{\citenamefont {Duan}\ \emph {et~al.}(2000)\citenamefont {Duan},
  \citenamefont {Giedke}, \citenamefont {Cirac},\ and\ \citenamefont
  {Zoller}}]{Duan_2000}%
  \BibitemOpen
  \bibfield  {author} {\bibinfo {author} {\bibfnamefont {L.-M.}\ \bibnamefont
  {Duan}}, \bibinfo {author} {\bibfnamefont {G.}~\bibnamefont {Giedke}},
  \bibinfo {author} {\bibfnamefont {J.~I.}\ \bibnamefont {Cirac}},\ and\
  \bibinfo {author} {\bibfnamefont {P.}~\bibnamefont {Zoller}},\ }\bibfield
  {title} {\bibinfo {title} {Inseparability criterion for continuous variable
  systems},\ }\href {https://doi.org/10.1103/physrevlett.84.2722} {\bibfield
  {journal} {\bibinfo  {journal} {Physical Review Letters}\ }\textbf {\bibinfo
  {volume} {84}},\ \bibinfo {pages} {2722–2725} (\bibinfo {year}
  {2000})}\BibitemShut {NoStop}%
\bibitem [{\citenamefont {{Giedke}}\ \emph {et~al.}(2000)\citenamefont
  {{Giedke}}, \citenamefont {{Duan}}, \citenamefont {{Cirac}},\ and\
  \citenamefont {{Zoller}}}]{2000quant.ph..7061G}%
  \BibitemOpen
  \bibfield  {author} {\bibinfo {author} {\bibfnamefont {G.}~\bibnamefont
  {{Giedke}}}, \bibinfo {author} {\bibfnamefont {L.-M.}\ \bibnamefont
  {{Duan}}}, \bibinfo {author} {\bibfnamefont {J.~I.}\ \bibnamefont
  {{Cirac}}},\ and\ \bibinfo {author} {\bibfnamefont {P.}~\bibnamefont
  {{Zoller}}},\ }\bibfield  {title} {\bibinfo {title} {{All inseparable
  two-mode Gaussian continuous variable states are distillable}},\ }\href@noop
  {} {\bibfield  {journal} {\bibinfo  {journal} {arXiv e-prints}\ ,\ \bibinfo
  {eid} {quant-ph/0007061}} (\bibinfo {year} {2000})},\ \Eprint
  {https://arxiv.org/abs/quant-ph/0007061} {arXiv:quant-ph/0007061 [quant-ph]}
  \BibitemShut {NoStop}%
\bibitem [{\citenamefont {Giedke}\ \emph
  {et~al.}(2001{\natexlab{a}})\citenamefont {Giedke}, \citenamefont {Duan},
  \citenamefont {Cirac},\ and\ \citenamefont
  {Zoller}}]{giedke2001distillability}%
  \BibitemOpen
  \bibfield  {author} {\bibinfo {author} {\bibfnamefont {G.}~\bibnamefont
  {Giedke}}, \bibinfo {author} {\bibfnamefont {L.-M.}\ \bibnamefont {Duan}},
  \bibinfo {author} {\bibfnamefont {J.~I.}\ \bibnamefont {Cirac}},\ and\
  \bibinfo {author} {\bibfnamefont {P.}~\bibnamefont {Zoller}},\ }\bibfield
  {title} {\bibinfo {title} {Distillability criterion for all bipartite
  gaussian states},\ }\href
  {https://citeseerx.ist.psu.edu/viewdoc/download?doi=10.1.1.252.68&rep=rep1&type=pdf}
  {\bibfield  {journal} {\bibinfo  {journal} {Quantum Information and
  Computation}\ }\textbf {\bibinfo {volume} {1}},\ \bibinfo {pages} {79}
  (\bibinfo {year} {2001}{\natexlab{a}})},\ \Eprint
  {https://arxiv.org/abs/quant-ph/0104072} {arXiv:quant-ph/0104072 [quant-ph]}
  \BibitemShut {NoStop}%
\bibitem [{\citenamefont {{Werner}}\ and\ \citenamefont
  {{Wolf}}(2001)}]{2001PhRvL..86.3658W}%
  \BibitemOpen
  \bibfield  {author} {\bibinfo {author} {\bibfnamefont {R.~F.}\ \bibnamefont
  {{Werner}}}\ and\ \bibinfo {author} {\bibfnamefont {M.~M.}\ \bibnamefont
  {{Wolf}}},\ }\bibfield  {title} {\bibinfo {title} {{Bound Entangled Gaussian
  States}},\ }\href {https://doi.org/10.1103/PhysRevLett.86.3658} {\bibfield
  {journal} {\bibinfo  {journal} {\prl}\ }\textbf {\bibinfo {volume} {86}},\
  \bibinfo {pages} {3658} (\bibinfo {year} {2001})},\ \Eprint
  {https://arxiv.org/abs/quant-ph/0009118} {arXiv:quant-ph/0009118 [quant-ph]}
  \BibitemShut {NoStop}%
\bibitem [{\citenamefont {{Botero}}\ and\ \citenamefont
  {{Reznik}}(2003)}]{2003PhRvA..67e2311B}%
  \BibitemOpen
  \bibfield  {author} {\bibinfo {author} {\bibfnamefont {A.}~\bibnamefont
  {{Botero}}}\ and\ \bibinfo {author} {\bibfnamefont {B.}~\bibnamefont
  {{Reznik}}},\ }\bibfield  {title} {\bibinfo {title} {{Modewise entanglement
  of Gaussian states}},\ }\href {https://doi.org/10.1103/PhysRevA.67.052311}
  {\bibfield  {journal} {\bibinfo  {journal} {\pra}\ }\textbf {\bibinfo
  {volume} {67}},\ \bibinfo {eid} {052311} (\bibinfo {year} {2003})},\ \Eprint
  {https://arxiv.org/abs/quant-ph/0209026} {arXiv:quant-ph/0209026 [quant-ph]}
  \BibitemShut {NoStop}%
\bibitem [{\citenamefont {Serafini}\ \emph {et~al.}(2005)\citenamefont
  {Serafini}, \citenamefont {Adesso},\ and\ \citenamefont
  {Illuminati}}]{Serafini_2005}%
  \BibitemOpen
  \bibfield  {author} {\bibinfo {author} {\bibfnamefont {A.}~\bibnamefont
  {Serafini}}, \bibinfo {author} {\bibfnamefont {G.}~\bibnamefont {Adesso}},\
  and\ \bibinfo {author} {\bibfnamefont {F.}~\bibnamefont {Illuminati}},\
  }\bibfield  {title} {\bibinfo {title} {Unitarily localizable entanglement of
  gaussian states},\ }\bibfield  {journal} {\bibinfo  {journal} {Physical
  Review A}\ }\textbf {\bibinfo {volume} {71}},\ \href
  {https://doi.org/10.1103/physreva.71.032349} {10.1103/physreva.71.032349}
  (\bibinfo {year} {2005})\BibitemShut {NoStop}%
\bibitem [{\citenamefont {{Wolf}}(2008)}]{2008PhRvL.100g0505W}%
  \BibitemOpen
  \bibfield  {author} {\bibinfo {author} {\bibfnamefont {M.~M.}\ \bibnamefont
  {{Wolf}}},\ }\bibfield  {title} {\bibinfo {title} {{Not-So-Normal Mode
  Decomposition}},\ }\href {https://doi.org/10.1103/PhysRevLett.100.070505}
  {\bibfield  {journal} {\bibinfo  {journal} {\prl}\ }\textbf {\bibinfo
  {volume} {100}},\ \bibinfo {eid} {070505} (\bibinfo {year} {2008})},\ \Eprint
  {https://arxiv.org/abs/0707.0604} {arXiv:0707.0604 [quant-ph]} \BibitemShut
  {NoStop}%
\bibitem [{\citenamefont {{Lami}}\ \emph {et~al.}(2018)\citenamefont {{Lami}},
  \citenamefont {{Serafini}},\ and\ \citenamefont
  {{Adesso}}}]{2016arXiv161205215L}%
  \BibitemOpen
  \bibfield  {author} {\bibinfo {author} {\bibfnamefont {L.}~\bibnamefont
  {{Lami}}}, \bibinfo {author} {\bibfnamefont {A.}~\bibnamefont {{Serafini}}},\
  and\ \bibinfo {author} {\bibfnamefont {G.}~\bibnamefont {{Adesso}}},\
  }\bibfield  {title} {\bibinfo {title} {{Gaussian entanglement revisited}},\
  }\href {https://doi.org/10.1088/1367-2630/aaa654} {\bibfield  {journal}
  {\bibinfo  {journal} {New Journal of Physics}\ }\textbf {\bibinfo {volume}
  {20}},\ \bibinfo {eid} {arXiv:1612.05215} (\bibinfo {year} {2018})},\ \Eprint
  {https://arxiv.org/abs/1612.05215} {arXiv:1612.05215 [quant-ph]} \BibitemShut
  {NoStop}%
\bibitem [{\citenamefont {Serafini}(2017)}]{Serafini2017}%
  \BibitemOpen
  \bibfield  {author} {\bibinfo {author} {\bibfnamefont {A.}~\bibnamefont
  {Serafini}},\ }\href@noop {} {\emph {\bibinfo {title} {Quantum Continuous
  Variables: A Primier of Theoretical Methods}}}\ (\bibinfo  {publisher} {CRC
  Press},\ \bibinfo {year} {2017})\BibitemShut {NoStop}%
\bibitem [{\citenamefont {Giedke}\ \emph
  {et~al.}(2001{\natexlab{b}})\citenamefont {Giedke}, \citenamefont {Kraus},
  \citenamefont {Lewenstein},\ and\ \citenamefont {Cirac}}]{Giedke_2001}%
  \BibitemOpen
  \bibfield  {author} {\bibinfo {author} {\bibfnamefont {G.}~\bibnamefont
  {Giedke}}, \bibinfo {author} {\bibfnamefont {B.}~\bibnamefont {Kraus}},
  \bibinfo {author} {\bibfnamefont {M.}~\bibnamefont {Lewenstein}},\ and\
  \bibinfo {author} {\bibfnamefont {J.~I.}\ \bibnamefont {Cirac}},\ }\bibfield
  {title} {\bibinfo {title} {Separability properties of three-mode gaussian
  states},\ }\bibfield  {journal} {\bibinfo  {journal} {Physical Review A}\
  }\textbf {\bibinfo {volume} {64}},\ \href
  {https://doi.org/10.1103/physreva.64.052303} {10.1103/physreva.64.052303}
  (\bibinfo {year} {2001}{\natexlab{b}})\BibitemShut {NoStop}%
\bibitem [{\citenamefont {Adesso}\ \emph {et~al.}(2006)\citenamefont {Adesso},
  \citenamefont {Serafini},\ and\ \citenamefont {Illuminati}}]{Adesso_2006}%
  \BibitemOpen
  \bibfield  {author} {\bibinfo {author} {\bibfnamefont {G.}~\bibnamefont
  {Adesso}}, \bibinfo {author} {\bibfnamefont {A.}~\bibnamefont {Serafini}},\
  and\ \bibinfo {author} {\bibfnamefont {F.}~\bibnamefont {Illuminati}},\
  }\bibfield  {title} {\bibinfo {title} {Multipartite entanglement in
  three-mode gaussian states of continuous-variable systems: Quantification,
  sharing structure, and decoherence},\ }\bibfield  {journal} {\bibinfo
  {journal} {Physical Review A}\ }\textbf {\bibinfo {volume} {73}},\ \href
  {https://doi.org/10.1103/physreva.73.032345} {10.1103/physreva.73.032345}
  (\bibinfo {year} {2006})\BibitemShut {NoStop}%
\bibitem [{\citenamefont {Cohen}(1998)}]{PhysRevLett.80.2493}%
  \BibitemOpen
  \bibfield  {author} {\bibinfo {author} {\bibfnamefont {O.}~\bibnamefont
  {Cohen}},\ }\bibfield  {title} {\bibinfo {title} {Unlocking hidden
  entanglement with classical information},\ }\href
  {https://doi.org/10.1103/PhysRevLett.80.2493} {\bibfield  {journal} {\bibinfo
   {journal} {Phys. Rev. Lett.}\ }\textbf {\bibinfo {volume} {80}},\ \bibinfo
  {pages} {2493} (\bibinfo {year} {1998})}\BibitemShut {NoStop}%
\bibitem [{\citenamefont {Sanpera}\ \emph {et~al.}(1998)\citenamefont
  {Sanpera}, \citenamefont {Tarrach},\ and\ \citenamefont
  {Vidal}}]{PhysRevA.58.826}%
  \BibitemOpen
  \bibfield  {author} {\bibinfo {author} {\bibfnamefont {A.}~\bibnamefont
  {Sanpera}}, \bibinfo {author} {\bibfnamefont {R.}~\bibnamefont {Tarrach}},\
  and\ \bibinfo {author} {\bibfnamefont {G.}~\bibnamefont {Vidal}},\ }\bibfield
   {title} {\bibinfo {title} {Local description of quantum inseparability},\
  }\href {https://doi.org/10.1103/PhysRevA.58.826} {\bibfield  {journal}
  {\bibinfo  {journal} {Phys. Rev. A}\ }\textbf {\bibinfo {volume} {58}},\
  \bibinfo {pages} {826} (\bibinfo {year} {1998})}\BibitemShut {NoStop}%
\bibitem [{\citenamefont {Vidal}\ and\ \citenamefont
  {Tarrach}(1999)}]{Vidal:1998ch}%
  \BibitemOpen
  \bibfield  {author} {\bibinfo {author} {\bibfnamefont {G.}~\bibnamefont
  {Vidal}}\ and\ \bibinfo {author} {\bibfnamefont {R.}~\bibnamefont
  {Tarrach}},\ }\bibfield  {title} {\bibinfo {title} {{Robustness of
  entanglement}},\ }\href {https://doi.org/10.1103/PhysRevA.59.141} {\bibfield
  {journal} {\bibinfo  {journal} {Phys. Rev. A}\ }\textbf {\bibinfo {volume}
  {59}},\ \bibinfo {pages} {141} (\bibinfo {year} {1999})},\ \Eprint
  {https://arxiv.org/abs/quant-ph/9806094} {arXiv:quant-ph/9806094}
  \BibitemShut {NoStop}%
\bibitem [{\citenamefont {Smolin}(2001)}]{PhysRevA.63.032306}%
  \BibitemOpen
  \bibfield  {author} {\bibinfo {author} {\bibfnamefont {J.~A.}\ \bibnamefont
  {Smolin}},\ }\bibfield  {title} {\bibinfo {title} {Four-party unlockable
  bound entangled state},\ }\href {https://doi.org/10.1103/PhysRevA.63.032306}
  {\bibfield  {journal} {\bibinfo  {journal} {Phys. Rev. A}\ }\textbf {\bibinfo
  {volume} {63}},\ \bibinfo {pages} {032306} (\bibinfo {year}
  {2001})}\BibitemShut {NoStop}%
\bibitem [{\citenamefont {Yu}\ and\ \citenamefont {Eberly}(2006)}]{Yu_2006}%
  \BibitemOpen
  \bibfield  {author} {\bibinfo {author} {\bibfnamefont {T.}~\bibnamefont
  {Yu}}\ and\ \bibinfo {author} {\bibfnamefont {J.}~\bibnamefont {Eberly}},\
  }\bibfield  {title} {\bibinfo {title} {Sudden death of entanglement:
  Classical noise effects},\ }\href
  {https://doi.org/10.1016/j.optcom.2006.01.061} {\bibfield  {journal}
  {\bibinfo  {journal} {Optics Communications}\ }\textbf {\bibinfo {volume}
  {264}},\ \bibinfo {pages} {393–397} (\bibinfo {year} {2006})}\BibitemShut
  {NoStop}%
\bibitem [{\citenamefont {Yu}\ and\ \citenamefont {Eberly}(2009)}]{Yu_2009}%
  \BibitemOpen
  \bibfield  {author} {\bibinfo {author} {\bibfnamefont {T.}~\bibnamefont
  {Yu}}\ and\ \bibinfo {author} {\bibfnamefont {J.~H.}\ \bibnamefont
  {Eberly}},\ }\bibfield  {title} {\bibinfo {title} {Sudden death of
  entanglement},\ }\href {https://doi.org/10.1126/science.1167343} {\bibfield
  {journal} {\bibinfo  {journal} {Science}\ }\textbf {\bibinfo {volume}
  {323}},\ \bibinfo {pages} {598–601} (\bibinfo {year} {2009})}\BibitemShut
  {NoStop}%
\bibitem [{\citenamefont {Retzker}\ \emph {et~al.}(2005)\citenamefont
  {Retzker}, \citenamefont {Cirac},\ and\ \citenamefont
  {Reznik}}]{Retzker_2005}%
  \BibitemOpen
  \bibfield  {author} {\bibinfo {author} {\bibfnamefont {A.}~\bibnamefont
  {Retzker}}, \bibinfo {author} {\bibfnamefont {J.~I.}\ \bibnamefont {Cirac}},\
  and\ \bibinfo {author} {\bibfnamefont {B.}~\bibnamefont {Reznik}},\
  }\bibfield  {title} {\bibinfo {title} {Detecting vacuum entanglement in a
  linear ion trap},\ }\bibfield  {journal} {\bibinfo  {journal} {Physical
  Review Letters}\ }\textbf {\bibinfo {volume} {94}},\ \href
  {https://doi.org/10.1103/physrevlett.94.050504}
  {10.1103/physrevlett.94.050504} (\bibinfo {year} {2005})\BibitemShut
  {NoStop}%
\bibitem [{\citenamefont {Martin-Martinez}\ \emph {et~al.}(2013)\citenamefont
  {Martin-Martinez}, \citenamefont {Brown}, \citenamefont {Donnelly},\ and\
  \citenamefont {Kempf}}]{MartinMartinez2013}%
  \BibitemOpen
  \bibfield  {author} {\bibinfo {author} {\bibfnamefont {E.}~\bibnamefont
  {Martin-Martinez}}, \bibinfo {author} {\bibfnamefont {E.~G.}\ \bibnamefont
  {Brown}}, \bibinfo {author} {\bibfnamefont {W.}~\bibnamefont {Donnelly}},\
  and\ \bibinfo {author} {\bibfnamefont {A.}~\bibnamefont {Kempf}},\ }\bibfield
   {title} {\bibinfo {title} {Sustainable entanglement production from a
  quantum field},\ }\bibfield  {journal} {\bibinfo  {journal} {Physical Review
  A}\ }\textbf {\bibinfo {volume} {88}},\ \href
  {https://doi.org/10.1103/physreva.88.052310} {10.1103/physreva.88.052310}
  (\bibinfo {year} {2013})\BibitemShut {NoStop}%
\bibitem [{\citenamefont {Pozas-Kerstjens}\ and\ \citenamefont
  {Martin-Martinez}(2016)}]{Pozas_Kerstjens_2016}%
  \BibitemOpen
  \bibfield  {author} {\bibinfo {author} {\bibfnamefont {A.}~\bibnamefont
  {Pozas-Kerstjens}}\ and\ \bibinfo {author} {\bibfnamefont {E.}~\bibnamefont
  {Martin-Martinez}},\ }\bibfield  {title} {\bibinfo {title} {Entanglement
  harvesting from the electromagnetic vacuum with hydrogenlike atoms},\
  }\bibfield  {journal} {\bibinfo  {journal} {Physical Review D}\ }\textbf
  {\bibinfo {volume} {94}},\ \href {https://doi.org/10.1103/physrevd.94.064074}
  {10.1103/physrevd.94.064074} (\bibinfo {year} {2016})\BibitemShut {NoStop}%
\bibitem [{\citenamefont {Pozas-Kerstjens}\ \emph {et~al.}(2017)\citenamefont
  {Pozas-Kerstjens}, \citenamefont {Louko},\ and\ \citenamefont
  {Martin-Martinez}}]{Pozas_Kerstjens_2017}%
  \BibitemOpen
  \bibfield  {author} {\bibinfo {author} {\bibfnamefont {A.}~\bibnamefont
  {Pozas-Kerstjens}}, \bibinfo {author} {\bibfnamefont {J.}~\bibnamefont
  {Louko}},\ and\ \bibinfo {author} {\bibfnamefont {E.}~\bibnamefont
  {Martin-Martinez}},\ }\bibfield  {title} {\bibinfo {title} {Degenerate
  detectors are unable to harvest spacelike entanglement},\ }\bibfield
  {journal} {\bibinfo  {journal} {Physical Review D}\ }\textbf {\bibinfo
  {volume} {95}},\ \href {https://doi.org/10.1103/physrevd.95.105009}
  {10.1103/physrevd.95.105009} (\bibinfo {year} {2017})\BibitemShut {NoStop}%
\bibitem [{\citenamefont {Simidzija}\ \emph {et~al.}(2018)\citenamefont
  {Simidzija}, \citenamefont {Jonsson},\ and\ \citenamefont
  {Martin-Martinez}}]{Simidzija_2018}%
  \BibitemOpen
  \bibfield  {author} {\bibinfo {author} {\bibfnamefont {P.}~\bibnamefont
  {Simidzija}}, \bibinfo {author} {\bibfnamefont {R.~H.}\ \bibnamefont
  {Jonsson}},\ and\ \bibinfo {author} {\bibfnamefont {E.}~\bibnamefont
  {Martin-Martinez}},\ }\bibfield  {title} {\bibinfo {title} {General no-go
  theorem for entanglement extraction},\ }\bibfield  {journal} {\bibinfo
  {journal} {Physical Review D}\ }\textbf {\bibinfo {volume} {97}},\ \href
  {https://doi.org/10.1103/physrevd.97.125002} {10.1103/physrevd.97.125002}
  (\bibinfo {year} {2018})\BibitemShut {NoStop}%
\bibitem [{\citenamefont {Unruh}(1976)}]{Unruh1976}%
  \BibitemOpen
  \bibfield  {author} {\bibinfo {author} {\bibfnamefont {W.~G.}\ \bibnamefont
  {Unruh}},\ }\bibfield  {title} {\bibinfo {title} {Notes on black-hole
  evaporation},\ }\href {https://doi.org/10.1103/PhysRevD.14.870} {\bibfield
  {journal} {\bibinfo  {journal} {Phys. Rev. D}\ }\textbf {\bibinfo {volume}
  {14}},\ \bibinfo {pages} {870} (\bibinfo {year} {1976})}\BibitemShut
  {NoStop}%
\bibitem [{\citenamefont {Salton}\ \emph {et~al.}(2015)\citenamefont {Salton},
  \citenamefont {Mann},\ and\ \citenamefont {Menicucci}}]{Salton_2015}%
  \BibitemOpen
  \bibfield  {author} {\bibinfo {author} {\bibfnamefont {G.}~\bibnamefont
  {Salton}}, \bibinfo {author} {\bibfnamefont {R.~B.}\ \bibnamefont {Mann}},\
  and\ \bibinfo {author} {\bibfnamefont {N.~C.}\ \bibnamefont {Menicucci}},\
  }\bibfield  {title} {\bibinfo {title} {Acceleration-assisted entanglement
  harvesting and rangefinding},\ }\href
  {https://doi.org/10.1088/1367-2630/17/3/035001} {\bibfield  {journal}
  {\bibinfo  {journal} {New Journal of Physics}\ }\textbf {\bibinfo {volume}
  {17}},\ \bibinfo {pages} {035001} (\bibinfo {year} {2015})}\BibitemShut
  {NoStop}%
\bibitem [{\citenamefont {Horodecki}\ \emph {et~al.}(1998)\citenamefont
  {Horodecki}, \citenamefont {Horodecki},\ and\ \citenamefont
  {Horodecki}}]{PhysRevLett.80.5239}%
  \BibitemOpen
  \bibfield  {author} {\bibinfo {author} {\bibfnamefont {M.}~\bibnamefont
  {Horodecki}}, \bibinfo {author} {\bibfnamefont {P.}~\bibnamefont
  {Horodecki}},\ and\ \bibinfo {author} {\bibfnamefont {R.}~\bibnamefont
  {Horodecki}},\ }\bibfield  {title} {\bibinfo {title} {Mixed-state
  entanglement and distillation: Is there a ``bound'' entanglement in
  nature?},\ }\href {https://doi.org/10.1103/PhysRevLett.80.5239} {\bibfield
  {journal} {\bibinfo  {journal} {Phys. Rev. Lett.}\ }\textbf {\bibinfo
  {volume} {80}},\ \bibinfo {pages} {5239} (\bibinfo {year}
  {1998})}\BibitemShut {NoStop}%
\bibitem [{\citenamefont {{Bennett}}\ \emph {et~al.}(1999)\citenamefont
  {{Bennett}}, \citenamefont {{Divincenzo}}, \citenamefont {{Mor}},
  \citenamefont {{Shor}}, \citenamefont {{Smolin}},\ and\ \citenamefont
  {{Terhal}}}]{1999PhRvL..82.5385B}%
  \BibitemOpen
  \bibfield  {author} {\bibinfo {author} {\bibfnamefont {C.~H.}\ \bibnamefont
  {{Bennett}}}, \bibinfo {author} {\bibfnamefont {D.~P.}\ \bibnamefont
  {{Divincenzo}}}, \bibinfo {author} {\bibfnamefont {T.}~\bibnamefont {{Mor}}},
  \bibinfo {author} {\bibfnamefont {P.~W.}\ \bibnamefont {{Shor}}}, \bibinfo
  {author} {\bibfnamefont {J.~A.}\ \bibnamefont {{Smolin}}},\ and\ \bibinfo
  {author} {\bibfnamefont {B.~M.}\ \bibnamefont {{Terhal}}},\ }\bibfield
  {title} {\bibinfo {title} {{Unextendible Product Bases and Bound
  Entanglement}},\ }\href {https://doi.org/10.1103/PhysRevLett.82.5385}
  {\bibfield  {journal} {\bibinfo  {journal} {\prl}\ }\textbf {\bibinfo
  {volume} {82}},\ \bibinfo {pages} {5385} (\bibinfo {year} {1999})},\ \Eprint
  {https://arxiv.org/abs/quant-ph/9808030} {arXiv:quant-ph/9808030 [quant-ph]}
  \BibitemShut {NoStop}%
\bibitem [{\citenamefont {{Bru{\ss}}}\ and\ \citenamefont
  {{Peres}}(2000)}]{2000PhRvA..61c0301B}%
  \BibitemOpen
  \bibfield  {author} {\bibinfo {author} {\bibfnamefont {D.}~\bibnamefont
  {{Bru{\ss}}}}\ and\ \bibinfo {author} {\bibfnamefont {A.}~\bibnamefont
  {{Peres}}},\ }\bibfield  {title} {\bibinfo {title} {{Construction of quantum
  states with bound entanglement}},\ }\href
  {https://doi.org/10.1103/PhysRevA.61.030301} {\bibfield  {journal} {\bibinfo
  {journal} {\pra}\ }\textbf {\bibinfo {volume} {61}},\ \bibinfo {eid} {030301}
  (\bibinfo {year} {2000})},\ \Eprint {https://arxiv.org/abs/quant-ph/9911056}
  {arXiv:quant-ph/9911056 [quant-ph]} \BibitemShut {NoStop}%
\bibitem [{\citenamefont {Horodecki}\ and\ \citenamefont
  {Lewenstein}(2000)}]{PhysRevLett.85.2657}%
  \BibitemOpen
  \bibfield  {author} {\bibinfo {author} {\bibfnamefont {P.}~\bibnamefont
  {Horodecki}}\ and\ \bibinfo {author} {\bibfnamefont {M.}~\bibnamefont
  {Lewenstein}},\ }\bibfield  {title} {\bibinfo {title} {Bound entanglement and
  continuous variables},\ }\href {https://doi.org/10.1103/PhysRevLett.85.2657}
  {\bibfield  {journal} {\bibinfo  {journal} {Phys. Rev. Lett.}\ }\textbf
  {\bibinfo {volume} {85}},\ \bibinfo {pages} {2657} (\bibinfo {year}
  {2000})}\BibitemShut {NoStop}%
\bibitem [{\citenamefont {{Smolin}}(2001)}]{2001PhRvA..63c2306S}%
  \BibitemOpen
  \bibfield  {author} {\bibinfo {author} {\bibfnamefont {J.~A.}\ \bibnamefont
  {{Smolin}}},\ }\bibfield  {title} {\bibinfo {title} {{Four-party unlockable
  bound entangled state}},\ }\href {https://doi.org/10.1103/PhysRevA.63.032306}
  {\bibfield  {journal} {\bibinfo  {journal} {\pra}\ }\textbf {\bibinfo
  {volume} {63}},\ \bibinfo {eid} {032306} (\bibinfo {year} {2001})},\ \Eprint
  {https://arxiv.org/abs/quant-ph/0001001} {arXiv:quant-ph/0001001 [quant-ph]}
  \BibitemShut {NoStop}%
\bibitem [{\citenamefont {{Breuer}}(2006)}]{2006PhRvL..97h0501B}%
  \BibitemOpen
  \bibfield  {author} {\bibinfo {author} {\bibfnamefont {H.-P.}\ \bibnamefont
  {{Breuer}}},\ }\bibfield  {title} {\bibinfo {title} {{Optimal Entanglement
  Criterion for Mixed Quantum States}},\ }\href
  {https://doi.org/10.1103/PhysRevLett.97.080501} {\bibfield  {journal}
  {\bibinfo  {journal} {\prl}\ }\textbf {\bibinfo {volume} {97}},\ \bibinfo
  {eid} {080501} (\bibinfo {year} {2006})},\ \Eprint
  {https://arxiv.org/abs/quant-ph/0605036} {arXiv:quant-ph/0605036 [quant-ph]}
  \BibitemShut {NoStop}%
\bibitem [{\citenamefont {Piani}\ and\ \citenamefont
  {Mora}(2007)}]{Piani_2007}%
  \BibitemOpen
  \bibfield  {author} {\bibinfo {author} {\bibfnamefont {M.}~\bibnamefont
  {Piani}}\ and\ \bibinfo {author} {\bibfnamefont {C.~E.}\ \bibnamefont
  {Mora}},\ }\bibfield  {title} {\bibinfo {title} {Class of
  positive-partial-transpose bound entangled states associated with almost any
  set of pure entangled states},\ }\bibfield  {journal} {\bibinfo  {journal}
  {Physical Review A}\ }\textbf {\bibinfo {volume} {75}},\ \href
  {https://doi.org/10.1103/physreva.75.012305} {10.1103/physreva.75.012305}
  (\bibinfo {year} {2007})\BibitemShut {NoStop}%
\bibitem [{\citenamefont {Verstraete}(2017)}]{VerstraeteQIP2017}%
  \BibitemOpen
  \bibfield  {author} {\bibinfo {author} {\bibfnamefont {F.}~\bibnamefont
  {Verstraete}},\ }\bibfield  {title} {\bibinfo {title} {The entanglement of
  distillation for gauge theories},\ }\href
  {https://www.microsoft.com/en-us/research/event/qip-2017/} {\  (\bibinfo
  {year} {2017})}\BibitemShut {NoStop}%
\bibitem [{\citenamefont {{Giedke}}\ \emph {et~al.}(2003)\citenamefont
  {{Giedke}}, \citenamefont {{Wolf}}, \citenamefont {{Kr{\"u}ger}},
  \citenamefont {{Werner}},\ and\ \citenamefont
  {{Cirac}}}]{2003PhRvL..91j7901G}%
  \BibitemOpen
  \bibfield  {author} {\bibinfo {author} {\bibfnamefont {G.}~\bibnamefont
  {{Giedke}}}, \bibinfo {author} {\bibfnamefont {M.~M.}\ \bibnamefont
  {{Wolf}}}, \bibinfo {author} {\bibfnamefont {O.}~\bibnamefont
  {{Kr{\"u}ger}}}, \bibinfo {author} {\bibfnamefont {R.~F.}\ \bibnamefont
  {{Werner}}},\ and\ \bibinfo {author} {\bibfnamefont {J.~I.}\ \bibnamefont
  {{Cirac}}},\ }\bibfield  {title} {\bibinfo {title} {{Entanglement of
  Formation for Symmetric Gaussian States}},\ }\href
  {https://doi.org/10.1103/PhysRevLett.91.107901} {\bibfield  {journal}
  {\bibinfo  {journal} {\prl}\ }\textbf {\bibinfo {volume} {91}},\ \bibinfo
  {eid} {107901} (\bibinfo {year} {2003})},\ \Eprint
  {https://arxiv.org/abs/quant-ph/0304042} {arXiv:quant-ph/0304042 [quant-ph]}
  \BibitemShut {NoStop}%
\bibitem [{\citenamefont {{Wolf}}\ \emph {et~al.}(2004)\citenamefont {{Wolf}},
  \citenamefont {{Giedke}}, \citenamefont {{Kr{\"u}ger}}, \citenamefont
  {{Werner}},\ and\ \citenamefont {{Cirac}}}]{2004PhRvA..69e2320W}%
  \BibitemOpen
  \bibfield  {author} {\bibinfo {author} {\bibfnamefont {M.~M.}\ \bibnamefont
  {{Wolf}}}, \bibinfo {author} {\bibfnamefont {G.}~\bibnamefont {{Giedke}}},
  \bibinfo {author} {\bibfnamefont {O.}~\bibnamefont {{Kr{\"u}ger}}}, \bibinfo
  {author} {\bibfnamefont {R.~F.}\ \bibnamefont {{Werner}}},\ and\ \bibinfo
  {author} {\bibfnamefont {J.~I.}\ \bibnamefont {{Cirac}}},\ }\bibfield
  {title} {\bibinfo {title} {{Gaussian entanglement of formation}},\ }\href
  {https://doi.org/10.1103/PhysRevA.69.052320} {\bibfield  {journal} {\bibinfo
  {journal} {\pra}\ }\textbf {\bibinfo {volume} {69}},\ \bibinfo {eid} {052320}
  (\bibinfo {year} {2004})},\ \Eprint {https://arxiv.org/abs/quant-ph/0306177}
  {arXiv:quant-ph/0306177 [quant-ph]} \BibitemShut {NoStop}%
\bibitem [{\citenamefont {Marian}\ and\ \citenamefont
  {Marian}(2008)}]{Marian:2008zza}%
  \BibitemOpen
  \bibfield  {author} {\bibinfo {author} {\bibfnamefont {P.}~\bibnamefont
  {Marian}}\ and\ \bibinfo {author} {\bibfnamefont {T.~A.}\ \bibnamefont
  {Marian}},\ }\bibfield  {title} {\bibinfo {title} {{Entanglement of Formation
  for an Arbitrary Two-Mode Gaussian State}},\ }\href
  {https://doi.org/10.1103/PhysRevLett.101.220403} {\bibfield  {journal}
  {\bibinfo  {journal} {Phys. Rev. Lett.}\ }\textbf {\bibinfo {volume} {101}},\
  \bibinfo {pages} {220403} (\bibinfo {year} {2008})},\ \Eprint
  {https://arxiv.org/abs/0809.0321} {arXiv:0809.0321 [quant-ph]} \BibitemShut
  {NoStop}%
\bibitem [{\citenamefont {{Tserkis}}\ \emph {et~al.}(2019)\citenamefont
  {{Tserkis}}, \citenamefont {{Onoe}},\ and\ \citenamefont
  {{Ralph}}}]{2019arXiv190309961T}%
  \BibitemOpen
  \bibfield  {author} {\bibinfo {author} {\bibfnamefont {S.}~\bibnamefont
  {{Tserkis}}}, \bibinfo {author} {\bibfnamefont {S.}~\bibnamefont {{Onoe}}},\
  and\ \bibinfo {author} {\bibfnamefont {T.~C.}\ \bibnamefont {{Ralph}}},\
  }\bibfield  {title} {\bibinfo {title} {{Quantifying entanglement of formation
  for two-mode Gaussian states: Analytical expressions for upper and lower
  bounds and numerical estimation of its exact value}},\ }\href@noop {}
  {\bibfield  {journal} {\bibinfo  {journal} {arXiv e-prints}\ ,\ \bibinfo
  {eid} {arXiv:1903.09961}} (\bibinfo {year} {2019})},\ \Eprint
  {https://arxiv.org/abs/1903.09961} {arXiv:1903.09961 [quant-ph]} \BibitemShut
  {NoStop}%
\bibitem [{\citenamefont {{D{\"u}r}}\ \emph {et~al.}(2000)\citenamefont
  {{D{\"u}r}}, \citenamefont {{Cirac}}, \citenamefont {{Lewenstein}},\ and\
  \citenamefont {{Bru{\ss}}}}]{2000PhRvA..61f2313D}%
  \BibitemOpen
  \bibfield  {author} {\bibinfo {author} {\bibfnamefont {W.}~\bibnamefont
  {{D{\"u}r}}}, \bibinfo {author} {\bibfnamefont {J.~I.}\ \bibnamefont
  {{Cirac}}}, \bibinfo {author} {\bibfnamefont {M.}~\bibnamefont
  {{Lewenstein}}},\ and\ \bibinfo {author} {\bibfnamefont {D.}~\bibnamefont
  {{Bru{\ss}}}},\ }\bibfield  {title} {\bibinfo {title} {{Distillability and
  partial transposition in bipartite systems}},\ }\href
  {https://doi.org/10.1103/PhysRevA.61.062313} {\bibfield  {journal} {\bibinfo
  {journal} {\pra}\ }\textbf {\bibinfo {volume} {61}},\ \bibinfo {eid} {062313}
  (\bibinfo {year} {2000})},\ \Eprint {https://arxiv.org/abs/quant-ph/9910022}
  {arXiv:quant-ph/9910022 [quant-ph]} \BibitemShut {NoStop}%
\bibitem [{\citenamefont {DiVincenzo}\ \emph {et~al.}(2000)\citenamefont
  {DiVincenzo}, \citenamefont {Shor}, \citenamefont {Smolin}, \citenamefont
  {Terhal},\ and\ \citenamefont {Thapliyal}}]{DiVincenzo_2000}%
  \BibitemOpen
  \bibfield  {author} {\bibinfo {author} {\bibfnamefont {D.~P.}\ \bibnamefont
  {DiVincenzo}}, \bibinfo {author} {\bibfnamefont {P.~W.}\ \bibnamefont
  {Shor}}, \bibinfo {author} {\bibfnamefont {J.~A.}\ \bibnamefont {Smolin}},
  \bibinfo {author} {\bibfnamefont {B.~M.}\ \bibnamefont {Terhal}},\ and\
  \bibinfo {author} {\bibfnamefont {A.~V.}\ \bibnamefont {Thapliyal}},\
  }\bibfield  {title} {\bibinfo {title} {Evidence for bound entangled states
  with negative partial transpose},\ }\bibfield  {journal} {\bibinfo  {journal}
  {Physical Review A}\ }\textbf {\bibinfo {volume} {61}},\ \href
  {https://doi.org/10.1103/physreva.61.062312} {10.1103/physreva.61.062312}
  (\bibinfo {year} {2000})\BibitemShut {NoStop}%
\bibitem [{\citenamefont {{Simon}}(2006)}]{2006quant.ph..8250S}%
  \BibitemOpen
  \bibfield  {author} {\bibinfo {author} {\bibfnamefont {R.}~\bibnamefont
  {{Simon}}},\ }\bibfield  {title} {\bibinfo {title} {{NPPT Bound Entanglement
  Exists}},\ }\href@noop {} {\bibfield  {journal} {\bibinfo  {journal} {arXiv
  e-prints}\ ,\ \bibinfo {eid} {quant-ph/0608250}} (\bibinfo {year} {2006})},\
  \Eprint {https://arxiv.org/abs/quant-ph/0608250} {arXiv:quant-ph/0608250
  [quant-ph]} \BibitemShut {NoStop}%
\bibitem [{\citenamefont {{Chattopadhyay}}\ and\ \citenamefont
  {{Sarkar}}(2006)}]{2006quant.ph..9050C}%
  \BibitemOpen
  \bibfield  {author} {\bibinfo {author} {\bibfnamefont {I.}~\bibnamefont
  {{Chattopadhyay}}}\ and\ \bibinfo {author} {\bibfnamefont {D.}~\bibnamefont
  {{Sarkar}}},\ }\bibfield  {title} {\bibinfo {title} {{NPT Bound Entanglement-
  The Problem Revisited}},\ }\href@noop {} {\bibfield  {journal} {\bibinfo
  {journal} {arXiv e-prints}\ ,\ \bibinfo {eid} {quant-ph/0609050}} (\bibinfo
  {year} {2006})},\ \Eprint {https://arxiv.org/abs/quant-ph/0609050}
  {arXiv:quant-ph/0609050 [quant-ph]} \BibitemShut {NoStop}%
\bibitem [{\citenamefont {Pankowski}\ \emph {et~al.}(2010)\citenamefont
  {Pankowski}, \citenamefont {Piani}, \citenamefont {Horodecki},\ and\
  \citenamefont {Horodecki}}]{Pankowski_2010}%
  \BibitemOpen
  \bibfield  {author} {\bibinfo {author} {\bibfnamefont {L.}~\bibnamefont
  {Pankowski}}, \bibinfo {author} {\bibfnamefont {M.}~\bibnamefont {Piani}},
  \bibinfo {author} {\bibfnamefont {M.}~\bibnamefont {Horodecki}},\ and\
  \bibinfo {author} {\bibfnamefont {P.}~\bibnamefont {Horodecki}},\ }\bibfield
  {title} {\bibinfo {title} {A few steps more towards npt bound entanglement},\
  }\href {https://doi.org/10.1109/tit.2010.2050810} {\bibfield  {journal}
  {\bibinfo  {journal} {IEEE Transactions on Information Theory}\ }\textbf
  {\bibinfo {volume} {56}},\ \bibinfo {pages} {4085–4100} (\bibinfo {year}
  {2010})}\BibitemShut {NoStop}%
\bibitem [{\citenamefont {Pastawski}\ \emph {et~al.}(2015)\citenamefont
  {Pastawski}, \citenamefont {Yoshida}, \citenamefont {Harlow},\ and\
  \citenamefont {Preskill}}]{Pastawski:2015qua}%
  \BibitemOpen
  \bibfield  {author} {\bibinfo {author} {\bibfnamefont {F.}~\bibnamefont
  {Pastawski}}, \bibinfo {author} {\bibfnamefont {B.}~\bibnamefont {Yoshida}},
  \bibinfo {author} {\bibfnamefont {D.}~\bibnamefont {Harlow}},\ and\ \bibinfo
  {author} {\bibfnamefont {J.}~\bibnamefont {Preskill}},\ }\bibfield  {title}
  {\bibinfo {title} {{Holographic quantum error-correcting codes: Toy models
  for the bulk/boundary correspondence}},\ }\href
  {https://doi.org/10.1007/JHEP06(2015)149} {\bibfield  {journal} {\bibinfo
  {journal} {JHEP}\ }\textbf {\bibinfo {volume} {06}},\ \bibinfo {pages}
  {149}},\ \Eprint {https://arxiv.org/abs/1503.06237} {arXiv:1503.06237
  [hep-th]} \BibitemShut {NoStop}%
\bibitem [{\citenamefont {Wagman}\ \emph {et~al.}(2017)\citenamefont {Wagman},
  \citenamefont {Winter}, \citenamefont {Chang}, \citenamefont {Davoudi},
  \citenamefont {Detmold}, \citenamefont {Orginos}, \citenamefont {Savage},\
  and\ \citenamefont {Shanahan}}]{Wagman:2017tmp}%
  \BibitemOpen
  \bibfield  {author} {\bibinfo {author} {\bibfnamefont {M.~L.}\ \bibnamefont
  {Wagman}}, \bibinfo {author} {\bibfnamefont {F.}~\bibnamefont {Winter}},
  \bibinfo {author} {\bibfnamefont {E.}~\bibnamefont {Chang}}, \bibinfo
  {author} {\bibfnamefont {Z.}~\bibnamefont {Davoudi}}, \bibinfo {author}
  {\bibfnamefont {W.}~\bibnamefont {Detmold}}, \bibinfo {author} {\bibfnamefont
  {K.}~\bibnamefont {Orginos}}, \bibinfo {author} {\bibfnamefont {M.~J.}\
  \bibnamefont {Savage}},\ and\ \bibinfo {author} {\bibfnamefont {P.~E.}\
  \bibnamefont {Shanahan}},\ }\bibfield  {title} {\bibinfo {title}
  {{Baryon-Baryon Interactions and Spin-Flavor Symmetry from Lattice Quantum
  Chromodynamics}},\ }\href {https://doi.org/10.1103/PhysRevD.96.114510}
  {\bibfield  {journal} {\bibinfo  {journal} {Phys. Rev. D}\ }\textbf {\bibinfo
  {volume} {96}},\ \bibinfo {pages} {114510} (\bibinfo {year} {2017})},\
  \Eprint {https://arxiv.org/abs/1706.06550} {arXiv:1706.06550 [hep-lat]}
  \BibitemShut {NoStop}%
\bibitem [{\citenamefont {{Wolfram Research{,} Inc.}}()}]{Mathematica}%
  \BibitemOpen
  \bibfield  {author} {\bibinfo {author} {\bibnamefont {{Wolfram Research{,}
  Inc.}}},\ }\href {https://www.wolfram.com/mathematica} {\bibinfo {title}
  {Mathematica, {V}ersion 11.1}},\ \bibinfo {note} {champaign, IL,
  2020}\BibitemShut {NoStop}%
\bibitem [{\citenamefont {Werner}(1989)}]{PhysRevA.40.4277}%
  \BibitemOpen
  \bibfield  {author} {\bibinfo {author} {\bibfnamefont {R.~F.}\ \bibnamefont
  {Werner}},\ }\bibfield  {title} {\bibinfo {title} {Quantum states with
  einstein-podolsky-rosen correlations admitting a hidden-variable model},\
  }\href {https://doi.org/10.1103/PhysRevA.40.4277} {\bibfield  {journal}
  {\bibinfo  {journal} {Phys. Rev. A}\ }\textbf {\bibinfo {volume} {40}},\
  \bibinfo {pages} {4277} (\bibinfo {year} {1989})}\BibitemShut {NoStop}%
\bibitem [{\citenamefont {Horodecki}\ and\ \citenamefont
  {Horodecki}(1999)}]{PhysRevA.59.4206}%
  \BibitemOpen
  \bibfield  {author} {\bibinfo {author} {\bibfnamefont {M.}~\bibnamefont
  {Horodecki}}\ and\ \bibinfo {author} {\bibfnamefont {P.}~\bibnamefont
  {Horodecki}},\ }\bibfield  {title} {\bibinfo {title} {Reduction criterion of
  separability and limits for a class of distillation protocols},\ }\href
  {https://doi.org/10.1103/PhysRevA.59.4206} {\bibfield  {journal} {\bibinfo
  {journal} {Phys. Rev. A}\ }\textbf {\bibinfo {volume} {59}},\ \bibinfo
  {pages} {4206} (\bibinfo {year} {1999})},\ \Eprint
  {https://arxiv.org/abs/quant-ph/9708015} {arXiv:quant-ph/9708015}
  \BibitemShut {NoStop}%
\bibitem [{\citenamefont {Ollivier}\ and\ \citenamefont
  {Zurek}(2001)}]{PhysRevLett.88.017901}%
  \BibitemOpen
  \bibfield  {author} {\bibinfo {author} {\bibfnamefont {H.}~\bibnamefont
  {Ollivier}}\ and\ \bibinfo {author} {\bibfnamefont {W.~H.}\ \bibnamefont
  {Zurek}},\ }\bibfield  {title} {\bibinfo {title} {Quantum discord: A measure
  of the quantumness of correlations},\ }\href
  {https://doi.org/10.1103/PhysRevLett.88.017901} {\bibfield  {journal}
  {\bibinfo  {journal} {Phys. Rev. Lett.}\ }\textbf {\bibinfo {volume} {88}},\
  \bibinfo {pages} {017901} (\bibinfo {year} {2001})}\BibitemShut {NoStop}%
\bibitem [{\citenamefont {Ferraro}\ \emph {et~al.}(2010)\citenamefont
  {Ferraro}, \citenamefont {Aolita}, \citenamefont {Cavalcanti}, \citenamefont
  {Cucchietti},\ and\ \citenamefont {Ac\'{\i}n}}]{PhysRevA.81.052318}%
  \BibitemOpen
  \bibfield  {author} {\bibinfo {author} {\bibfnamefont {A.}~\bibnamefont
  {Ferraro}}, \bibinfo {author} {\bibfnamefont {L.}~\bibnamefont {Aolita}},
  \bibinfo {author} {\bibfnamefont {D.}~\bibnamefont {Cavalcanti}}, \bibinfo
  {author} {\bibfnamefont {F.~M.}\ \bibnamefont {Cucchietti}},\ and\ \bibinfo
  {author} {\bibfnamefont {A.}~\bibnamefont {Ac\'{\i}n}},\ }\bibfield  {title}
  {\bibinfo {title} {Almost all quantum states have nonclassical
  correlations},\ }\href {https://doi.org/10.1103/PhysRevA.81.052318}
  {\bibfield  {journal} {\bibinfo  {journal} {Phys. Rev. A}\ }\textbf {\bibinfo
  {volume} {81}},\ \bibinfo {pages} {052318} (\bibinfo {year}
  {2010})}\BibitemShut {NoStop}%
\bibitem [{\citenamefont {{Poxleitner}}\ and\ \citenamefont
  {{Hinrichsen}}(2021)}]{poxleitner2021gaussian}%
  \BibitemOpen
  \bibfield  {author} {\bibinfo {author} {\bibfnamefont {M.}~\bibnamefont
  {{Poxleitner}}}\ and\ \bibinfo {author} {\bibfnamefont {H.}~\bibnamefont
  {{Hinrichsen}}},\ }\bibfield  {title} {\bibinfo {title} {{Gaussian
  continuous-variable isotropic state}},\ }\href
  {https://doi.org/10.1103/PhysRevA.104.032423} {\bibfield  {journal} {\bibinfo
   {journal} {\pra}\ }\textbf {\bibinfo {volume} {104}},\ \bibinfo {eid}
  {032423} (\bibinfo {year} {2021})},\ \Eprint
  {https://arxiv.org/abs/2105.03141} {arXiv:2105.03141 [quant-ph]} \BibitemShut
  {NoStop}%
\end{thebibliography}%

\onecolumngrid

\clearpage
\appendix

\vspace{-1cm}


\section{Symplectic Gram Schmidt}
\label{sec:symplecticGS}

The symplectic Gram-Schmidt (sGS) procedure aims to produce a set of symplectically orthogonal vectors seeded by an initial basis of interest, $\{ \mathbf{\bar{v}} \}$.
The symplectic inner product is written as,
\begin{equation}
  \langle \mathbf{u}, \mathbf{v}\rangle_{\Omega} = \mathbf{u} {\Omega} \mathbf{v}^T \ \ \ ,
\end{equation}
which vanishes by symmetry for any real vector with itself, $\langle \mathbf{v},\mathbf{v}\rangle_{\Omega} = 0$.
The aim is to produce a basis for which $\langle \mathbf{v}_i, \mathbf{v}_j\rangle_\Omega = \Omega_{i,j}$.
Subsequently constructing an operator, $ {S}$, through the stacking of these basis vectors in rows will produce a symplectic transformation preserving the canonical commutation relations, $ {S} {\Omega} {S}^T =  {\Omega}$, by design.

At each step, vectors in the basis must be symplectically orthogonalized and normalized.
Consider the first two vectors proposed $\mathbf{\bar{v}}_{x,1}, \mathbf{\bar{v}}_{p,1}$ corresponding to the two quadratures of the first mode.
By symmetry, the diagonal elements of the intended inner product vanish such that only the normalization requires modification,
\begin{equation}
  \mathbf{v}_{x,1} = \frac{\mathbf{\bar{v}}_{x,1}}{\sqrt{ |\mathcal{A} |}} \qquad ,\qquad \mathbf{v}_{p,1} =   \frac{\mathbf{\bar{v}}_{p,1}}{\text{sign}\left(\mathcal{A} \right)\sqrt{|\mathcal{A}|}} \qquad , \qquad \mathcal{A} = \langle \mathbf{\bar{v}}_{x,1}, \mathbf{\bar{v}}_{p,1}\rangle_{\Omega} \ \ \ .
\end{equation}
Though alternate distributions of the normalization lead also to  effective symplectic transformations for entanglement consolidation, the chosen symmetric distribution between the position and momentum vectors is found to have advantageous properties of numerical stability.
For the incorporation of further vectors into the symplectically orthogonal basis, a procedure of removing projective contributions must first be performed,
\begin{align}
  \mathbf{v}'_{z,k} &= \mathbf{\bar{v}}_{z,k} - \sum_{j = 1}^{k-1} \Big[ \langle \mathbf{v}_{x,j}, \mathbf{\bar{v}}_{z,k}\rangle_{\Omega} \ \langle \mathbf{v}_{x,j}, \mathbf{v}_{p,j}\rangle_{\Omega} \mathbf{v}_{p,j} + \langle \mathbf{v}_{p,j}, \mathbf{\bar{v}}_{z,k}\rangle_{\Omega}\ \langle \mathbf{v}_{p,j}, \mathbf{v}_{x,j}\rangle_{\Omega} \mathbf{v}_{x,j}\Big] \label{eq:sGSstep} \ \ \ ,  \\
  &=  \mathbf{\bar{v}}_{z,k} - \sum_{j = 1}^{k-1} \Big[ \langle \mathbf{v}_{x,j}, \mathbf{\bar{v}}_{z,k}\rangle_{\Omega} \ \mathbf{v}_{p,j} - \langle \mathbf{v}_{p,j}, \mathbf{\bar{v}}_{z,k}\rangle_{\Omega}\  \mathbf{v}_{x,j}\Big] \ \ \ ,
\end{align}
with $z \in \{ x, p\}$, indicating that the symplectic orthogonalization applies equivalently for the position and momentum vectors.
The first inner product in each term of Eq.~\eqref{eq:sGSstep} captures the projection magnitude, while the second inner product and vector set the contribution in a direction that will have $+1$ symplectic overlap with the component being removed.
After a final normalization procedure,
\begin{equation}
  \mathbf{v}_{x,k} = \frac{\mathbf{v}'_{x,k}}{\sqrt{\mathcal{A}}} \quad , \quad \mathbf{v}_{p,k} = \frac{\mathbf{v}'_{p,k}}{\text{sign}(\mathcal{A}) \sqrt{|\mathcal{A}|}} \quad , \quad \mathcal{A} = \langle \mathbf{v}'_{x,k}, \mathbf{v}'_{p,k}\rangle_{\Omega} \ \ \ ,
\end{equation}
the collection of symplectically orthogonal vectors may be organized row-wise to produce a symplectic transformation.
For example, in the alternating position-momentum basis described above, the symplectic transformation may be constructed as,
\begin{equation}
 {S}_{i,j} =  \begin{cases} \left(\mathbf{v}_{x,\frac{i+1}{2}}\right)_{\frac{j+1}{2}}  & i,j  \text{ odd} \\
\left(\mathbf{v}_{p,\frac{i}{2}}\right)_{\frac{j}{2}} & i,j \text{ even} \\
0 & \text{otherwise}
\end{cases} \ \ \ .
\end{equation}

\section{Examples of Entanglement Consolidation}
\label{app:consolidationEX}
\subsection{\texorpdfstring{$GH^\Gamma$}{GHgamma} Eigenvectors}
\label{app:GHgammaEx}

Consider the vacuum of one-dimensional massive non-interacting lattice scalar field theory.
With two modes per region at zero lattice separation ($\tilde{r} = 0$) and mass arbitrarily chosen to be $m = 0.003$, the CM written in the basis of alternating $\phi$ and $\pi$ coordinates is,
\begin{equation}
  \sigma = \left(
\begin{array}{cccccccc}
 2.51 & 0 & 1.87 & 0 & 1.66 & 0 & 1.53 & 0 \\
 0 & 1.27 & 0 & -0.424 & 0 & -0.0849 & 0 & -0.0364 \\
 1.87 & 0 & 2.51 & 0 & 1.87 & 0 & 1.66 & 0 \\
 0 & -0.424 & 0 & 1.27 & 0 & -0.424 & 0 & -0.0849 \\
 1.66 & 0 & 1.87 & 0 & 2.51 & 0 & 1.87 & 0 \\
 0 & -0.0849 & 0 & -0.424 & 0 & 1.27 & 0 & -0.424 \\
 1.53 & 0 & 1.66 & 0 & 1.87 & 0 & 2.51 & 0 \\
 0 & -0.0364 & 0 & -0.0849 & 0 & -0.424 & 0 & 1.27 \\
\end{array}
\right) \ \ \ .
\label{eq:sigmaEX2mode}
\end{equation}
Modes have been ordered from left to right such that the inner two modes correspond to the third through sixth row and column indices.
In this alternating basis, the matrix $G_{i,j} = \langle \phi_i \phi_j\rangle$ of correlation functions is composed of the odd-odd coordinates and the matrix $H_{i,j} = \langle \pi_i \pi_j \rangle$ is composed of the even-even coordinates of the CM, with a factor of 2 as shown in Eq.~\eqref{eq:cm2exp}.
There is one symplectic eigenvalue contributing to the negativity for a total negativity between the two regions of $\mathcal{N} = 0.654$.
Before and after symplectic orthogonalization, the local transformation is determined by the left and right eigenvectors of the product $GH^\Gamma$ as,
\begin{equation}
  \bar{S}_A =   \left(
\begin{array}{cccc}
 0.0607 & 0 & 0.683 & 0 \\
 0 & 0.321 & 0 & 0.704 \\
 0.762 & 0 & -0.347 & 0 \\
 0 & 0.631 & 0 & -0.0560 \\
\end{array}
\right)
\quad , \quad S_A =  \left(
\begin{array}{cccc}
 0.0858 & 0 & 0.965 & 0 \\
 0 & 0.454 & 0 & 0.995 \\
 1.08 & 0 & -0.491 & 0 \\
 0 & 0.892 & 0 & -0.0793 \\
\end{array}
\right) \ \ \ .
\end{equation}
The transformed CM,
\begin{equation}
  \sigma' = S \sigma S^T = \scalemath{0.8}{\left(
\begin{array}{cccccccc}
 2.67 & 0 & 0.912 & 0 & 0.912 & 0 & 2.03 & 0 \\
 0 & 1.14 & 0 & 0.0536 & 0 & -0.0536 & 0 & -0.505 \\
 0.912 & 0 & 1.54 & 0 & 0.474 & 0 & 0.912 & 0 \\
 0 & 0.0536 & 0 & 1.08 & 0 & -0.0196 & 0 & -0.0536 \\
 0.912 & 0 & 0.474 & 0 & 1.54 & 0 & 0.912 & 0 \\
 0 & -0.0536 & 0 & -0.0196 & 0 & 1.08 & 0 & 0.0536 \\
 2.03 & 0 & 0.912 & 0 & 0.912 & 0 & 2.67 & 0 \\
 0 & -0.505 & 0 & -0.0536 & 0 & 0.0536 & 0 & 1.14 \\
\end{array}
\right)
  }  \ \ \ ,
  \label{eq:sigprimeGHex}
\end{equation}
while not visually exhibiting special entanglement structure, captures the negativity core.
The negativity that had been distributed nonlocally within the $(2\times 2)_{\rm mixed}$ CM is now localized to the $(1\times 1)_{\rm mixed}$ mode pair of the outer two modes.
More precisely, the negativity present in the reduced CM of the first and fourth mode (one from each local region), is equal to the negativity of the full system $\mathcal{N} = 0.654$.
Though it does not visually appear to be so, an application of separability flow developed in Ref.~\cite{2001PhRvL..87p7904G} determines that the $(1\times 1)_{\rm mixed}$ pair of outer modes 1 and 4 is separable from the remaining $(1 \times 1)_{\rm mixed}$ pair of inner modes 2 and 3.

While for two lattice sites per region there is only one PT symplectic eigenvalue contributing to the negativity, one can find a set of additive negativity contributions as the pixelation of each region is increased.
For example, the same system at zero separation $(\tilde{r} = 0)$ and mass $(m = 0.003)$ but with four modes per region exhibits three separate contributions to the negativity:  ${0.858, 0.0230, 0.000298}$ for a total of $\mathcal{N} = 0.888$.
The local symplectic transformation designed to localize this negativity into a core of local modes is,
\begin{equation}
 S_A = \left(
\begin{array}{cccccccc}
 0.0286 & 0 & -0.0736 & 0 & -0.190 & 0 & -0.818 & 0 \\
 0 & -0.244 & 0 & -0.460 & 0 & -0.693 & 0 & -1.03 \\
 0.373 & 0 & -0.433 & 0 & -0.803 & 0 & 0.645 & 0 \\
 0 & 0.0505 & 0 & -0.559 & 0 & -0.741 & 0 & 0.224 \\
 0.386 & 0 & -1.03 & 0 & 0.759 & 0 & -0.143 & 0 \\
 0 & 0.112 & 0 & -0.646 & 0 & 0.379 & 0 & -0.0260 \\
 0.984 & 0 & 0.124 & 0 & -0.0943 & 0 & -0.225 & 0 \\
 0 & 0.960 & 0 & 0.479 & 0 & 0.152 & 0 & -0.0448 \\
\end{array}
\right) \ \ \ .
\end{equation}
After this transformation, the pair of outer modes in each region carries an amount of negativity equal to that of the dominant contribution, $\mathcal{N}_{1,8} = 0.858$.  The second $(1\times 1)_{\rm mixed}$ pair of modes 2 and 7 carry negativity equal to that of the following symplectic eigenvalue, $\mathcal{N}_{2,7} = 0.0230$.  The remaining pairs continue this pattern governed by the symplectic eigenvalue contributions, $\mathcal{N}_{3,6} =  0.000298$ and $\mathcal{N}_{4,5} = 0$.
As the separation between the regions is increased, the number of negativity contributions rapidly diminishes and becomes dominated by a single eigenvalue before vanishing completely.  By calculating the local symplectic transformations in this manner, the consolidation of entanglement mimics the structure of symplectic eigenvalues contributing to the negativity.

\subsection{\texorpdfstring{$i \Omega \sigma^\Gamma $}{} Eigenvectors }
\label{app:iOmegaSigmaGammaEx}

The local symplectic operators designed to consolidate the vacuum entanglement into a core was above (and in the main text) informed by the left and right eigenvectors of $GH^\Gamma$.
Naturally, such a transformation may also be designed from symplectic eigenvectors themselves.
Consider the right eigenvectors of the product $ \Omega {\sigma}^\Gamma  \mathbf{\bar{v}}_j = \lambda_j \mathbf{\bar{v}}_j$.  The eigenvalues, $\lambda_j$, are the PT symplectic eigenvalues from which the logarithmic negativity may be calculated additively,
\begin{equation}
  \mathcal{N} =  \sum_{j=1}^{2d} \mathcal{N}_j \qquad , \qquad  \mathcal{N}_j = - \log_2 \left[ \min \left(|\lambda_j|, 1\right) \right] \ \ \ .
\end{equation}
Note that the symplectic eigenvalues appear in $\pm$ pairs through this method of calculation, though do not contribute twice to the negativity.
The eigenvectors with paired eigenvalues are related by a conjugation, if $\lambda_j = - \lambda_k$ then $\mathbf{\bar{v}}_j = \mathbf{\bar{v}}_k^*$, up to arbitrary global phases.
If ordered from smallest to largest, ignoring $\mathbf{\bar{v}}_j$ for which $\lambda_j < 0$,  the initial basis of the local consolidating symplectic transformation can be chosen as,
\begin{equation}
\bar{S}_A =
  \begin{pmatrix}
  \begin{pmatrix}
    & \text{Re}\mathbf{\bar{v}}_{1} &
  \end{pmatrix}_A \\
  \begin{pmatrix}
    & \text{Im}\mathbf{\bar{v}}_{1} &
  \end{pmatrix}_A \\ \vdots \\
  \begin{pmatrix}
    & \text{Re}\mathbf{\bar{v}}_{d} &
  \end{pmatrix}_A \\
  \begin{pmatrix}
    & \text{Im}\mathbf{\bar{v}}_{d} &
  \end{pmatrix}_A
  \end{pmatrix}
  \ \ \ .
  \label{eq:sAbar}
\end{equation}
Separating the independent real and imaginary components of each symplectic eigenvector into two basis vectors of the CM informs the two quadratures of each Gaussian mode~\footnote{Note that this basis is distinct from that utilized by Ref.~\cite{giedke2001distillability} to demonstrate that non-zero negativity can always be transformed into a $(1\times 1)$-mode subsystem through local symplectic operations, and thus any NPT entangled Gaussian state provides non-zero distillable entanglement.  In particular, the eigenvectors chosen to seed this transformation to the negativity core satisfies $\mathbf{\bar{v}}^\dagger \left( \sigma - i \Omega^\Gamma \right) \mathbf{\bar{v}} \geq 0$.}.
In the desired transformation, the eigenvectors $\bar{\mathbf{v}}_j$ have been each globally phased so that their first elements are real, allowing the resulting symplectic transformation to act separately on the $(\phi,\pi)$ quadratures without mixture.
This separation provides commutation with the PT operation, and thus allows the negativity information in the PT space to be utilized in defining a realizable symplectic transformation on the physical CM, as discussed in Section~\ref{sec:broaderapp}.
For the purposes of locally transforming two regions of the scalar vacuum into a basis with a manifest negativity core, this choice of information distribution is not unique, e.g., a choice of $\text{Re}\mathbf{\bar{v}}_1 \pm \text{Im}\mathbf{\bar{v}}_1$ would have a similar effect.
The source of arbitrariness in this distribution reflects the arbitrary set of single mode transformations that do not affect the negativity structure of the final CM.
After an application of the symplectic Gram-Schmit procedure, the symplectic operator generated from this basis acts upon the CM, ${\sigma}' =  S {\sigma}  S^{T}$ with $ {S} =  {S}_A \oplus  {S}_B$, and consolidates the entanglement between regions of the field into a number of $(1\times 1)_{\rm mixed}$ mode pairs equal to the number of PT symplectic eigenvalues less than unity.

Consider the example of Section~\ref{app:GHgammaEx} with two modes per region at zero separation ($\tilde{r} = 0$) and mass of $m = 0.003$.
The CM written in the basis of alternating $\phi$ and $\pi$ coordinates is provided in Eq.~\eqref{eq:sigmaEX2mode}.
Before and after symplectic orthogonalization, the local transformation is determined by the eigenvectors of the product $i \Omega \sigma^\Gamma$ as,
\begin{equation}
  \bar{S}_A = \left(
\begin{array}{cccc}
 0.041 & 0 & 0.467 & 0 \\
 0 & -0.220 & 0 & -0.482 \\
 0.513 & 0 & -0.234 & 0 \\
 0 & -0.425 & 0 & 0.038 \\
\end{array}
\right)
  \qquad , \quad
  S_A =
\left(
\begin{array}{cccc}
 0.086 & 0 & 0.965 & 0 \\
 0 & 0.454 & 0 & 0.995 \\
 1.077 & 0 & -0.491 & 0 \\
 0 & 0.892 & 0 & -0.079 \\
\end{array}
\right) \ \ \ .
\end{equation}
The CM transformed by local consolidating symplectic operation, $S = S_A \oplus S_B$, becomes,
\begin{equation}
  \sigma' = S \sigma S^T = \left(
\begin{array}{cccccccc}
 2.670 & 0 & 0.912 & 0 & 0.912 & 0 & 2.034 & 0 \\
 0 & 1.140 & 0 & 0.054 & 0 & -0.054 & 0 & -0.505 \\
 0.912 & 0 & 1.536 & 0 & 0.474 & 0 & 0.912 & 0 \\
 0 & 0.054 & 0 & 1.081 & 0 & -0.020 & 0 & -0.054 \\
 0.912 & 0 & 0.474 & 0 & 1.536 & 0 & 0.912 & 0 \\
 0 & -0.054 & 0 & -0.020 & 0 & 1.081 & 0 & 0.054 \\
 2.034 & 0 & 0.912 & 0 & 0.912 & 0 & 2.670 & 0 \\
 0 & -0.505 & 0 & -0.054 & 0 & 0.054 & 0 & 1.140 \\
\end{array}
\right) \ \ \ .
\end{equation}
The negativity that had been distributed nonlocally within the $(2\times 2)_{\rm mixed}$ system is again localized to the $(1\times 1)_{\rm mixed}$ mode pair of the outer two modes.

\section{The Gaussian Separability Criterion}
\label{app:sepcriterion}
A necessary and sufficient condition for a CM, $\sigma$, to be separable is the existence of two valid CMs, $\sigma_a, \sigma_b$  such that~\cite{2001PhRvL..86.3658W},
\begin{equation}
  \sigma \geq \sigma_a \oplus \sigma_b \qquad , \qquad \sigma_{a,b} \geq i \Omega \ \ \ .
\end{equation}
Reference~\cite{2001PhRvL..87p7904G} describes not only how to numerically identify separability/inseparability for mixed Gaussian states, but also provides a constructive method for calculating a viable tensor-product state and classical noise contribution if separability is determined.
For example, applied to the core-halo separability, these techniques identify an explicit decomposition in the form of Eq.~\eqref{eq:core-halosep}.
For a transformed CM, $\sigma'$, in a basis ordered as core modes followed by halo modes,
\begin{equation}
  \sigma' = \begin{pmatrix}
    \sigma'_{cc} & \sigma'_{ch} \\
    \left(\sigma'_{ch}\right)^T & \sigma'_{hh}
  \end{pmatrix} \ \ \ ,
\end{equation}
the flow of Ref.~\cite{2001PhRvL..87p7904G} begins with the creation of a symmetric object (of modified dimension if the core and halo have different numbers of modes, $\dim\left(\sigma'_{cc}\right) \neq \dim\left( \sigma'_{hh}\right)$) as,
\begin{equation}
  \sigma'_1 = \begin{pmatrix}
    \sigma'_{cc}-\text{Re} \left(X\right) & -\text{Im}(X) \\
    -\text{Im}\left(X^T\right) & \sigma'_{cc}-\text{Re} \left(X\right)
  \end{pmatrix} \ \ \ ,
\end{equation}
with $X = \sigma'_{ch} \left( \sigma'_{hh}-i \Omega\right)^{-1} \left(\sigma'_{ch}\right)^T$.
This transformation describes a flow, $\sigma'_k$ for increasing $k$, that can be iterated until separability/inseparability is determined.
After the first step (and generalized to further steps if necessary), the condition for separability is
\begin{equation}
  \sigma_c = \sigma'_{cc}-\text{Re}(X) - ||\text{Im}(X)||_{op} \mathbb{I} \geq i \Omega \ \ \ ,
\end{equation}
where the operator norm $||\xi||_{op}$ is the maximum eigenvalue of $\sqrt{\xi^\dagger \xi}$.
This condition means that separability between the core and halo is identified if $\sigma_c$ is a valid CM.
A constructive relationship is then found to be,
\begin{equation}
  \sigma'_1 = \sigma_c \otimes \sigma_c + Y_1 \ \ \ .
\end{equation}
This information can be transferred back through the flow to inform a decomposition of the physical $\sigma'$ as,
\begin{equation}
  \sigma' = \sigma_c \oplus \sigma_h + Y \quad , \quad \sigma_h = \sigma'_{hh}-\left(\sigma'_{ch}\right)^T \left(\sigma'_{cc} - \sigma_c\right)^{-1} \sigma'_{ch} \ \ \ .
\end{equation}
When $(\sigma'_{cc}-\sigma_c)$ is singular, as is often the case in the lattice scalar field vacuum, the pseudoinverse can be utilized for successful identification of a valid tensor-product and noise decomposition.
However, the associated poor condition numbers often require high precision (hundreds of digits for applications in the main text) to maintain numerical stability.
Conveniently, the core-halo separability of the field is observed to be determined after a single step in the flow, and thus $\sigma'_k$ for $k = 1$ is sufficient for quantifying the inaccessable entanglement of Fig.~\ref{fig:boundent}.
For constructive decompositions at higher flow depth, an additional step of symmetrization is required as separable states underlying the noise are tracked backwards through the flow to the original physical CM.
Reference~\cite{2001PhRvL..87p7904G} provides further details.

The following is an example of sequentially applying this separability criterion to establish simultaneous separability among the core pairs.
If the number of core mode pairs, $n_c$, is four, one possible two-level splitting procedure that may yield a fully separable pair decomposition in the core is,
\begin{align}
  \sigma_c &= \sigma_{c,12} \oplus \sigma_{c,34} + Y_1 \ \ \ ,\nonumber \\
  \sigma_{c,12} &= \sigma_{c,1} \oplus \sigma_{c,2} + Y_2 \ \ \ , \\
  \sigma_{c,34} &= \sigma_{c,3} \oplus \sigma_{c,4} + Y_3  \ \ \ , \nonumber
\end{align}
where the $Y_i$ are all PSD matrices representing additional local noise contributions.
With such sequential bipartitions, the simultaneous separability of the core would be identified as,
\begin{equation}
  \sigma' = \sigma_{c,1} \oplus \sigma_{c,2} \oplus \sigma_{c,3} \oplus \sigma_{c,4} \oplus \sigma_h + Y + \sum_i Y_i \ \ \ ,
\end{equation}
where the $Y_i$'s are understood to be placed in their relevant Hilbert spaces completed by $\mathbf{0}$ block matrices.
For $n_c = 2$, this sequential procedure is achieved in one level with no ambiguity.
For $n_c > 2$, this sequential procedure requires multiple levels with an array of different possible splitting structures at each level.
A search over splitting sequences allows full simultaneous separability in the core to be reliably identified.
The rapid loss of numerical precision through this technique with matrix inversions at each level has required utilization of precision with several hundreds of digits to calculate the simultaneously separable decomposition of Eq.~\eqref{eq:simultaneouscoresep}, exacerbated for fixed $n_c$ at large number of lattice sites per region, $d$, and associated spatial separations.

\section{Separability-Obscured Entanglement}
\label{app:sepobscent}

For mixed states, the designation of Hilbert space separability implies the existence of a convex density matrix decomposition of only tensor-product pure states,
\begin{equation}
  \rho_{\rm separable} = \sum_i p_i \ \rho^A_i \otimes \rho^B_i \ \ \ ,
\end{equation}
with positive, real weights $\sum_i p_i=1$.
However, it is possible for both tensor-product and entangled convex decompositions to exist as compatible underlying descriptions of a separable mixed-state density matrix.
For a concrete two-qubit example in which a separable state can be produced with distributed entanglement, consider the mixture of a Bell state with the maximally mixed state,
\begin{equation}
  \rho(\eta) = \eta \ \rho\left(\frac{|00\rangle + |11\rangle}{\sqrt{2}}\right) + (1-\eta) \frac{\mathbb{I}}{4} \ \ \ ,
  \label{eq:isotropic}
\end{equation}
known as an isotropic state or two-qubit Werner state up to local transformation~\cite{PhysRevA.40.4277,PhysRevA.59.4206}.
In two-qubit systems, negativity is both necessary and sufficient for the identification of separability.
With this quantity, it can be determined that the mixture in Eq.~\eqref{eq:isotropic} is separable in the regime $\eta \leq 1/3$.
Explicitly, one viable tensor-product decomposition of this density matrix is,
\begin{align}
  \rho(\eta) &= \frac{\eta}{2} \left[ \rho\left( |++\rangle_x \right) + \rho\left(|--\rangle_x\right)  + \rho\left(|+-\rangle_y\right) + \rho\left(|-+\rangle_y\right) \right] \nonumber \\ & \qquad  + \frac{1 - \eta}{4} \left[ \rho\left( |00\rangle\right) + \rho\left( |11\rangle \right) \right]
  + \frac{1-3 \eta}{4} \left[ \rho\left( |01\rangle \right) + \rho\left( |10\rangle \right) \right] \ \ \ ,
  \label{eq:TPconvexdecomp}
\end{align}
where $\rho(|\psi\rangle) = |\psi\rangle\langle\psi|$ is the density matrix associated with pure state $|\psi\rangle$,  $|\pm\rangle_x = \frac{|0\rangle \pm |1\rangle}{\sqrt{2}}$, and  $|\pm\rangle_y = \frac{|0\rangle \pm i|1\rangle}{\sqrt{2}}$.
This decomposition is convex only for $\eta \leq 1/3$, beyond which the state is entangled and the last coefficient becomes negative.
Thus, it is possible that the regime identified as separable can be physically produced through Eq.~\eqref{eq:TPconvexdecomp} with solely classical correlations or through Eq.~\eqref{eq:isotropic} as a mixed ensemble of entangled pairs and unentangled basis states, dominated by the latter.

Two parties, A and B, receiving distributed samples of this separable state ($\eta \leq 1/3$) from a source will not be able to distinguish between the entangled and unentangled production mechanisms, even with classical communication and a complete tomography of the mixed density matrix that they share.
However, with the addition of classical communication from the source to parties A and B, e.g., indexing the pure state from which each provided sample was generated as in Eq.~\eqref{eq:isotropic}, parties A,B may be able to extract entangled pairs.
For states with separability-obscured entanglement, it is possible to identify distributed entanglement provided only additional classical communication from the source.
This is a physical mechanism underlying many observations of entanglement structure, e.g., Refs.~\cite{PhysRevLett.80.2493,PhysRevA.58.826,Vidal:1998ch,PhysRevA.63.032306,PhysRevLett.88.017901,PhysRevA.81.052318,poxleitner2021gaussian}.

\subsection{Example: Lattice Scalar Field Vacuum}
\label{app:sepobsEntEX}

The following is an explicit example of a mixed state of four CV quantum modes, transformed from a pair of low-pixelation vacuum regions, that exhibits both core-halo and halo$_A$-halo$_B$ separability but is not simultaneously separable along these two partitions.
In particular, despite this pair of separability observations, the distillable entanglement between, for example, halo$_A$ and the rest of the system, will be non-zero.

Consider the CM associated with two sites per region, shown in Eq.~\eqref{eq:sigprimeGHex}.
Having been transformed through local symplectics, modes 1 and 4 comprise core$_A$ and core$_B$, while modes 2 and 3 comprise halo$_A$ and halo$_B$, where $A,B$ indicates the associated region of the scalar field vacuum.
Through the techniques described in Ref.~\cite{2001PhRvL..87p7904G}, core-halo separability is identified through the inequality of Eq.~\eqref{eq:core-halosep} with the following decomposition,
\hfsetfillcolor{BlueGreen!10}
\hfsetbordercolor{BlueGreen}
\begin{equation}
  \sigma_c = \scalemath{0.8}{\left(
\begin{array}{cccc}
 1.1143 & 0 & 0.4785 & 0 \\
 0 & 1.1044 & 0 & -0.4687 \\
 0.4785 & 0 & 1.1143 & 0 \\
 0 & -0.4687 & 0 & 1.1044 \\
\end{array}
\right)} \quad , \quad
\sigma_h  = \scalemath{0.8}{\left(
\begin{array}{cccc}
 1.0018 & 0 & -0.0600 & 0 \\
 0 & 1.0018 & 0 & 0.0600 \\
 -0.0600 & 0 & 1.0018 & 0 \\
 0 & 0.0600 & 0 & 1.0018 \\
\end{array}
\right)} \  \ \ ,
\label{eq:corehaloEX}
\end{equation}
\begin{equation}
  Y = \scalemath{0.6}{
  \left(
\begin{array}{cccccccc}
 1.5553 & 0 & 1.5553 & 0 & 0.9115 & 0 & 0.9115 & 0 \\
 0 & 0.0360 & 0 & -0.0360 & 0 & 0.0536 & 0 & -0.0536 \\
 1.5553 & 0 & 1.5553 & 0 & 0.9115 & 0 & 0.9115 & 0 \\
 0 & -0.0360 & 0 & 0.0360 & 0 & -0.0536 & 0 & 0.0536 \\
 0.9115 & 0 & 0.9115 & 0 & \tikzmarkin{a}(0.2,-0.1)(-0.1,0.35)0.5343 & 0 & 0.5343 & 0 \\
 0 & 0.0536 & 0 & -0.0536 & 0 & 0.0796 & 0 & -0.0796 \\
 0.9115 & 0 & 0.9115 & 0 & 0.5343 & 0 & 0.5343 & 0 \\
 0 & -0.0536 & 0 & 0.0536 & 0 & -0.0796 & 0 & 0.0796\tikzmarkend{a} \\
\end{array}
\right)
} \ \ \ ,
\end{equation}
in the basis of $\{ \rm{core}_{A}, \rm{core}_{B}, \rm{halo}_{A}, \rm{halo}_{B}\}$.

For Gaussian states with density matrix $\rho$, the purity may be calculated from the associated CM, $\sigma$, as $\Tr\left(\rho^2\right) = 1/\sqrt{\det(\sigma)}$.
A decomposition with $\sigma_c$ and $\sigma_h$ pure can always be created as mixed Gaussian states may be described as randomly displaced pure states.
One method for separating the pure contribution is to utilize the Williamson normal decomposition in which a CM is diagonalized through global symplectic operations, $\sigma = S_W \Sigma S_W^T$, where $\Sigma$ is diagonal and $S_W$ is symplectic.
After identification of the symplectic operation that transforms to the Williamson normal form, the CM can be separated into a pure state (with $\Sigma = \mathbb{I}$) and a PSD noise contribution describing classically distributed displacements,
\begin{equation}
\sigma = S_WS^T_W + S_W (\Sigma-\mathbb{I}) S_W^T = \sigma^{(p)} + Y^{(p)} \ \ \ .
\label{eq:purify}
\end{equation}
When applied to the core-halo decomposition,
\begin{align}
\sigma' = \sigma_c \oplus \sigma_h + Y &= \sigma_c^{(p)} \oplus \sigma_h^{(p)} + Y^{(p)}_c \oplus Y^{(p)}_h + Y \ \ \ , \nonumber \\ &= \sigma_c^{(p)} \oplus \sigma_h^{(p)} + Y^{(p)} \ \ \ ,
\end{align}
the explicit CM and noise matrices become,
\begin{equation}
  \sigma_c^{(p)}  = \scalemath{0.8}{\left(
\begin{array}{cccc}
 1.1093 & 0 & 0.4736 & 0 \\
 0 & 1.1024 & 0 & -0.4706 \\
 0.4736 & 0 & 1.1093 & 0 \\
 0 & -0.4706 & 0 & 1.1024 \\
\end{array}
\right)} \quad , \quad
\sigma_h^{(p)} = \scalemath{0.8}{\left(
\begin{array}{cccc}
 1.0018 & 0 & -0.0600 & 0 \\
 0 & 1.0018 & 0 & 0.0600 \\
 -0.0600 & 0 & 1.0018 & 0 \\
 0 & 0.0600 & 0 & 1.0018 \\
\end{array}
\right)} \ \ \ ,
\label{eq:sigcorepure}
\end{equation}
\begin{equation}
Y^{(p)} = \scalemath{0.6}{
\left(
\begin{array}{cccccccc}
 1.5602 & 0 & 1.5602 & 0 & 0.9115 & 0 & 0.9115 & 0 \\
 0 & 0.0381 & 0 & -0.0341 & 0 & 0.0536 & 0 & -0.0536 \\
 1.5602 & 0 & 1.5602 & 0 & 0.9115 & 0 & 0.9115 & 0 \\
 0 & -0.0341 & 0 & 0.0381 & 0 & -0.0536 & 0 & 0.0536 \\
 0.9115 & 0 & 0.9115 & 0 & \tikzmarkin{b}(0.2,-0.1)(-0.1,0.35)0.5343 & 0 & 0.5343 & 0 \\
 0 & 0.0536 & 0 & -0.0536 & 0 & 0.0796 & 0 & -0.0796 \\
 0.9115 & 0 & 0.9115 & 0 & 0.5343 & 0 & 0.5343 & 0 \\
 0 & -0.0536 & 0 & 0.0536 & 0 & -0.0796 & 0 & 0.0796\tikzmarkend{b} \\
\end{array}
\right)
} \ \ \ .
\end{equation}
Written in this form, a preparation procedure can be imagined for two-site pixelated regions of the vacuum through distribution of two two-mode pure entangled CV states, $\sigma_{c,h}$, and the addition of classical correlations.
The added noise causes $\sigma_h$ to exhibit only separability-obscured entanglement, and thus the accessible entanglement of the second pair to vanish.

The shaded sub-blocks in the noise matrices indicate $\delta$ of Eq.~\eqref{eq:core-halosep}, where the PSD property after replacement of $\delta$ by the noise structure of halo separability, $Y_h$, will determine whether the core-halo and halo$_A$-halo$_B$ can be simultaneously separable.
Similar techniques to those above allow the halo$_A$-halo$_B$ separability to be identified through the inequality of Eq.~\eqref{eq:halo-halosep} with the following decomposition,
\begin{equation}
  \sigma_{h,A}^{(p)} = \sigma_{h,B}^{(p)} = \left(
\begin{array}{cc}
 0.9951 & 0 \\
 0 & 1.0050 \\
\end{array}
\right) \quad , \quad
  Y_h^{(p)} = \left(
\begin{array}{cccc}
 0.5410 & 0 & 0.4743 & 0 \\
 0 & 0.0765 & 0 & -0.0196 \\
 0.4743 & 0 & 0.5410 & 0 \\
 0 & -0.0196 & 0 & 0.0765 \\
\end{array}
\right)  \ \ \ .
\label{eq:Yh}
\end{equation}
For clarity, a symmetric decomposition with $\sigma^{(p)}_{h,A} = \sigma^{(p)}_{h,B}$ has been chosen.
The simultaneous separability condition may be explored through replacement of the lower block of $Y$ with $Y_h$, creating $Y'$ as discussed in Eq.~\eqref{eq:simsepineq}.
The resulting absence of PSD structure (negative $Y'$-eigenvalues of $-0.037$ and $-0.024$) indicates that these particular separability observations are not simultaneously realizable.
Furthermore, this is consistent with the presence of non-zero negativity between halo$_A$ and the rest of the system $\mathcal{N}_{\rm{halo}_A|\rm{halo}_B,\rm{core}} = 0.022$.
Note that this incompatibility is also heralded by $\sigma_h$ alone, which is found to have negativity $\mathcal{N}_{A,B}(\sigma_h)= 0.087$.

\section{Numerical Tables}
In this section, we provide numerical values appearing in main text figures.
Table~\ref{tab:fig2tab} corresponds to Fig.~\ref{fig:corestructure} presenting the logarithmic negativity contributions organized by PT symplectic eigenvalues for a variety of region separations in the vacuum.
Tables~\ref{tab:fig3d10}-\ref{tab:fig3d20halo} correspond to Fig.~\ref{fig:boundent} presenting the region-region logarithmic negativity, $\mathcal{N}_{A|B}(\sigma)$, and the inaccessible logarithmic negativity between regions in the halo, $\mathcal{N}_{A|B}(\sigma_h)$ for $d= 10, 20$.  Further numerical values are available upon request.

\begin{table}[ht]
    \centering
    \begin{tabular}{c|c}
    \hline
    \hline
      $\tilde{r}$  & $\mathcal{N}_j$ \\
      \hline
      \hline
      \multirow{20}{*}{0}& $1.360$ \\
& $2.143\times 10^{-1}$ \\
& $2.775\times 10^{-2}$ \\
& $2.650\times 10^{-3}$ \\
& $2.117\times 10^{-4}$ \\
& $1.252\times 10^{-5}$ \\
& $6.810\times 10^{-7}$ \\
& $2.581\times 10^{-8}$ \\
& $9.840\times 10^{-10}$ \\
& $2.352\times 10^{-11}$ \\
& $6.284\times 10^{-13}$ \\
& $9.023\times 10^{-15}$ \\
& $1.653\times 10^{-16}$ \\
& $1.306\times 10^{-18}$ \\
& $1.560\times 10^{-20}$ \\
& $5.761\times 10^{-23}$ \\
& $4.035\times 10^{-25}$ \\
& $4.925\times 10^{-28}$ \\
& $1.530\times 10^{-30}$ \\
& $1.739\times 10^{-34}$ \\
\hline
\hline
\end{tabular}
\begin{tabular}{c|c}
    \hline
    \hline
      $\tilde{r}$  & $\mathcal{N}_j$ \\
      \hline
      \hline
\multirow{17}{*}{5}& $1.857\times 10^{-1}$ \\
& $5.312\times 10^{-3}$ \\
& $1.761\times 10^{-4}$ \\
& $4.880\times 10^{-6}$ \\
& $7.475\times 10^{-8}$ \\
& $4.128\times 10^{-10}$ \\
& $5.458\times 10^{-12}$ \\
& $6.857\times 10^{-14}$ \\
& $1.055\times 10^{-15}$ \\
& $8.574\times 10^{-18}$ \\
& $1.017\times 10^{-19}$ \\
& $4.554\times 10^{-22}$ \\
& $3.610\times 10^{-24}$ \\
& $7.300\times 10^{-27}$ \\
& $3.207\times 10^{-29}$ \\
& $1.831\times 10^{-32}$ \\
& $2.765\times 10^{-35}$ \\
\hline
\hline
\end{tabular}
\begin{tabular}{c|c}
    \hline
    \hline
      $\tilde{r}$  & $\mathcal{N}_j$ \\
      \hline
      \hline
\multirow{15}{*}{10}& $9.067\times 10^{-2}$ \\
& $1.204\times 10^{-3}$ \\
& $4.939\times 10^{-6}$ \\
& $4.399\times 10^{-8}$ \\
& $3.939\times 10^{-10}$ \\
& $5.584\times 10^{-13}$ \\
& $3.079\times 10^{-15}$ \\
& $1.580\times 10^{-17}$ \\
& $1.220\times 10^{-19}$ \\
& $4.124\times 10^{-22}$ \\
& $7.510\times 10^{-25}$ \\
& $6.752\times 10^{-29}$ \\
& $3.924\times 10^{-32}$ \\
& $8.617\times 10^{-36}$ \\
& $4.019\times 10^{-39}$ \\
\hline
\hline
\end{tabular}
\begin{tabular}{c|c}
    \hline
    \hline
      $\tilde{r}$  & $\mathcal{N}_j$ \\
      \hline
      \hline
\multirow{12}{*}{20}& $3.295\times 10^{-2}$ \\
& $3.902\times 10^{-5}$ \\
& $2.799\times 10^{-8}$ \\
& $4.473\times 10^{-11}$ \\
& $1.319\times 10^{-14}$ \\
& $1.971\times 10^{-17}$ \\
& $1.976\times 10^{-21}$ \\
& $1.412\times 10^{-24}$ \\
& $1.524\times 10^{-27}$ \\
& $1.283\times 10^{-32}$ \\
& $3.177\times 10^{-36}$ \\
& $7.773\times 10^{-41}$ \\
\hline
\hline
\end{tabular}
\begin{tabular}{c|c}
    \hline
    \hline
      $\tilde{r}$  & $\mathcal{N}_j$ \\
      \hline
      \hline
\multirow{8}{*}{40}& $4.314\times 10^{-3}$ \\
& $1.538\times 10^{-7}$ \\
& $5.242\times 10^{-12}$ \\
& $9.355\times 10^{-17}$ \\
& $1.350\times 10^{-21}$ \\
& $3.085\times 10^{-27}$ \\
& $5.506\times 10^{-33}$ \\
& $7.941\times 10^{-38}$ \\
\hline
\multirow{7}{*}{60}& $6.551\times 10^{-4}$ \\
& $6.670\times 10^{-10}$ \\
& $2.723\times 10^{-16}$ \\
& $1.264\times 10^{-22}$ \\
& $7.052\times 10^{-29}$ \\
& $3.273\times 10^{-36}$ \\
& $1.524\times 10^{-44}$ \\
\hline
\hline
\end{tabular}
\begin{tabular}{c|c}
    \hline
    \hline
      $\tilde{r}$  & $\mathcal{N}_j$ \\
      \hline
      \hline
\multirow{4}{*}{100}& $1.189\times 10^{-5}$ \\
& $1.196\times 10^{-14}$ \\
& $1.242\times 10^{-23}$ \\
& $4.019\times 10^{-34}$ \\
\hline
\multirow{2}{*}{200}& $4.092\times 10^{-10}$ \\
& $3.596\times 10^{-26}$ \\
\hline
\multirow{1}{*}{400}& $8.640\times 10^{-20}$ \\
\hline
    \end{tabular}
    \caption{Numerical values of the logarithmic negativity contributions from PT symplectic eigenvalues at region separation, $\tilde{r}$, presented in Fig.~\ref{fig:corestructure}.}
    \label{tab:fig2tab}
\end{table}

\begin{table}
  \centering
  \begin{tabular}{ccc}
  \multicolumn{3}{c}{d = 10}\\
  \hline
  \hline
  $\tilde{r}$ & $\mathcal{N}_{A|B}$ & $\mathcal{N}(\sigma_h)_{A|B}$ \\
  \hline
  \hline
0& $1.214$ & $3.575\times 10^{-1}$\\
1& $3.915\times 10^{-1}$ & $3.659\times 10^{-1}$\\
2& $1.907\times 10^{-1}$ & $3.867\times 10^{-1}$\\
3& $1.079\times 10^{-1}$ & $4.306\times 10^{-1}$\\
4& $6.725\times 10^{-2}$ & $4.773\times 10^{-1}$\\
5& $4.597\times 10^{-2}$ & $1.267\times 10^{-1}$\\
6& $3.353\times 10^{-2}$ & $1.406\times 10^{-1}$\\
7& $2.546\times 10^{-2}$ & $1.036\times 10^{-1}$\\
8& $1.981\times 10^{-2}$ & $3.465\times 10^{-2}$\\
9& $1.563\times 10^{-2}$ & $8.211\times 10^{-3}$\\
10& $1.243\times 10^{-2}$ & $3.010\times 10^{-3}$\\
11& $9.888\times 10^{-3}$ & $1.910\times 10^{-3}$\\
12& $7.836\times 10^{-3}$ & $1.381\times 10^{-3}$\\
13& $6.154\times 10^{-3}$ & $1.111\times 10^{-3}$\\
14& $4.765\times 10^{-3}$ & $9.363\times 10^{-4}$\\
15& $3.618\times 10^{-3}$ & $8.215\times 10^{-4}$\\
16& $2.680\times 10^{-3}$ & $7.479\times 10^{-4}$\\
17& $1.926\times 10^{-3}$ & $7.086\times 10^{-4}$\\
18& $1.339\times 10^{-3}$ & $5.565\times 10^{-4}$\\
19& $9.043\times 10^{-4}$ & $5.930\times 10^{-4}$\\
20& $6.022\times 10^{-4}$ & $6.732\times 10^{-4}$\\
21& $4.060\times 10^{-4}$ & $7.785\times 10^{-4}$\\
22& $2.835\times 10^{-4}$ & $8.734\times 10^{-4}$\\
23& $2.064\times 10^{-4}$ & $9.132\times 10^{-4}$\\
24& $1.560\times 10^{-4}$ & $8.631\times 10^{-4}$\\
25& $1.212\times 10^{-4}$ & $7.098\times 10^{-4}$\\
26& $9.621\times 10^{-5}$ & $4.701\times 10^{-4}$\\
27& $7.744\times 10^{-5}$ & $1.997\times 10^{-4}$\\
28& $6.289\times 10^{-5}$ & $3.456\times 10^{-5}$\\
29& $5.132\times 10^{-5}$ & $1.090\times 10^{-5}$\\
30& $4.192\times 10^{-5}$ & $4.814\times 10^{-6}$\\
31& $3.414\times 10^{-5}$ & $3.173\times 10^{-6}$\\
\end{tabular}
\begin{tabular}{ccc}
32& $2.762\times 10^{-5}$ & $2.343\times 10^{-6}$\\
33& $2.211\times 10^{-5}$ & $1.898\times 10^{-6}$\\
34& $1.743\times 10^{-5}$ & $1.663\times 10^{-6}$\\
35& $1.345\times 10^{-5}$ & $1.553\times 10^{-6}$\\
36& $1.010\times 10^{-5}$ & $1.512\times 10^{-6}$\\
37& $7.342\times 10^{-6}$ & $1.491\times 10^{-6}$\\
38& $5.136\times 10^{-6}$ & $1.437\times 10^{-6}$\\
39& $3.454\times 10^{-6}$ & $1.328\times 10^{-6}$\\
40& $2.237\times 10^{-6}$ & $1.216\times 10^{-6}$\\
41& $1.405\times 10^{-6}$ & $1.194\times 10^{-6}$\\
42& $8.728\times 10^{-7}$ & $1.323\times 10^{-6}$\\
43& $5.554\times 10^{-7}$ & $1.604\times 10^{-6}$\\
44& $3.730\times 10^{-7}$ & $1.977\times 10^{-6}$\\
45& $2.655\times 10^{-7}$ & $2.338\times 10^{-6}$\\
46& $1.985\times 10^{-7}$ & $2.576\times 10^{-6}$\\
47& $1.538\times 10^{-7}$ & $2.609\times 10^{-6}$\\
48& $1.223\times 10^{-7}$ & $2.396\times 10^{-6}$\\
49& $9.902\times 10^{-8}$ & $1.946\times 10^{-6}$\\
50& $8.124\times 10^{-8}$ & $1.320\times 10^{-6}$\\
51& $6.723\times 10^{-8}$ & $6.198\times 10^{-7}$\\
52& $5.591\times 10^{-8}$ & $4.730\times 10^{-8}$\\
53& $4.659\times 10^{-8}$ & $7.074\times 10^{-9}$\\
54& $3.878\times 10^{-8}$ & $3.502\times 10^{-9}$\\
55& $3.216\times 10^{-8}$ & $2.261\times 10^{-9}$\\
56& $2.648\times 10^{-8}$ & $1.668\times 10^{-9}$\\
57& $2.157\times 10^{-8}$ & $1.354\times 10^{-9}$\\
58& $1.729\times 10^{-8}$ & $1.192\times 10^{-9}$\\
59& $1.357\times 10^{-8}$ & $1.133\times 10^{-9}$\\
60& $1.033\times 10^{-8}$ & $1.155\times 10^{-9}$\\
61& $7.568\times 10^{-9}$ & $1.244\times 10^{-9}$\\
62& $5.273\times 10^{-9}$ & $1.374\times 10^{-9}$\\
63& $3.479\times 10^{-9}$ & $1.452\times 10^{-9}$\\
64& $2.187\times 10^{-9}$ & $1.280\times 10^{-9}$\\
\end{tabular}
\begin{tabular}{ccc}
65& $1.326\times 10^{-9}$ & $7.757\times 10^{-10}$\\
66& $7.760\times 10^{-10}$ & $4.197\times 10^{-10}$\\
67& $4.345\times 10^{-10}$ & $3.444\times 10^{-10}$\\
68& $2.362\times 10^{-10}$ & $4.111\times 10^{-10}$\\
69& $1.345\times 10^{-10}$ & $6.025\times 10^{-10}$\\
70& $8.489\times 10^{-11}$ & $9.183\times 10^{-10}$\\
71& $5.886\times 10^{-11}$ & $1.350\times 10^{-9}$\\
72& $4.359\times 10^{-11}$ & $1.875\times 10^{-9}$\\
73& $3.374\times 10^{-11}$ & $2.442\times 10^{-9}$\\
74& $2.692\times 10^{-11}$ & $2.971\times 10^{-9}$\\
75& $2.194\times 10^{-11}$ & $3.374\times 10^{-9}$\\
76& $1.816\times 10^{-11}$ & $3.573\times 10^{-9}$\\
77& $1.518\times 10^{-11}$ & $3.509\times 10^{-9}$\\
78& $1.278\times 10^{-11}$ & $3.161\times 10^{-9}$\\
79& $1.081\times 10^{-11}$ & $2.551\times 10^{-9}$\\
80& $9.162\times 10^{-12}$ & $1.771\times 10^{-9}$\\
81& $7.760\times 10^{-12}$ & $9.780\times 10^{-10}$\\
82& $6.555\times 10^{-12}$ & $3.280\times 10^{-10}$\\
83& $5.508\times 10^{-12}$ & $2.552\times 10^{-12}$\\
84& $4.590\times 10^{-12}$ & $6.526\times 10^{-13}$\\
85& $3.779\times 10^{-12}$ & $3.864\times 10^{-13}$\\
86& $3.059\times 10^{-12}$ & $2.865\times 10^{-13}$\\
87& $2.417\times 10^{-12}$ & $2.403\times 10^{-13}$\\
88& $1.843\times 10^{-12}$ & $2.215\times 10^{-13}$\\
89& $1.333\times 10^{-12}$ & $2.240\times 10^{-13}$\\
90& $8.883\times 10^{-13}$ & $2.531\times 10^{-13}$\\
91& $5.245\times 10^{-13}$ & $3.309\times 10^{-13}$\\
92& $2.731\times 10^{-13}$ & $5.014\times 10^{-13}$\\
93& $1.382\times 10^{-13}$ & $7.867\times 10^{-13}$\\
94& $7.269\times 10^{-14}$ & $1.152\times 10^{-12}$\\
95& $3.805\times 10^{-14}$ & $1.562\times 10^{-12}$\\
96& $1.747\times 10^{-14}$ & $2.002\times 10^{-12}$\\
97& $4.040\times 10^{-15}$ & $2.465\times 10^{-12}$\\
\hline
\hline
  \end{tabular}
  \caption{Numerical values of region-region logarithmic negativity $\mathcal{N}_{A|B}(\sigma)$ and the inaccessible negativity in the halo $\mathcal{N}_{A|B}(\sigma_h)$ for $m = 3\times 10^{-3}/d$ presented in Fig.~\ref{fig:boundent}.}
  \label{tab:fig3d10}
\end{table}

\begin{table}
  \centering
  \tiny
  \begin{tabular}{cc}
  \multicolumn{2}{c}{d = 20}\\
  \hline
  \hline
  $\tilde{r}$ & $\mathcal{N}_{A|B} (\sigma)$ \\
  \hline
  \hline
0& $1.463$\\
1& $5.438\times 10^{-1}$\\
2& $3.187\times 10^{-1}$\\
3& $2.209\times 10^{-1}$\\
4& $1.648\times 10^{-1}$\\
5& $1.283\times 10^{-1}$\\
6& $1.032\times 10^{-1}$\\
7& $8.490\times 10^{-2}$\\
8& $7.111\times 10^{-2}$\\
9& $6.034\times 10^{-2}$\\
10& $5.171\times 10^{-2}$\\
11& $4.466\times 10^{-2}$\\
12& $3.881\times 10^{-2}$\\
13& $3.388\times 10^{-2}$\\
14& $2.966\times 10^{-2}$\\
15& $2.602\times 10^{-2}$\\
16& $2.285\times 10^{-2}$\\
17& $2.006\times 10^{-2}$\\
18& $1.759\times 10^{-2}$\\
19& $1.541\times 10^{-2}$\\
20& $1.346\times 10^{-2}$\\
21& $1.173\times 10^{-2}$\\
22& $1.018\times 10^{-2}$\\
23& $8.805\times 10^{-3}$\\
24& $7.579\times 10^{-3}$\\
25& $6.493\times 10^{-3}$\\
26& $5.536\times 10^{-3}$\\
27& $4.700\times 10^{-3}$\\
28& $3.975\times 10^{-3}$\\
29& $3.353\times 10^{-3}$\\
30& $2.826\times 10^{-3}$\\
31& $2.383\times 10^{-3}$\\
32& $2.014\times 10^{-3}$\\
33& $1.710\times 10^{-3}$\\
34& $1.458\times 10^{-3}$\\
35& $1.251\times 10^{-3}$\\
36& $1.079\times 10^{-3}$\\
37& $9.367\times 10^{-4}$\\
38& $8.170\times 10^{-4}$\\
39& $7.160\times 10^{-4}$\\
40& $6.299\times 10^{-4}$\\
41& $5.561\times 10^{-4}$\\
42& $4.922\times 10^{-4}$\\
43& $4.366\times 10^{-4}$\\
44& $3.878\times 10^{-4}$\\
45& $3.447\times 10^{-4}$\\
46& $3.065\times 10^{-4}$\\
47& $2.725\times 10^{-4}$\\
48& $2.420\times 10^{-4}$\\
49& $2.147\times 10^{-4}$\\
50& $1.901\times 10^{-4}$\\
51& $1.678\times 10^{-4}$\\
52& $1.477\times 10^{-4}$\\
53& $1.296\times 10^{-4}$\\
54& $1.131\times 10^{-4}$\\
55& $9.833\times 10^{-5}$\\
56& $8.501\times 10^{-5}$\\
57& $7.308\times 10^{-5}$\\
58& $6.245\times 10^{-5}$\\
59& $5.305\times 10^{-5}$\\
\end{tabular}
\begin{tabular}{cc}
60& $4.480\times 10^{-5}$\\
61& $3.764\times 10^{-5}$\\
62& $3.149\times 10^{-5}$\\
63& $2.628\times 10^{-5}$\\
64& $2.191\times 10^{-5}$\\
65& $1.828\times 10^{-5}$\\
66& $1.530\times 10^{-5}$\\
67& $1.288\times 10^{-5}$\\
68& $1.090\times 10^{-5}$\\
69& $9.288\times 10^{-6}$\\
70& $7.972\times 10^{-6}$\\
71& $6.889\times 10^{-6}$\\
72& $5.992\times 10^{-6}$\\
73& $5.242\times 10^{-6}$\\
74& $4.610\times 10^{-6}$\\
75& $4.072\times 10^{-6}$\\
76& $3.610\times 10^{-6}$\\
77& $3.211\times 10^{-6}$\\
78& $2.863\times 10^{-6}$\\
79& $2.557\times 10^{-6}$\\
80& $2.288\times 10^{-6}$\\
81& $2.048\times 10^{-6}$\\
82& $1.834\times 10^{-6}$\\
83& $1.642\times 10^{-6}$\\
84& $1.469\times 10^{-6}$\\
85& $1.313\times 10^{-6}$\\
86& $1.171\times 10^{-6}$\\
87& $1.043\times 10^{-6}$\\
88& $9.253\times 10^{-7}$\\
89& $8.185\times 10^{-7}$\\
90& $7.210\times 10^{-7}$\\
91& $6.323\times 10^{-7}$\\
92& $5.515\times 10^{-7}$\\
93& $4.783\times 10^{-7}$\\
94& $4.121\times 10^{-7}$\\
95& $3.526\times 10^{-7}$\\
96& $2.996\times 10^{-7}$\\
97& $2.527\times 10^{-7}$\\
98& $2.116\times 10^{-7}$\\
99& $1.761\times 10^{-7}$\\
100& $1.457\times 10^{-7}$\\
101& $1.201\times 10^{-7}$\\
102& $9.870\times 10^{-8}$\\
103& $8.113\times 10^{-8}$\\
104& $6.683\times 10^{-8}$\\
105& $5.531\times 10^{-8}$\\
106& $4.608\times 10^{-8}$\\
107& $3.869\times 10^{-8}$\\
108& $3.277\times 10^{-8}$\\
109& $2.799\times 10^{-8}$\\
110& $2.411\times 10^{-8}$\\
111& $2.093\times 10^{-8}$\\
112& $1.829\times 10^{-8}$\\
113& $1.609\times 10^{-8}$\\
114& $1.422\times 10^{-8}$\\
115& $1.263\times 10^{-8}$\\
116& $1.126\times 10^{-8}$\\
117& $1.007\times 10^{-8}$\\
118& $9.026\times 10^{-9}$\\
119& $8.109\times 10^{-9}$\\
\end{tabular}
\begin{tabular}{cc}
120& $7.296\times 10^{-9}$\\
121& $6.571\times 10^{-9}$\\
122& $5.921\times 10^{-9}$\\
123& $5.337\times 10^{-9}$\\
124& $4.808\times 10^{-9}$\\
125& $4.328\times 10^{-9}$\\
126& $3.891\times 10^{-9}$\\
127& $3.491\times 10^{-9}$\\
128& $3.126\times 10^{-9}$\\
129& $2.791\times 10^{-9}$\\
130& $2.483\times 10^{-9}$\\
131& $2.201\times 10^{-9}$\\
132& $1.941\times 10^{-9}$\\
133& $1.703\times 10^{-9}$\\
134& $1.484\times 10^{-9}$\\
135& $1.285\times 10^{-9}$\\
136& $1.104\times 10^{-9}$\\
137& $9.414\times 10^{-10}$\\
138& $7.956\times 10^{-10}$\\
139& $6.665\times 10^{-10}$\\
140& $5.537\times 10^{-10}$\\
141& $4.563\times 10^{-10}$\\
142& $3.734\times 10^{-10}$\\
143& $3.038\times 10^{-10}$\\
144& $2.461\times 10^{-10}$\\
145& $1.990\times 10^{-10}$\\
146& $1.608\times 10^{-10}$\\
147& $1.304\times 10^{-10}$\\
148& $1.063\times 10^{-10}$\\
149& $8.742\times 10^{-11}$\\
150& $7.258\times 10^{-11}$\\
151& $6.090\times 10^{-11}$\\
152& $5.164\times 10^{-11}$\\
153& $4.424\times 10^{-11}$\\
154& $3.825\times 10^{-11}$\\
155& $3.334\times 10^{-11}$\\
156& $2.927\times 10^{-11}$\\
157& $2.586\times 10^{-11}$\\
158& $2.297\times 10^{-11}$\\
159& $2.050\times 10^{-11}$\\
160& $1.836\times 10^{-11}$\\
161& $1.650\times 10^{-11}$\\
162& $1.487\times 10^{-11}$\\
163& $1.342\times 10^{-11}$\\
164& $1.214\times 10^{-11}$\\
165& $1.099\times 10^{-11}$\\
166& $9.959\times 10^{-12}$\\
167& $9.027\times 10^{-12}$\\
168& $8.181\times 10^{-12}$\\
169& $7.411\times 10^{-12}$\\
170& $6.707\times 10^{-12}$\\
171& $6.062\times 10^{-12}$\\
172& $5.470\times 10^{-12}$\\
173& $4.924\times 10^{-12}$\\
174& $4.420\times 10^{-12}$\\
175& $3.954\times 10^{-12}$\\
176& $3.524\times 10^{-12}$\\
177& $3.125\times 10^{-12}$\\
178& $2.756\times 10^{-12}$\\
179& $2.416\times 10^{-12}$\\
\end{tabular}
\begin{tabular}{cc}
180& $2.103\times 10^{-12}$\\
181& $1.815\times 10^{-12}$\\
182& $1.553\times 10^{-12}$\\
183& $1.315\times 10^{-12}$\\
184& $1.103\times 10^{-12}$\\
185& $9.152\times 10^{-13}$\\
186& $7.518\times 10^{-13}$\\
187& $6.119\times 10^{-13}$\\
188& $4.940\times 10^{-13}$\\
189& $3.961\times 10^{-13}$\\
190& $3.159\times 10^{-13}$\\
191& $2.509\times 10^{-13}$\\
192& $1.988\times 10^{-13}$\\
193& $1.576\times 10^{-13}$\\
194& $1.254\times 10^{-13}$\\
195& $1.005\times 10^{-13}$\\
196& $8.138\times 10^{-14}$\\
197& $6.672\times 10^{-14}$\\
198& $5.543\times 10^{-14}$\\
199& $4.665\times 10^{-14}$\\
200& $3.975\times 10^{-14}$\\
201& $3.424\times 10^{-14}$\\
202& $2.978\times 10^{-14}$\\
203& $2.611\times 10^{-14}$\\
204& $2.306\times 10^{-14}$\\
205& $2.049\times 10^{-14}$\\
206& $1.830\times 10^{-14}$\\
207& $1.642\times 10^{-14}$\\
208& $1.479\times 10^{-14}$\\
209& $1.335\times 10^{-14}$\\
210& $1.209\times 10^{-14}$\\
211& $1.097\times 10^{-14}$\\
212& $9.972\times 10^{-15}$\\
213& $9.074\times 10^{-15}$\\
214& $8.263\times 10^{-15}$\\
215& $7.527\times 10^{-15}$\\
216& $6.858\times 10^{-15}$\\
217& $6.246\times 10^{-15}$\\
218& $5.685\times 10^{-15}$\\
219& $5.169\times 10^{-15}$\\
220& $4.693\times 10^{-15}$\\
221& $4.252\times 10^{-15}$\\
222& $3.844\times 10^{-15}$\\
223& $3.466\times 10^{-15}$\\
224& $3.114\times 10^{-15}$\\
225& $2.786\times 10^{-15}$\\
226& $2.481\times 10^{-15}$\\
227& $2.197\times 10^{-15}$\\
228& $1.932\times 10^{-15}$\\
229& $1.687\times 10^{-15}$\\
230& $1.460\times 10^{-15}$\\
231& $1.250\times 10^{-15}$\\
232& $1.059\times 10^{-15}$\\
233& $8.865\times 10^{-16}$\\
234& $7.324\times 10^{-16}$\\
235& $5.974\times 10^{-16}$\\
236& $4.816\times 10^{-16}$\\
237& $3.843\times 10^{-16}$\\
238& $3.042\times 10^{-16}$\\
239& $2.392\times 10^{-16}$\\
\end{tabular}
\begin{tabular}{cc}
240& $1.870\times 10^{-16}$\\
241& $1.456\times 10^{-16}$\\
242& $1.129\times 10^{-16}$\\
243& $8.731\times 10^{-17}$\\
244& $6.756\times 10^{-17}$\\
245& $5.253\times 10^{-17}$\\
246& $4.126\times 10^{-17}$\\
247& $3.288\times 10^{-17}$\\
248& $2.664\times 10^{-17}$\\
249& $2.197\times 10^{-17}$\\
250& $1.841\times 10^{-17}$\\
251& $1.565\times 10^{-17}$\\
252& $1.347\times 10^{-17}$\\
253& $1.172\times 10^{-17}$\\
254& $1.029\times 10^{-17}$\\
255& $9.099\times 10^{-18}$\\
256& $8.101\times 10^{-18}$\\
257& $7.253\times 10^{-18}$\\
258& $6.523\times 10^{-18}$\\
259& $5.890\times 10^{-18}$\\
260& $5.336\times 10^{-18}$\\
261& $4.847\times 10^{-18}$\\
262& $4.413\times 10^{-18}$\\
263& $4.025\times 10^{-18}$\\
264& $3.676\times 10^{-18}$\\
265& $3.361\times 10^{-18}$\\
266& $3.075\times 10^{-18}$\\
267& $2.814\times 10^{-18}$\\
268& $2.576\times 10^{-18}$\\
269& $2.357\times 10^{-18}$\\
270& $2.155\times 10^{-18}$\\
271& $1.969\times 10^{-18}$\\
272& $1.797\times 10^{-18}$\\
273& $1.637\times 10^{-18}$\\
274& $1.488\times 10^{-18}$\\
275& $1.349\times 10^{-18}$\\
276& $1.219\times 10^{-18}$\\
277& $1.097\times 10^{-18}$\\
278& $9.839\times 10^{-19}$\\
279& $8.774\times 10^{-19}$\\
280& $7.777\times 10^{-19}$\\
281& $6.844\times 10^{-19}$\\
282& $5.971\times 10^{-19}$\\
283& $5.157\times 10^{-19}$\\
284& $4.402\times 10^{-19}$\\
285& $3.706\times 10^{-19}$\\
286& $3.073\times 10^{-19}$\\
287& $2.506\times 10^{-19}$\\
288& $2.009\times 10^{-19}$\\
289& $1.585\times 10^{-19}$\\
290& $1.235\times 10^{-19}$\\
291& $9.538\times 10^{-20}$\\
292& $7.328\times 10^{-20}$\\
293& $5.610\times 10^{-20}$\\
294& $4.280\times 10^{-20}$\\
295& $3.250\times 10^{-20}$\\
296& $2.451\times 10^{-20}$\\
297& $1.835\times 10^{-20}$\\
298& $1.367\times 10^{-20}$\\
299& $1.019\times 10^{-20}$\\
\end{tabular}
\begin{tabular}{cc}
300& $7.682\times 10^{-21}$\\
301& $5.902\times 10^{-21}$\\
302& $4.645\times 10^{-21}$\\
303& $3.746\times 10^{-21}$\\
304& $3.088\times 10^{-21}$\\
305& $2.595\times 10^{-21}$\\
306& $2.215\times 10^{-21}$\\
307& $1.916\times 10^{-21}$\\
308& $1.675\times 10^{-21}$\\
309& $1.478\times 10^{-21}$\\
310& $1.314\times 10^{-21}$\\
311& $1.175\times 10^{-21}$\\
312& $1.057\times 10^{-21}$\\
313& $9.546\times 10^{-22}$\\
314& $8.656\times 10^{-22}$\\
315& $7.873\times 10^{-22}$\\
316& $7.181\times 10^{-22}$\\
317& $6.563\times 10^{-22}$\\
318& $6.009\times 10^{-22}$\\
319& $5.510\times 10^{-22}$\\
320& $5.058\times 10^{-22}$\\
321& $4.646\times 10^{-22}$\\
322& $4.270\times 10^{-22}$\\
323& $3.925\times 10^{-22}$\\
324& $3.607\times 10^{-22}$\\
325& $3.314\times 10^{-22}$\\
326& $3.042\times 10^{-22}$\\
327& $2.790\times 10^{-22}$\\
328& $2.555\times 10^{-22}$\\
329& $2.336\times 10^{-22}$\\
330& $2.132\times 10^{-22}$\\
331& $1.940\times 10^{-22}$\\
332& $1.760\times 10^{-22}$\\
333& $1.591\times 10^{-22}$\\
334& $1.432\times 10^{-22}$\\
335& $1.282\times 10^{-22}$\\
336& $1.141\times 10^{-22}$\\
337& $1.008\times 10^{-22}$\\
338& $8.823\times 10^{-23}$\\
339& $7.644\times 10^{-23}$\\
340& $6.536\times 10^{-23}$\\
341& $5.503\times 10^{-23}$\\
342& $4.545\times 10^{-23}$\\
343& $3.670\times 10^{-23}$\\
344& $2.889\times 10^{-23}$\\
345& $2.214\times 10^{-23}$\\
346& $1.659\times 10^{-23}$\\
347& $1.225\times 10^{-23}$\\
348& $9.012\times 10^{-24}$\\
349& $6.647\times 10^{-24}$\\
350& $4.925\times 10^{-24}$\\
351& $3.656\times 10^{-24}$\\
352& $2.705\times 10^{-24}$\\
353& $1.980\times 10^{-24}$\\
354& $1.424\times 10^{-24}$\\
355& $1.000\times 10^{-24}$\\
356& $6.867\times 10^{-25}$\\
357& $4.688\times 10^{-25}$\\
358& $3.272\times 10^{-25}$\\
359& $2.381\times 10^{-25}$\\
\end{tabular}
\begin{tabular}{cc}
360& $1.813\times 10^{-25}$\\
361& $1.434\times 10^{-25}$\\
362& $1.171\times 10^{-25}$\\
363& $9.786\times 10^{-26}$\\
364& $8.336\times 10^{-26}$\\
365& $7.207\times 10^{-26}$\\
366& $6.305\times 10^{-26}$\\
367& $5.570\times 10^{-26}$\\
368& $4.960\times 10^{-26}$\\
369& $4.446\times 10^{-26}$\\
370& $4.007\times 10^{-26}$\\
371& $3.628\times 10^{-26}$\\
372& $3.297\times 10^{-26}$\\
373& $3.007\times 10^{-26}$\\
374& $2.750\times 10^{-26}$\\
375& $2.520\times 10^{-26}$\\
376& $2.314\times 10^{-26}$\\
377& $2.128\times 10^{-26}$\\
378& $1.960\times 10^{-26}$\\
379& $1.806\times 10^{-26}$\\
380& $1.666\times 10^{-26}$\\
381& $1.537\times 10^{-26}$\\
382& $1.418\times 10^{-26}$\\
383& $1.308\times 10^{-26}$\\
384& $1.207\times 10^{-26}$\\
385& $1.112\times 10^{-26}$\\
386& $1.024\times 10^{-26}$\\
387& $9.416\times 10^{-27}$\\
388& $8.645\times 10^{-27}$\\
389& $7.922\times 10^{-27}$\\
390& $7.243\times 10^{-27}$\\
391& $6.603\times 10^{-27}$\\
392& $5.999\times 10^{-27}$\\
393& $5.430\times 10^{-27}$\\
394& $4.891\times 10^{-27}$\\
395& $4.380\times 10^{-27}$\\
396& $3.897\times 10^{-27}$\\
397& $3.438\times 10^{-27}$\\
398& $3.002\times 10^{-27}$\\
399& $2.589\times 10^{-27}$\\
400& $2.196\times 10^{-27}$\\
401& $1.825\times 10^{-27}$\\
402& $1.475\times 10^{-27}$\\
403& $1.148\times 10^{-27}$\\
404& $8.493\times 10^{-28}$\\
405& $5.878\times 10^{-28}$\\
406& $3.797\times 10^{-28}$\\
407& $2.373\times 10^{-28}$\\
408& $1.516\times 10^{-28}$\\
409& $1.014\times 10^{-28}$\\
410& $7.052\times 10^{-29}$\\
411& $5.024\times 10^{-29}$\\
412& $3.610\times 10^{-29}$\\
413& $2.575\times 10^{-29}$\\
414& $1.789\times 10^{-29}$\\
415& $1.172\times 10^{-29}$\\
416& $6.766\times 10^{-30}$\\
417& $2.700\times 10^{-30}$\\
\hline
\hline
  \end{tabular}
  \caption{Numerical values of region-region logarithmic negativity $\mathcal{N}_{A|B}(\sigma)$  for $m = 3\times 10^{-3}/d$ presented in Fig.~\ref{fig:boundent}.}
  \label{tab:fig3d20full}
\end{table}

\begin{table}
  \centering
  \tiny
  \begin{tabular}{cc}
  \multicolumn{2}{c}{d = 20}\\
  \hline
  \hline
  $\tilde{r}$ & $\mathcal{N}(\sigma_h)_{A|B}$ \\
  \hline
  \hline
  0& $4.818\times 10^{-1}$\\
1& $4.862\times 10^{-1}$\\
2& $5.027\times 10^{-1}$\\
3& $5.444\times 10^{-1}$\\
4& $5.852\times 10^{-1}$\\
5& $6.224\times 10^{-1}$\\
6& $6.562\times 10^{-1}$\\
7& $1.953\times 10^{-1}$\\
8& $1.733\times 10^{-1}$\\
9& $1.522\times 10^{-1}$\\
10& $1.250\times 10^{-1}$\\
11& $4.682\times 10^{-2}$\\
12& $2.720\times 10^{-2}$\\
13& $1.697\times 10^{-2}$\\
14& $1.144\times 10^{-2}$\\
15& $8.260\times 10^{-3}$\\
16& $6.366\times 10^{-3}$\\
17& $5.167\times 10^{-3}$\\
18& $4.262\times 10^{-3}$\\
19& $3.578\times 10^{-3}$\\
20& $3.097\times 10^{-3}$\\
21& $2.783\times 10^{-3}$\\
22& $2.541\times 10^{-3}$\\
23& $2.414\times 10^{-3}$\\
24& $2.339\times 10^{-3}$\\
25& $2.303\times 10^{-3}$\\
26& $2.315\times 10^{-3}$\\
27& $2.356\times 10^{-3}$\\
28& $2.416\times 10^{-3}$\\
29& $2.488\times 10^{-3}$\\
30& $2.569\times 10^{-3}$\\
31& $2.649\times 10^{-3}$\\
32& $2.720\times 10^{-3}$\\
33& $2.768\times 10^{-3}$\\
34& $2.782\times 10^{-3}$\\
35& $2.746\times 10^{-3}$\\
36& $2.631\times 10^{-3}$\\
37& $2.483\times 10^{-3}$\\
38& $2.255\times 10^{-3}$\\
39& $1.947\times 10^{-3}$\\
40& $1.571\times 10^{-3}$\\
41& $1.152\times 10^{-3}$\\
42& $7.358\times 10^{-4}$\\
43& $4.036\times 10^{-4}$\\
44& $2.179\times 10^{-4}$\\
45& $1.333\times 10^{-4}$\\
46& $9.177\times 10^{-5}$\\
47& $6.856\times 10^{-5}$\\
48& $5.454\times 10^{-5}$\\
49& $4.523\times 10^{-5}$\\
50& $3.871\times 10^{-5}$\\
51& $3.401\times 10^{-5}$\\
52& $3.056\times 10^{-5}$\\
53& $2.802\times 10^{-5}$\\
54& $2.615\times 10^{-5}$\\
55& $2.474\times 10^{-5}$\\
56& $2.378\times 10^{-5}$\\
57& $2.313\times 10^{-5}$\\
58& $2.274\times 10^{-5}$\\
59& $2.255\times 10^{-5}$\\
\end{tabular}
\begin{tabular}{cc}
60& $2.262\times 10^{-5}$\\
61& $2.290\times 10^{-5}$\\
62& $2.334\times 10^{-5}$\\
63& $2.395\times 10^{-5}$\\
64& $2.470\times 10^{-5}$\\
65& $2.555\times 10^{-5}$\\
66& $2.644\times 10^{-5}$\\
67& $2.726\times 10^{-5}$\\
68& $2.790\times 10^{-5}$\\
69& $2.823\times 10^{-5}$\\
70& $2.815\times 10^{-5}$\\
71& $2.759\times 10^{-5}$\\
72& $2.640\times 10^{-5}$\\
73& $2.463\times 10^{-5}$\\
74& $2.229\times 10^{-5}$\\
75& $1.937\times 10^{-5}$\\
76& $1.598\times 10^{-5}$\\
77& $1.225\times 10^{-5}$\\
78& $8.371\times 10^{-6}$\\
79& $4.698\times 10^{-6}$\\
80& $2.065\times 10^{-6}$\\
81& $9.659\times 10^{-7}$\\
82& $5.701\times 10^{-7}$\\
83& $3.910\times 10^{-7}$\\
84& $2.933\times 10^{-7}$\\
85& $1.883\times 10^{-7}$\\
86& $1.594\times 10^{-7}$\\
87& $1.392\times 10^{-7}$\\
88& $1.249\times 10^{-7}$\\
89& $1.148\times 10^{-7}$\\
90& $1.075\times 10^{-7}$\\
91& $1.025\times 10^{-7}$\\
92& $9.904\times 10^{-8}$\\
93& $9.684\times 10^{-8}$\\
94& $9.553\times 10^{-8}$\\
95& $9.476\times 10^{-8}$\\
96& $9.426\times 10^{-8}$\\
97& $9.390\times 10^{-8}$\\
98& $9.382\times 10^{-8}$\\
99& $9.411\times 10^{-8}$\\
100& $9.479\times 10^{-8}$\\
101& $9.599\times 10^{-8}$\\
102& $9.790\times 10^{-8}$\\
103& $1.006\times 10^{-7}$\\
104& $1.042\times 10^{-7}$\\
105& $1.084\times 10^{-7}$\\
106& $1.129\times 10^{-7}$\\
107& $1.171\times 10^{-7}$\\
108& $1.205\times 10^{-7}$\\
109& $1.226\times 10^{-7}$\\
110& $1.227\times 10^{-7}$\\
111& $1.206\times 10^{-7}$\\
112& $1.159\times 10^{-7}$\\
113& $1.085\times 10^{-7}$\\
114& $9.860\times 10^{-8}$\\
115& $8.632\times 10^{-8}$\\
116& $7.209\times 10^{-8}$\\
117& $5.646\times 10^{-8}$\\
118& $4.024\times 10^{-8}$\\
119& $2.433\times 10^{-8}$\\
\end{tabular}
\begin{tabular}{cc}
120& $1.049\times 10^{-8}$\\
121& $3.449\times 10^{-9}$\\
122& $1.625\times 10^{-9}$\\
123& $1.005\times 10^{-9}$\\
124& $7.117\times 10^{-10}$\\
125& $5.454\times 10^{-10}$\\
126& $4.412\times 10^{-10}$\\
127& $3.718\times 10^{-10}$\\
128& $3.239\times 10^{-10}$\\
129& $2.903\times 10^{-10}$\\
130& $2.667\times 10^{-10}$\\
131& $2.504\times 10^{-10}$\\
132& $2.396\times 10^{-10}$\\
133& $2.330\times 10^{-10}$\\
134& $2.296\times 10^{-10}$\\
135& $2.286\times 10^{-10}$\\
136& $2.291\times 10^{-10}$\\
137& $2.304\times 10^{-10}$\\
138& $2.314\times 10^{-10}$\\
139& $2.317\times 10^{-10}$\\
140& $2.309\times 10^{-10}$\\
141& $2.285\times 10^{-10}$\\
142& $2.091\times 10^{-10}$\\
143& $2.060\times 10^{-10}$\\
144& $2.040\times 10^{-10}$\\
145& $2.044\times 10^{-10}$\\
146& $2.080\times 10^{-10}$\\
147& $2.151\times 10^{-10}$\\
148& $2.254\times 10^{-10}$\\
149& $2.382\times 10^{-10}$\\
150& $2.522\times 10^{-10}$\\
151& $2.660\times 10^{-10}$\\
152& $2.782\times 10^{-10}$\\
153& $2.873\times 10^{-10}$\\
154& $2.919\times 10^{-10}$\\
155& $2.912\times 10^{-10}$\\
156& $2.844\times 10^{-10}$\\
157& $2.713\times 10^{-10}$\\
158& $2.519\times 10^{-10}$\\
159& $2.267\times 10^{-10}$\\
160& $1.965\times 10^{-10}$\\
161& $1.624\times 10^{-10}$\\
162& $1.258\times 10^{-10}$\\
163& $8.813\times 10^{-11}$\\
164& $5.128\times 10^{-11}$\\
165& $1.900\times 10^{-11}$\\
166& $4.610\times 10^{-12}$\\
167& $2.033\times 10^{-12}$\\
168& $1.239\times 10^{-12}$\\
169& $8.700\times 10^{-13}$\\
170& $6.615\times 10^{-13}$\\
171& $5.304\times 10^{-13}$\\
172& $4.427\times 10^{-13}$\\
173& $3.818\times 10^{-13}$\\
174& $3.388\times 10^{-13}$\\
175& $3.084\times 10^{-13}$\\
176& $2.872\times 10^{-13}$\\
177& $2.731\times 10^{-13}$\\
178& $2.648\times 10^{-13}$\\
179& $2.610\times 10^{-13}$\\
\end{tabular}
\begin{tabular}{cc}
180& $2.611\times 10^{-13}$\\
181& $2.643\times 10^{-13}$\\
182& $2.697\times 10^{-13}$\\
183& $2.762\times 10^{-13}$\\
184& $2.822\times 10^{-13}$\\
185& $2.859\times 10^{-13}$\\
186& $2.853\times 10^{-13}$\\
187& $2.787\times 10^{-13}$\\
188& $2.661\times 10^{-13}$\\
189& $2.496\times 10^{-13}$\\
190& $2.329\times 10^{-13}$\\
191& $2.197\times 10^{-13}$\\
192& $2.126\times 10^{-13}$\\
193& $2.123\times 10^{-13}$\\
194& $2.190\times 10^{-13}$\\
195& $2.318\times 10^{-13}$\\
196& $2.495\times 10^{-13}$\\
197& $2.706\times 10^{-13}$\\
198& $2.933\times 10^{-13}$\\
199& $3.156\times 10^{-13}$\\
200& $3.356\times 10^{-13}$\\
201& $3.515\times 10^{-13}$\\
202& $3.616\times 10^{-13}$\\
203& $3.648\times 10^{-13}$\\
204& $3.602\times 10^{-13}$\\
205& $3.474\times 10^{-13}$\\
206& $3.266\times 10^{-13}$\\
207& $2.984\times 10^{-13}$\\
208& $2.638\times 10^{-13}$\\
209& $2.242\times 10^{-13}$\\
210& $1.813\times 10^{-13}$\\
211& $1.369\times 10^{-13}$\\
212& $9.274\times 10^{-14}$\\
213& $5.057\times 10^{-14}$\\
214& $1.436\times 10^{-14}$\\
215& $2.509\times 10^{-15}$\\
216& $1.125\times 10^{-15}$\\
217& $6.987\times 10^{-16}$\\
218& $4.971\times 10^{-16}$\\
219& $3.817\times 10^{-16}$\\
220& $3.084\times 10^{-16}$\\
221& $2.590\times 10^{-16}$\\
222& $2.244\times 10^{-16}$\\
223& $2.000\times 10^{-16}$\\
224& $1.827\times 10^{-16}$\\
225& $1.708\times 10^{-16}$\\
226& $1.633\times 10^{-16}$\\
227& $1.594\times 10^{-16}$\\
228& $1.586\times 10^{-16}$\\
229& $1.606\times 10^{-16}$\\
230& $1.652\times 10^{-16}$\\
231& $1.721\times 10^{-16}$\\
232& $1.810\times 10^{-16}$\\
233& $1.909\times 10^{-16}$\\
234& $2.006\times 10^{-16}$\\
235& $2.078\times 10^{-16}$\\
236& $2.096\times 10^{-16}$\\
237& $2.029\times 10^{-16}$\\
238& $1.861\times 10^{-16}$\\
239& $1.614\times 10^{-16}$\\
\end{tabular}
\begin{tabular}{cc}
240& $1.351\times 10^{-16}$\\
241& $1.139\times 10^{-16}$\\
242& $1.006\times 10^{-16}$\\
243& $9.483\times 10^{-17}$\\
244& $9.515\times 10^{-17}$\\
245& $1.005\times 10^{-16}$\\
246& $1.101\times 10^{-16}$\\
247& $1.232\times 10^{-16}$\\
248& $1.389\times 10^{-16}$\\
249& $1.562\times 10^{-16}$\\
250& $1.739\times 10^{-16}$\\
251& $1.910\times 10^{-16}$\\
252& $2.062\times 10^{-16}$\\
253& $2.185\times 10^{-16}$\\
254& $2.268\times 10^{-16}$\\
255& $2.305\times 10^{-16}$\\
256& $2.290\times 10^{-16}$\\
257& $2.221\times 10^{-16}$\\
258& $2.099\times 10^{-16}$\\
259& $1.930\times 10^{-16}$\\
260& $1.720\times 10^{-16}$\\
261& $1.480\times 10^{-16}$\\
262& $1.220\times 10^{-16}$\\
263& $9.527\times 10^{-17}$\\
264& $6.880\times 10^{-17}$\\
265& $4.349\times 10^{-17}$\\
266& $2.008\times 10^{-17}$\\
267& $2.482\times 10^{-18}$\\
268& $4.809\times 10^{-19}$\\
269& $2.475\times 10^{-19}$\\
270& $1.628\times 10^{-19}$\\
271& $1.198\times 10^{-19}$\\
272& $9.413\times 10^{-20}$\\
273& $7.742\times 10^{-20}$\\
274& $6.596\times 10^{-20}$\\
275& $5.787\times 10^{-20}$\\
276& $5.213\times 10^{-20}$\\
277& $4.811\times 10^{-20}$\\
278& $4.545\times 10^{-20}$\\
279& $4.390\times 10^{-20}$\\
280& $4.335\times 10^{-20}$\\
281& $4.372\times 10^{-20}$\\
282& $4.500\times 10^{-20}$\\
283& $4.721\times 10^{-20}$\\
284& $5.041\times 10^{-20}$\\
285& $5.462\times 10^{-20}$\\
286& $5.982\times 10^{-20}$\\
287& $6.576\times 10^{-20}$\\
288& $7.180\times 10^{-20}$\\
289& $7.671\times 10^{-20}$\\
290& $7.855\times 10^{-20}$\\
291& $7.503\times 10^{-20}$\\
292& $6.456\times 10^{-20}$\\
293& $4.816\times 10^{-20}$\\
294& $3.176\times 10^{-20}$\\
295& $2.161\times 10^{-20}$\\
296& $1.699\times 10^{-20}$\\
297& $1.534\times 10^{-20}$\\
298& $1.542\times 10^{-20}$\\
299& $1.680\times 10^{-20}$\\
\end{tabular}
\begin{tabular}{cc}
300& $1.932\times 10^{-20}$\\
301& $2.287\times 10^{-20}$\\
302& $2.734\times 10^{-20}$\\
303& $3.255\times 10^{-20}$\\
304& $3.834\times 10^{-20}$\\
305& $4.446\times 10^{-20}$\\
306& $5.067\times 10^{-20}$\\
307& $5.664\times 10^{-20}$\\
308& $6.205\times 10^{-20}$\\
309& $6.655\times 10^{-20}$\\
310& $6.980\times 10^{-20}$\\
311& $7.155\times 10^{-20}$\\
312& $7.161\times 10^{-20}$\\
313& $6.990\times 10^{-20}$\\
314& $6.646\times 10^{-20}$\\
315& $6.146\times 10^{-20}$\\
316& $5.516\times 10^{-20}$\\
317& $4.791\times 10^{-20}$\\
318& $4.011\times 10^{-20}$\\
319& $3.215\times 10^{-20}$\\
320& $2.436\times 10^{-20}$\\
321& $1.700\times 10^{-20}$\\
322& $1.024\times 10^{-20}$\\
323& $4.186\times 10^{-21}$\\
324& $1.796\times 10^{-22}$\\
325& $3.953\times 10^{-23}$\\
326& $2.149\times 10^{-23}$\\
327& $1.456\times 10^{-23}$\\
328& $1.094\times 10^{-23}$\\
329& $8.744\times 10^{-24}$\\
330& $7.301\times 10^{-24}$\\
331& $6.305\times 10^{-24}$\\
332& $5.602\times 10^{-24}$\\
333& $5.105\times 10^{-24}$\\
334& $4.763\times 10^{-24}$\\
335& $4.548\times 10^{-24}$\\
336& $4.441\times 10^{-24}$\\
337& $4.437\times 10^{-24}$\\
338& $4.534\times 10^{-24}$\\
339& $4.742\times 10^{-24}$\\
340& $5.075\times 10^{-24}$\\
341& $5.560\times 10^{-24}$\\
342& $6.229\times 10^{-24}$\\
343& $7.124\times 10^{-24}$\\
344& $8.275\times 10^{-24}$\\
345& $9.658\times 10^{-24}$\\
346& $1.111\times 10^{-23}$\\
347& $1.226\times 10^{-23}$\\
348& $1.252\times 10^{-23}$\\
349& $1.131\times 10^{-23}$\\
350& $8.285\times 10^{-24}$\\
351& $3.912\times 10^{-24}$\\
352& $1.327\times 10^{-24}$\\
353& $7.591\times 10^{-25}$\\
354& $6.104\times 10^{-25}$\\
355& $5.952\times 10^{-25}$\\
356& $6.676\times 10^{-25}$\\
357& $8.299\times 10^{-25}$\\
358& $1.092\times 10^{-24}$\\
359& $1.460\times 10^{-24}$\\
\end{tabular}
\begin{tabular}{cc}
360& $1.939\times 10^{-24}$\\
361& $2.537\times 10^{-24}$\\
362& $3.275\times 10^{-24}$\\
363& $4.177\times 10^{-24}$\\
364& $5.274\times 10^{-24}$\\
365& $6.595\times 10^{-24}$\\
366& $8.162\times 10^{-24}$\\
367& $9.978\times 10^{-24}$\\
368& $1.201\times 10^{-23}$\\
369& $1.421\times 10^{-23}$\\
370& $1.647\times 10^{-23}$\\
371& $1.869\times 10^{-23}$\\
372& $2.074\times 10^{-23}$\\
373& $2.251\times 10^{-23}$\\
374& $2.390\times 10^{-23}$\\
375& $2.480\times 10^{-23}$\\
376& $2.516\times 10^{-23}$\\
377& $2.491\times 10^{-23}$\\
378& $2.402\times 10^{-23}$\\
379& $2.250\times 10^{-23}$\\
380& $2.037\times 10^{-23}$\\
381& $1.773\times 10^{-23}$\\
382& $1.475\times 10^{-23}$\\
383& $1.166\times 10^{-23}$\\
384& $8.733\times 10^{-24}$\\
385& $6.184\times 10^{-24}$\\
386& $4.125\times 10^{-24}$\\
387& $2.536\times 10^{-24}$\\
388& $1.337\times 10^{-24}$\\
389& $4.339\times 10^{-25}$\\
390& $2.655\times 10^{-27}$\\
391& $7.780\times 10^{-28}$\\
392& $4.614\times 10^{-28}$\\
393& $3.325\times 10^{-28}$\\
394& $2.637\times 10^{-28}$\\
395& $2.220\times 10^{-28}$\\
396& $1.950\times 10^{-28}$\\
397& $1.773\times 10^{-28}$\\
398& $1.660\times 10^{-28}$\\
399& $1.598\times 10^{-28}$\\
400& $1.583\times 10^{-28}$\\
401& $1.618\times 10^{-28}$\\
402& $1.714\times 10^{-28}$\\
403& $1.902\times 10^{-28}$\\
404& $2.237\times 10^{-28}$\\
405& $2.836\times 10^{-28}$\\
406& $3.892\times 10^{-28}$\\
407& $5.582\times 10^{-28}$\\
408& $7.878\times 10^{-28}$\\
409& $1.060\times 10^{-27}$\\
410& $1.361\times 10^{-27}$\\
411& $1.682\times 10^{-27}$\\
412& $2.019\times 10^{-27}$\\
413& $2.371\times 10^{-27}$\\
414& $2.737\times 10^{-27}$\\
415& $3.116\times 10^{-27}$\\
416& $3.509\times 10^{-27}$\\
417& $3.916\times 10^{-27}$\\
\hline
\hline
  \end{tabular}
  \caption{Numerical values of region-region inaccessible negativity in the halo $\mathcal{N}_{A|B}(\sigma_h)$ for $m = 3\times 10^{-3}/d$ presented in Fig.~\ref{fig:boundent}.}
  \label{tab:fig3d20halo}
\end{table}

\end{document}